\documentclass[aps,prx,reprint,superscriptaddress,longbibliography,nofootinbib,floatfix]{revtex4-2}
\usepackage{graphicx,amsmath,amssymb,bm,booktabs,array}
\usepackage{microtype,placeins}
\usepackage[colorlinks=true,allcolors=blue]{hyperref}

\newcommand{\diag}{\operatorname{diag}}
\newcommand{\Gl}{G_{\rm lat}}
\newcommand{\kk}{\mathbf k}
\newcommand{\I}{\mathbb I}
\graphicspath{{figures/}}
\begin{document}
\raggedbottom
\title{Electronic symmetry beyond atomic space groups: A geometric criterion for breaking lattice and hidden symmetries}
\author{Qiu-Shi Huang}
\affiliation{Eastern Institute of Technology, Ningbo 315200, China}
\author{Xin-Gao Gong}
\affiliation{Key Laboratory for Computational Physical Sciences (MOE), State Key Laboratory of Surface Physics, Department of Physics, Fudan University, Shanghai 200433, China}
\author{Su-Huai Wei}
\email{suhuaiwei@eitech.edu.cn}
\affiliation{Eastern Institute of Technology, Ningbo 315200, China}
\begin{abstract}
The atomic space group \(G_{\rm lat}\) underlies the classification of electronic states, band degeneracies, and phase transitions. To what extent does atomic geometry determine electronic site equivalence? We combine group-theoretical analysis with the Landau free energy to describe how interactions control electronic site equivalence. Equivalent sites can spontaneously develop charge imbalance on a fixed lattice, removing the site-exchange generator \(g\) from the electronic symmetry and yielding the charge-domain group \(G_{\rm dom}\), with \(G_{\rm lat}=\langle G_{\rm dom},g\rangle\) (Type I). Conversely, crystallographically inequivalent sites can carry equal projected charges through a hidden unitary exchange generator \(g\); together with the lattice symmetries, it generates the supergroup \(G=\langle G_{\rm lat},g\rangle\) (Type II). Based on these generator relations, we have classified the charge-domain groups formed by the loss of generators (Type I) and the supergroups formed by adding generators (Type II) across all 230 space groups. To determine when charge equivalence becomes unstable in both types, we balance onsite charging cost against intersite Coulomb gain and derive the geometric instability criterion \(r<MR\), with intersite separation \(r\), Madelung constant \(M\), and charging radius \(R\) estimated from tabulated covalent radii. Using first-principles calculations, we verify Type I and Type II in alkali metals with body-centered cubic (BCC) and four-layer close-packed hP4 structures, respectively, and demonstrate the general applicability of the geometric instability criterion \(r<MR\) across both types.
\end{abstract}
\maketitle

\section{Introduction}
\label{sec:introduction}
Crystallographic space-group symmetry \(G_{\rm lat}\) organizes the classification of electronic bands, symmetry-enforced degeneracies, topological indicators, phase transitions, and order parameters \cite{ref34,ref35,ref36,ref37,ref11,bradlyn2016fermions,po2017indicators,kruthoff2017combinatorics}. Atomic species and their spatial arrangement determine this symmetry. The sites related to a given atom by operations of \(G_{\rm lat}\) form its Wyckoff orbit (WO). Wyckoff positions (WPs) classify these WOs and provide the real-space starting point for band-representation theory \cite{ref38,bradlyn2017tqc,zak1980bands}. The usual expectation is therefore that symmetry-related sites are electronically identical and inequivalent sites are electronically distinct.

The electronic symmetry of a crystal usually reproduces the site-equivalence relations of \(G_{\rm lat}\). For a nondegenerate ground state, observables such as the charge density inherit the crystal symmetry, consistent with Neumann's principle \cite{ref39}. However, the electronic symmetry realized by a crystal also depends on the electronic interactions between its atomic sites~\cite{perdew2021symmetry,wigner1934electrons,varignon2019gaps,trimarchi2018polymorphous}. Thus, the symmetries of \(G_{\rm lat}\) alone do not determine which sites are electronically equivalent.

We find that electronic equivalence can depart from crystallographic equivalence in two opposite directions. In Type I, charge transfer on the fixed atomic framework makes formerly equivalent sites carry unequal charges, spontaneously breaking the symmetry associated with the site-exchange generator \(g\). The resulting charge-domain group (CDG) satisfies \(G_{\rm lat}=\langle G_{\rm dom},g\rangle\) \cite{ref31,ref32,overhauser1968,ref1,wang2014honeycomb,chen2023nickelate}. In Type II, a hidden exchange generator \(g\) absent from \(G_{\rm lat}\) relates crystallographically inequivalent sites. Together with the lattice symmetries, it generates the supergroup \(G=\langle G_{\rm lat},g\rangle\). For example, in four-layer ABAC stacking, sites belonging to distinct WOs carry equal charges, while a band doublet at \(\Gamma\) appears accidentally degenerate when classified using \(G_{\rm lat}\) alone.

Previous work has established systematic classifications linking crystal symmetry to electronic properties \cite{bradlyn2017tqc,po2017indicators,kruthoff2017combinatorics,ref34,song2018mappings,vergniory2019catalogue,tang2019search,chen2024spingroups,xiao2024spingroups,liu2026oriented}. However, an atomic space-group classification alone cannot determine whether unequal charges on symmetry-related sites lower the energy or whether hidden symmetries within an orbital manifold enforce equal charges on inequivalent sites and protect band doublets~\cite{rontgen2021,hou2013hidden,hou2015honeycomb}.

To address this gap, we \textit{(i)} have systematically classified the relations across all 230 lattice space groups by the types of the parent group, resulting group, and broken or added generator, obtaining 3746 classes of CDG relations for Type I and 3721 classes of supergroup relations for Type II. For each class, we provide the explicit generator, the embedding between the parent and resulting groups, and the group index (Supplementary Tables~S5 and S6 and Supplementary Data~1 and 2). \textit{(ii)} To determine when charge equivalence becomes unstable in both types, we derive the geometric criterion \(r<MR\) for the onset of charge transfer. With the atomic positions fixed, the intersite Coulomb gain exceeds the onsite charging cost when the interatomic separation falls below the product of the Madelung constant of the charge pattern and the effective charging radius. The equal-charge state then becomes unstable, spontaneously breaking the exchange symmetry \(g\), a lattice symmetry in Type I and a hidden symmetry in Type II. \textit{(iii)} We use first-principles calculations on body-centered cubic (BCC) alkali metals to test Type I through pressure-induced charge transfer and the associated breaking of the body-centering translation symmetry \(t_c\). The calculated onset separations agree with the boundary \(r=MR\) of this criterion to within \(6\%\). For Type II, we explain equal site charges and band degeneracy in hP4 (ABAC) alkali metals by extending the layer gauge transformation for close-packed stackings from the short-range, single-\(s\)-orbital model of He \textit{et al.}~\cite{ref10} to models that include axial orbitals of arbitrary angular momentum \((s,p_z,d_{z^2},f_{z^3},\ldots)\) and their hybridization. Using this transformation, we transport symmetry operations from the ABAB parent structure to ABAC and explicitly construct two hidden symmetries~\cite{ref47}: the exchange \(g\), which enforces equal charges on distinct WOs, and the antiunitary symmetry \(\widetilde\Theta\), which protects band doublets. We test the same criterion \(r<MR\) for the loss of this charge equivalence with first-principles calculations. For ABAC-stacked Li, Na, and K, the calculated separations at the onset of charge transfer under compression differ from the predicted critical separation \(MR\) by at most \(7.1\%\) (Supplementary Table~S2). Charge transfer breaks these symmetries and splits the doublets, reducing the overlap between valence and conduction bands and providing a symmetry explanation for the well-known metal--insulator transition in sodium at high pressure~\cite{ref9}.

\section{Results}
\label{sec:results}

\subsection{Charge transfer and the stability of electronic equivalence}
\label{sec:charging}
\begin{figure}[!htbp]
 \centering
 \includegraphics[width=\linewidth]{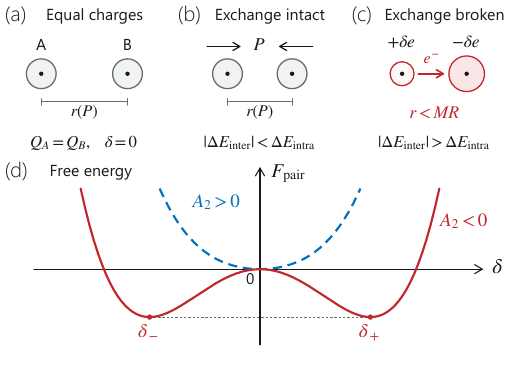}
 \caption{\textbf{Loss of site equivalence through charge transfer.} (a) Two sites with equal charges. (b,c) Onsite charging cost and intersite Coulomb gain before and after the instability. In (c), the circle sizes illustrate electron loss and gain. Both sites have the same charging radius $R$ in the energy estimate. (d) Schematic free energy per pair relative to the charge-equivalent state at each pressure $P$. The site separation is $r(P)$.}
 \label{fig:model}
\end{figure}

Consider two sites $A$ and $B$ exchanged by a symmetry of the charge-equivalent state. Their net charges are equal [Fig.~\ref{fig:model}(a)]. With the nuclei fixed, we consider electron transfer from $A$ to $B$. The net charge difference is $\Delta Q=Q_A-Q_B$. In units of the elementary charge $e>0$, the transfer per site is
\begin{equation}
 \delta=\frac{\Delta Q}{2e}.
 \label{eq:charge}
\end{equation}

The charge-equivalent state has $\delta=0$, and its stability is set by the competition between the onsite charging cost from the Coulomb interaction $U$ and the energy gain from the intersite Coulomb interaction between opposite charge deviations~\cite{hirsch1984cdw} [Fig.~\ref{fig:model}(b,c)]. At pressure $P$, the free-energy change per pair has the fourth-order expansion \cite{ref31,ref32,toledano1987landau,mcmillan1975cdw}
\begin{equation}
 F_{\rm pair}(\delta,P)\simeq A_2\delta^2+A_4\delta^4,\qquad A_4>0.
 \label{eq:landau}
\end{equation}
When the intersite gain overtakes the onsite cost, $A_2$ changes from positive to negative and the minimum at $\delta=0$ splits into two equal-depth minima at opposite charge transfers $\delta_\pm$. Selecting either minimum spontaneously breaks the site-exchange symmetry [Fig.~\ref{fig:model}(d)].

To estimate the quadratic coefficient $A_2$, we approximate the valence charge cloud on each site by a spherical shell of charging radius $R$. Let $r$ be the nearest separation between sites with opposite charge deviations and $M$ the Madelung constant of their arrangement. The net quadratic coefficient per pair is $A_2\simeq\kappa e^2(1/R-M/r)$, where $\kappa$ is the Coulomb constant (Supplementary Information, Note~3). Introducing the dimensionless intersite separation $f$, we derive the geometric criterion for instability of the charge-equivalent state,
\begin{equation}
 r<MR \quad\Longleftrightarrow\quad f<1,\qquad f\equiv\frac{r}{MR}.
 \label{eq:criterion}
\end{equation}
At $f=1$ ($r=MR$), the onsite charging cost and the intersite Coulomb gain cancel to quadratic order in $\delta$. Further compression reduces the site separation $r$, so that $f<1$ and $A_2<0$. The system then evolves from the charge-equivalent state in Fig.~\ref{fig:model}(b) to the charge-transfer state in Fig.~\ref{fig:model}(c), corresponding to the double-well free energy in Fig.~\ref{fig:model}(d). Our density functional theory (DFT)~\cite{hohenberg1964,kohn1965} calculations below show that charge imbalance sets in near $f=1$ ($0.94$--$1.07$) in the five BCC alkali metals and in four-layer close-packed Li, Na, and K, validating the predictive power of this simple geometric criterion.

\subsection{Type I: lattice symmetry breaking}
\label{sec:typeI}
In Type I, a site-exchange generator of the atomic space group $\Gl$ relates sites with equal charges. When $r<MR$, charge transfer breaks this symmetry at fixed atomic positions, making the sites electronically inequivalent.

\begin{table*}[t]
 \caption{\textbf{Generators of the BCC lattice group and its charge-domain group.} The rotations and inversion are centered at the corner site $A$. A check mark denotes a selected generator. Charge order breaks the body-centering translation $t_c$ while preserving the other listed generators.}
 \label{tab:bcc-generators}
 \centering
 \begin{ruledtabular}
 \begin{tabular}{lccccc}
 Group & Lattice translations & $C_{4z}$ & $C_{3[111]}$ & $\mathcal I_0$ & $t_c$\\
 \hline
 BCC lattice, $\Gl=Im\bar3m$ & $T_x,T_y,T_z$ & $\checkmark$ & $\checkmark$ & $\checkmark$ & $\checkmark$\\
 Charge order, $G_{\rm dom}=Pm\bar3m$ & $T_x,T_y,T_z$ & $\checkmark$ & $\checkmark$ & $\checkmark$ & ---\\
 \end{tabular}
 \end{ruledtabular}
\end{table*}

\subsubsection{Group-theoretical classification}
\label{sec:classification}
Let $i$ and $j$ lie on the same WO of the atomic space group $\Gl$. A symmetry operation $g=\{\mathcal R\mid\mathbf t\}\in\Gl$ relates their positions~\cite{ref38},
\[
 \mathbf r_j=g\mathbf r_i=\mathcal R\mathbf r_i+\mathbf t .
\]
Here $\mathcal R$ is the point operation and $\mathbf t$ its accompanying translation. In the charge-equivalent state the two sites carry equal charges. When the system selects one of the free-energy minima at $\delta\ne0$, charge transfer breaks the site-exchange symmetry $g$. The operations that preserve the charge pattern form the CDG $G_{\rm dom}$~\cite{stokes1988isotropy}. Adding the broken generator $g$ restores the parent group,
\begin{equation}
 \Gl=\langle G_{\rm dom},g\rangle,\qquad g\notin G_{\rm dom}.
 \label{eq:state-group}
\end{equation}
For all 230 atomic space groups, we classify the relations in Eq.~\eqref{eq:state-group} by the broken generator $g$ and the resulting CDG $G_{\rm dom}$ (Supplementary Table~S5).

\subsubsection{BCC alkali metals}
\label{sec:bcc}
\begin{figure}[!htbp]
 \centering
 \includegraphics[width=\linewidth]{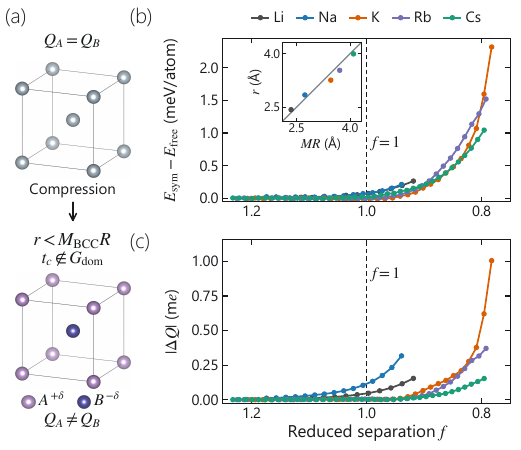}
 \caption{\textbf{Charge order in BCC alkali metals.} (a) The corner $A$ and body center $B$ carry equal charges before the instability (top) and unequal charges after it (bottom). (b) Energy gain per atom upon releasing the body-centering constraint, $E_{\rm sym}-E_{\rm free}$. (c) Charge difference $|\Delta Q|$ between corner and body-center sites, in units of $10^{-3}e$. Both quantities are plotted against $f=r/(M_{\rm BCC}R)$. The dashed line marks the predicted boundary $f=1$. The diagonal in the inset of (b) marks equality between the calculated onset distance $r$ and $M_{\rm BCC}R$.}
 \label{fig:bcc}
\end{figure}

The BCC alkali metals illustrate how breaking the body-centering translation generator $g=t_c$ removes charge equivalence between sites on the same WO. In $\Gl=Im\bar3m$, the corner $A$ and body center $B$ of the conventional cubic cell belong to the same $2a$ WO. In conventional fractional coordinates, the body-centering translation $t_c$ that exchanges $A$ and $B$ acts on their charge difference $\Delta Q=Q_A-Q_B$ as
\begin{equation}
 t_c=\left\{E\,\middle|\,\tfrac12,\tfrac12,\tfrac12\right\},\qquad
 t_c:\Delta Q\mapsto-\Delta Q.
 \label{eq:centering}
\end{equation}
Here $E$ is the identity operation. According to Eq.~\eqref{eq:criterion}, $\Delta Q$ becomes nonzero when compression brings the system to $f<1$. The corner and body-center sublattices then carry opposite charge deviations $+e\delta$ and $-e\delta$. Since $t_c$ reverses $\delta$, selecting either charge-transfer state breaks this translation symmetry and doubles the primitive-cell volume [Fig.~\ref{fig:bcc}(a)]. The system retains the cubic point group, and the parent $2a$ WO splits into a WO of corner sites and a WO of body-center sites. The retained generators form the index-two CDG $G_{\rm dom}=Pm\bar3m$ (Table~\ref{tab:bcc-generators}), with
\begin{equation}
 \Gl=\langle G_{\rm dom},t_c\rangle.
 \label{eq:bccgroup}
\end{equation}

To apply Eq.~\eqref{eq:criterion} to the alternating corner and body-center charge pattern, we use its CsCl-type Madelung constant $M_{\rm BCC}=1.762675$ \cite{ref16,ref26,ref27} and estimate $R$ from tabulated covalent radii \cite{ref14,ref15}. For Li, Na, K, Rb, and Cs, we express the corner--body-center separation and the dimensionless separation $f$ in terms of the conventional cubic lattice constant $a$,
\begin{equation}
 r=\frac{\sqrt3}{2}a,\qquad f=\frac{\sqrt3\,a}{2M_{\rm BCC}R}.
 \label{eq:f}
\end{equation}

We compress each BCC lattice in steps and compare the DFT energies per atom with and without the $t_c$ constraint at the same ionic positions. The one-atom primitive cell enforces the body-centering translation $t_c$, giving $Q_A=Q_B$ and energy $E_{\rm sym}$. In the conventional two-atom cubic cell, we release the $t_c$ constraint and allow the two site charges to relax independently, obtaining $E_{\rm free}$. At large $f$ the relaxed charges are equal within numerical resolution. Near $f=1$, we find a finite $|\Delta Q|$ in all five metals [Fig.~\ref{fig:bcc}(c)] and a lower energy for the charge-ordered state, $E_{\rm free}<E_{\rm sym}$ [Fig.~\ref{fig:bcc}(b)].

\subsection{Type II: hidden symmetry and site equivalence}
\label{sec:typeII}
In Type I, charge transfer removes a site-exchange generator from the electronic symmetry, as expressed by Eq.~\eqref{eq:state-group}. In Type II, adjoining a new generator $g\notin\Gl$ to $\Gl$ gives the supergroup
\begin{equation}
 G=\left\langle\Gl,g\right\rangle.
 \label{eq:electronic-supergroup}
\end{equation}
For all 230 lattice space groups, we have classified the generator relations in Eq.~\eqref{eq:electronic-supergroup} by the added spatial generator $g$ and the resulting supergroup $G$ (Supplementary Table~S6). When the added operation $g$ is a unitary symmetry of the electronic Hamiltonian, this generator relation also defines the electronic supergroup. For site equivalence, we consider the case in which $g$ exchanges two distinct WOs, $\mathcal O_1$ and $\mathcal O_2$:
\[
 g\mathcal O_1=\mathcal O_2,\qquad g\mathcal O_2=\mathcal O_1.
\]
In an electronic state that preserves this exchange, sites on the two WOs have equal projected charges.

\subsubsection{Gauge equivalence of ABAB and ABAC}
\label{sec:multiorbital}
\begin{figure*}[tbp]
 \centering
 \includegraphics[width=0.984\textwidth]{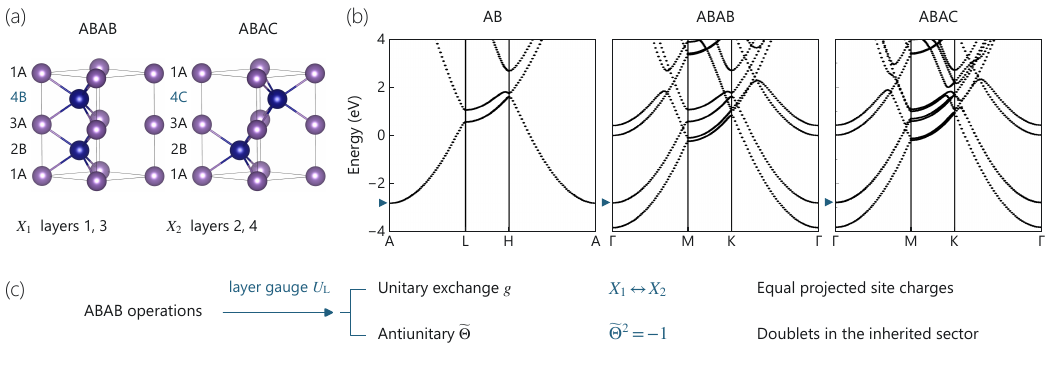}
 \caption{\textbf{Band folding and hidden symmetries in four-layer close packing.} (a) ABAB and ABAC cells. In ABAC, purple marks $X_1$ (layers 1 and 3) and blue marks $X_2$ (layers 2 and 4). The top A layer starts the next period. (b) Calculated Na bands at 0~GPa along A--L--H--A in AB and $\Gamma$--M--K--$\Gamma$ in ABAB and ABAC. Arrowheads mark the AB doublet at A and its counterparts at $\Gamma$. The pressure--distance correspondence follows from the equation of state, $P=-\partial E/\partial V$, and the four-layer geometry, $r(P)=[a(P)^2/3+c_4(P)^2/16]^{1/2}$, where $a(P)$ and $c_4(P)$ are the basal lattice constant and four-layer period at pressure $P$. (c) In the axial multiorbital model, ABAC inherits the unitary exchange $g$ and the antiunitary symmetry $\widetilde\Theta$ from ABAB through the layer gauge $U_{\rm L}$.}
 \label{fig:mechanism}
\end{figure*}

Four-layer close packing in alkali metals illustrates how a hidden symmetry can relate sites on distinct WOs. A, B, and C denote the three lateral positions of a close-packed layer. The two-layer AB stack repeats as ABAB in a four-layer cell, with all sites on one WO. Moving the fourth layer from B to C gives the ABAC structure (hP4) reported in compressed Na and K~\cite{ref9,ref44,ref25} [Fig.~\ref{fig:mechanism}(a)]. In its atomic space group $\Gl=P6_3/mmc$, layers 1 and 3 form the WO represented by $X_1=(0,0,0)$, and layers 2 and 4 form the WO represented by $X_2=(1/3,2/3,1/4)$.

Our calculations for hP4 Na reveal three features that the atomic space group $\Gl$ alone cannot explain. At low pressure, (i)~the band dispersions of ABAC and ABAB are essentially the same [Fig.~\ref{fig:mechanism}(b)]. (ii)~The sites $X_1$ and $X_2$ belong to distinct WOs, yet carry equal charges. (iii)~In the ABAC band structure, the $\Gamma_3^-$ and $\Gamma_4^+$ states appear degenerate at $\Gamma$ [Fig.~\ref{fig:pressure}(a,b)]. They belong to different one-dimensional irreducible representations, so their degeneracy appears accidental when classified by $\Gl$. To trace these features to their common origin, we examine how shifting layer 4 from B to C (as shown in Fig.~\ref{fig:mechanism}(a)) changes the hopping between adjacent layers.

Let $\mathbf a_1$ and $\mathbf a_2$ be basal lattice vectors of length $a$ at $120^\circ$. The lateral shifts from A to B and from A to C are $\mathbf b=(\mathbf a_1+2\mathbf a_2)/3$ and $-\mathbf b$, modulo lattice translations. These two stacking senses define the chirality $+$ or $-$ of an upward bond. Including the bond to the next cell, the signs are $(+,-,+,-)$ in ABAB and $(+,-,-,+)$ in ABAC. Both have zero net chirality.

He \textit{et al.} showed that stackings of equal net chirality are gauge equivalent for one $s$ orbital per site with short-range hopping~\cite{ref10}. We extend this result to a gauge-equivalence theorem for multiple orbitals per site. For identical layers coupled only to adjacent layers, two stackings of equal net chirality in a common cell are gauge equivalent if reversing the stacking sense multiplies every element of the interlayer hopping matrix by the same phase factor. At Bloch wavevector $\kk$, this condition on the two forward hopping matrices reads
\begin{equation}
 T_+(\kk)=e^{2i\phi(\kk)}T_-(\kk).
 \label{eq:common-phase}
\end{equation}

Axial orbitals have angular form $Y_\ell^0$ and are invariant under rotation about the stacking axis. The three nearest bonds to the next layer differ only in azimuth, so they have the same two-center hopping matrix $\mathsf B$~\cite{slater1954}. Their lateral displacements $\boldsymbol\rho_\alpha$ enter through the in-plane wavevector $\kk_\parallel$. For layer spacing $h$, the Bloch sums are
\begin{equation}
 \begin{aligned}
 T_+(\kk)&=e^{ik_zh}S(\kk_\parallel)\mathsf B,\\
 T_-(\kk)&=e^{ik_zh}S^*(\kk_\parallel)\mathsf B,\\
 S(\kk_\parallel)&=\sum_{\alpha=1}^{3}e^{i\kk_\parallel\cdot\boldsymbol\rho_\alpha}
                  =|S|e^{i\phi}.
 \end{aligned}
 \label{eq:factorization}
\end{equation}
Reversing the lateral displacements conjugates $S$ and leaves $\mathsf B$ unchanged. We thus establish the phase relation in Eq.~\eqref{eq:common-phase} for axial orbitals of any angular momentum, $(s,p_z,d_{z^2},f_{z^3},\ldots)$, including hopping between different orbitals. Supplementary Note~6 gives the full derivation.

For stackings of equal net chirality, the bond phase differences sum to zero over a period and the layer phases can be chosen periodically. In ABAB and ABAC the two bonds that change meet at layer 4, so rephasing that layer alone relates the Hamiltonians. For $N$ axial orbitals per site, the layer gauge $U_{\rm L}$ gives
\begin{equation}
 \begin{aligned}
 U_{\rm L}(\kk)&=\diag(1,1,1,e^{-2i\phi(\kk)})\otimes\I_N,\\
 H_{\rm ABAC}(\kk)&=U_{\rm L}^\dagger(\kk)H_{\rm ABAB}(\kk)U_{\rm L}(\kk).
 \end{aligned}
 \label{eq:gauge}
\end{equation}
Here $\I_N$ acts on the $N$ axial orbitals at one site. The two stackings thus have identical model bands at every $\kk$ (Supplementary Note~6). This gauge equivalence explains why the calculated ABAC and ABAB bands in observation~(i) are so close.

\subsubsection{Hidden site exchange and charge equivalence}
\label{sec:gauge-map}
To explain equal charges on distinct WOs, we use the layer gauge in Eq.~\eqref{eq:gauge} to carry the inversion $\mathcal I_{12}$ that exchanges adjacent layers in ABAB into an electronic symmetry of ABAC. The two stackings have the same space-group type $P6_3/mmc$ but different symmetry generators (Table~\ref{tab:stack-generators}). With the origin on layer~1, ABAB is represented in a four-layer cell but its lattice repeats after $c_2=2h$, whereas ABAC repeats after $c_4=4h$. In ABAB, inversion $\mathcal I_{12}$ about the midpoint of neighboring sites on layer~1 ($X_1$) and layer~2 ($X_2$) exchanges layers $1(X_1)\leftrightarrow2(X_2)$ and $3(X_1)\leftrightarrow4(X_2)$. It reverses displacements from its center, so its point matrix is $-\I_3$, with $\I_3$ the three-dimensional identity. In four-layer coordinates,
\begin{equation}
 \mathcal I_{12}=\{-\I_3\mid\mathbf b+h\hat{\mathbf z}\}
   =\left\{-\I_3\,\middle|\,\tfrac13,\tfrac23,\tfrac14\right\}.
 \label{eq:g}
\end{equation}

Moving layer 4 from B to C produces ABAC, whose atomic space group $\Gl=P6_3/mmc$ lacks $\mathcal I_{12}$. Its inversion about the origin, $\mathcal I_0=\{-\I_3\mid\mathbf 0\}$, preserves the two WOs separately and cannot exchange them. The ABAC entry in Supplementary Table~S6 classifies the group extension by the spatial operation $\mathcal I_{12}$. To construct the full electronic exchange, we retain the layer-gauge phases: the transformation $U_{\rm L}$ in Eq.~\eqref{eq:gauge} carries $\mathcal I_{12}$ from ABAB into the hidden unitary exchange $g$ acting on ABAC electronic states [Fig.~\ref{fig:mechanism}(c)]. Writing the inversion matrix in the layer-orbital basis as $\mathcal I_{12}(\kk)$, we obtain
\begin{equation}
 \begin{aligned}
 g(\kk)&=U_{\rm L}^\dagger(-\kk)\mathcal I_{12}(\kk)U_{\rm L}(\kk),\\
 g(\kk)H_{\rm ABAC}(\kk)g^\dagger(\kk)&=H_{\rm ABAC}(-\kk).
 \end{aligned}
 \label{eq:exchange}
\end{equation}

The layer gauge $U_{\rm L}$ changes only the phases of the electronic wavefunctions and preserves their weight on each layer. The exchange $g$ inherited from $\mathcal I_{12}$ therefore interchanges the electronic states on the WOs of $X_1$ and $X_2$. Equation~\eqref{eq:exchange} maps a state at $\kk$ to an equal-energy state at $-\kk$ with the two WO weights exchanged. In a state that preserves $g$, these corresponding states have equal occupations. Summing over all occupied states across the Brillouin zone (BZ) gives equal total charges on the two WOs. Since each WO contains two sites, the projected charges per site satisfy
\begin{equation}
 Q_1^{\rm proj}=Q_2^{\rm proj}.
 \label{eq:projected-charges}
\end{equation}
The hidden exchange $g$ thus explains why the two atomic spheres in observation~(ii) carry equal charges.
Combining the hidden exchange $g$ with the lattice generators in Table~\ref{tab:stack-generators} gives the electronic supergroup of Eq.~\eqref{eq:electronic-supergroup},
\begin{equation}
 \begin{aligned}
 G&=\left\langle T_{\mathbf a_1},T_{\mathbf a_2},T_{c_4},
 C_{3z},m_v,m_h,\mathcal I_0,g\right\rangle\\
 &=\langle\Gl,g\rangle.
 \end{aligned}
 \label{eq:group-extension}
\end{equation}

\begin{table*}[t]
 \caption{\textbf{Generators of the ABAB and ABAC lattice groups and the ABAC electronic supergroup.} A check mark denotes a selected generator. $C_{3z}$ passes through the origin, $m_v$ contains $\mathbf b$ and $\hat{\mathbf z}$, and $m_h$ passes through layer~2.}
 \label{tab:stack-generators}
 \centering
 \begin{ruledtabular}
 \begin{tabular}{lccccccc}
 Group & Lattice translations & $C_{3z}$ & $m_v$ & $m_h$ & $\mathcal I_0$ & $\mathcal I_{12}$ & $g$\\
 \hline
 ABAB lattice, $P6_3/mmc$ & $T_{\mathbf a_1},T_{\mathbf a_2},T_{c_2}$ & $\checkmark$ & $\checkmark$ & $\checkmark$ & --- & $\checkmark$ & ---\\
 ABAC lattice, $P6_3/mmc$ & $T_{\mathbf a_1},T_{\mathbf a_2},T_{c_4}$ & $\checkmark$ & $\checkmark$ & $\checkmark$ & $\checkmark$ & --- & ---\\
 ABAC electronic, $G$ & $T_{\mathbf a_1},T_{\mathbf a_2},T_{c_4}$ & $\checkmark$ & $\checkmark$ & $\checkmark$ & $\checkmark$ & --- & $\checkmark$\\
 \end{tabular}
 \end{ruledtabular}
\end{table*}

\subsubsection{Symmetry protection of band doublets}
\label{sec:doublet-protection}
The degenerate pair at $\Gamma$ in ABAC can be traced to a doublet at the zone-boundary point A of two-layer AB. In the four-layer ABAB cell, A folds to $\Gamma$ [Fig.~\ref{fig:mechanism}(b)].
The horizontal mirror $m_h$ through layer~2 sends $z$ to $2h-z$ in both stackings. In ABAB, $m_h\mathcal I_{12}$ gives the screw $S_{2z}=\{C_{2z}\mid\mathbf b+h\hat{\mathbf z}\}$, which rotates by $\pi$ about the stacking axis and advances by one layer. Together with spinless time reversal $\mathcal T$, $S_{2z}$ gives the antiunitary symmetry $\Theta$,
\begin{equation}
 \Theta=S_{2z}\mathcal T,\qquad \Theta^2=T_{c_2},\qquad c_2=2h,
 \label{eq:theta-parent}
\end{equation}
where $T_{c_2}$ translates by one AB period. On the boundary plane $k_z=\pi/c_2$, $\Theta$ preserves each momentum modulo a reciprocal lattice vector and $T_{c_2}$ has eigenvalue $e^{-ik_zc_2}=-1$. Each boundary state therefore has an orthogonal partner at the same energy in a single spin channel~\cite{wu2018nodalsurface,young2015dirac,parameswaran2013nonsymmorphic,wieder2018wallpaper,herring1937time}.

The layer gauge $U_{\rm L}$ in Eq.~\eqref{eq:gauge} maps $\mathcal I_{12}$ to $g$ while preserving $m_h$ and $\mathcal T$. It therefore carries the ABAB symmetry $\Theta=m_h\mathcal I_{12}\mathcal T$ into the ABAC antiunitary symmetry $\widetilde\Theta$. Since $\Theta^2=T_{c_2}$, the same transformation gives $\widetilde\Theta^2=\widetilde T_{c_2}$, where $\widetilde T_{c_2}$ is the hidden operation inherited from the AB two-layer translation $T_{c_2}$.
Expressing AB in the four-layer cell with $c_4=2c_2$ folds its boundary path A--L--H--A onto the basal path $\Gamma$--M--K--$\Gamma$ [Fig.~\ref{fig:mechanism}(b)]. This change of unit cell relabels momentum without changing the eigenvalue $-1$ under $T_{c_2}$. The unitary gauge $U_{\rm L}$ then maps both the states and the translation to ABAC while preserving that eigenvalue. We denote these ABAC states by $|u_-\rangle$, with the subscript indicating their eigenvalue $-1$ under $\widetilde T_{c_2}$, and obtain
\begin{equation}
 \begin{gathered}
 \widetilde\Theta=m_hg\mathcal T\in\langle G,\mathcal T\rangle,\\
 \widetilde\Theta^2|u_-\rangle=\widetilde T_{c_2}|u_-\rangle=-|u_-\rangle.
 \end{gathered}
 \label{eq:sector}
\end{equation}
On the basal plane of the four-layer BZ, $|u_-\rangle$ and $\widetilde\Theta|u_-\rangle$ have the same momentum and energy (Supplementary Note~8). Antiunitarity and Eq.~\eqref{eq:sector} make them orthogonal,
\[
 \langle u_-|\widetilde\Theta u_-\rangle=0.
\]
The antiunitary symmetry $\widetilde\Theta$ thus protects the ABAC doublet inherited from the AB zone boundary, explaining the seemingly accidental degeneracy at $\Gamma$ in observation~(iii).

\subsubsection{Charge transfer and doublet splitting under pressure}
\label{sec:breaking}
When compression brings the system to $r<MR$ ($f<1$), Eq.~\eqref{eq:criterion} predicts charge transfer between $X_1$ and $X_2$. The charge redistribution changes the local effective potentials at the two sites. To first order in the charge imbalance, their onsite energy difference is
\begin{equation}
 \Delta\varepsilon\simeq\lambda\left(Q_1^{\rm proj}-Q_2^{\rm proj}\right),
 \label{eq:charge-potential}
\end{equation}
where $\lambda$ is the linear response coefficient relating the onsite energy difference to the charge imbalance. Both the site exchange $g$ and the antiunitary symmetry $\widetilde\Theta=m_hg\mathcal T$ exchange $X_1$ and $X_2$. A nonzero onsite potential difference $\Delta\varepsilon$ therefore breaks both symmetries, lifting the protection of the folded doublets (Supplementary Note~9). Charge transfer preserves inversion $\mathcal I_0$ about the ABAC origin. Under this potential, the opposite-parity states $\Gamma_3^-$ and $\Gamma_4^+$ shift in energy without mixing at $\Gamma$. Their weights on $X_1$ and $X_2$ determine their first-order energy shifts, whose difference gives the doublet splitting (Supplementary Note~9).

\begin{figure}[!htbp]
 \centering
 \includegraphics[width=\linewidth]{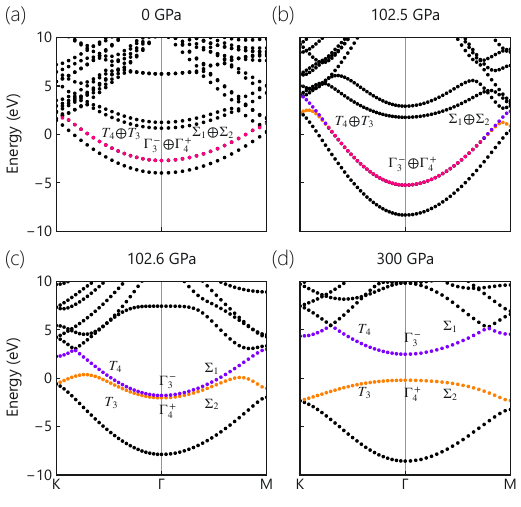}
 \caption{\textbf{Doublet splitting in hP4 Na under pressure.} (a--d) Calculated bands along K--$\Gamma$--M at 0, 102.5, 102.6, and 300~GPa. Zero energy is the Fermi level at each pressure. Magenta marks the band pair corresponding to the model doublet. Purple and orange trace the two branches after splitting. $T$, $\Gamma$, and $\Sigma$ label the irreducible representations along K--$\Gamma$, at $\Gamma$, and along $\Gamma$--M, respectively.}
 \label{fig:pressure}
\end{figure}
\FloatBarrier

Figure~\ref{fig:pressure} follows this evolution in hP4 Na under compression. At 102.5~GPa, the charge difference between the two atomic spheres is $1.33\times10^{-4}e$, and $\Gamma_3^-$ and $\Gamma_4^+$ appear degenerate [Fig.~\ref{fig:pressure}(b)]. At 102.6~GPa, the charge difference rises sharply to $1.22\times10^{-2}e$ and the splitting at $\Gamma$ increases. Both branches still cross the Fermi level along K--$\Gamma$--M [Fig.~\ref{fig:pressure}(c)]. By 300~GPa the branch containing $\Gamma_4^+$ lies below the Fermi level and the one containing $\Gamma_3^-$ lies above it throughout this path [Fig.~\ref{fig:pressure}(d)]. This evolution provides a symmetry-based explanation of the well-known metal--insulator transition in high-pressure Na~\cite{ref9,ref23,ref24}. This symmetry mechanism also suggests that other hexagonal close-packed structures may undergo similar electronic symmetry transitions at high pressure.

\section{Discussion}
\label{sec:discussion}
Our work shows that electronic interactions can change which sites of a crystal are equivalent even when the atoms remain fixed. Using generator relations, we have systematically classified the charge-domain groups and supergroups associated with electronic symmetry reduction and extension, respectively. We derive the geometric criterion $r<MR$ for the instability of the charge-equivalent state [Eq.~\eqref{eq:criterion}]. When $r<MR$, the intersite Coulomb gain exceeds the onsite charging cost and charge transfer lowers the energy. The charge-transfer onset separations obtained from DFT under compression differ from the predictions of this criterion by at most $5.8\%$ for the five BCC alkali metals and $7.1\%$ for Li, Na, and K in the hP4 (ABAC) structure. The same criterion applies to lattice-enforced site equivalence (Type I) and hidden site equivalence (Type II). In Type I, exemplified by BCC, we find that spontaneous symmetry breaking under high pressure removes the generator $t_c$ from the electronic symmetry, producing a charge-domain group $G_{\rm dom}$ with unequal charges on the corner and body center, where $\Gl=\langle G_{\rm dom},t_c\rangle$. In Type II, exemplified by ABAC, we find a hidden unitary exchange generator $g$ between different layers of the close-packed structure. The exchange $g$ enforces equal site charges and, together with $\Gl$, generates the electronic supergroup $G=\langle\Gl,g\rangle$. Combining $g$ with the lattice mirror $m_h$ and spinless time reversal $\mathcal T$ gives the antiunitary symmetry $\widetilde\Theta=m_hg\mathcal T\in\langle G,\mathcal T\rangle$, which protects the band doublet. These symmetries explain the charge equivalence and band degeneracy that appear accidental under $\Gl$. This implies that relaxing charge-density symmetry constraints in first-principles calculations when $r<MR$ can reveal lower-energy charge-ordered states, as in BCC, while including hidden symmetries beyond $\Gl$ in band classification can identify protected multiplets, as in ABAC. Atomic positions set the crystallographic baseline. Interactions determine whether the realized electronic symmetry preserves, reduces, or extends it. Applying this geometric criterion and hidden-symmetry analysis to materials beyond alkali metals may reveal new charge-ordered phases and band multiplets protected by hidden symmetries.

\section{Methods}
\label{sec:methods}
\subsection{Electronic-structure calculations}
We used the full-potential (linearized) augmented plane wave plus local orbitals method~\cite{andersen1975linear,sjostedt2000apw,madsen2001apw} in WIEN2k~\cite{ref17,ref18} with the Perdew--Burke--Ernzerhof functional~\cite{ref46}. The all-electron calculations were nonmagnetic and scalar-relativistic, with spin--orbit coupling omitted. Site populations include both spin channels. For both structures the basis cutoff was $R_{\rm MT}K_{\max}=7$, where $R_{\rm MT}$ is the muffin-tin (MT) radius and $K_{\max}$ is the cutoff wavevector for the plane-wave expansion of the wavefunctions. The charge-density Fourier expansion used the cutoff wavevector $G_{\max}=16~\mathrm{bohr}^{-1}$. The energy and charge convergence thresholds were $10^{-8}$~Ry and $10^{-7}e$.

For each BCC metal the constrained and unconstrained calculations used the same ionic positions, MT radii, and basis settings. The MT radii were 2.0~bohr for Li, 2.4~bohr for Na, and 2.5~bohr for K, Rb, and Cs. Exact folding of a mesh with 16 subdivisions gives 8192 points in the full BZ of the one-atom cell and 4096 in the two-atom cell. We took $r$ at the first compression step with $E_{\rm sym}-E_{\rm free}>0.03~\mathrm{meV/atom}$ and checked representative onset points with 24 subdivisions.

For hP4 the self-consistent meshes targeted 2000 points in the full BZ and 2500 near the transition. We obtained the pressures in Fig.~\ref{fig:pressure} from fits to the Na energy--volume curve, $P=-\partial E/\partial V$~\cite{ref19,ref20}. For the Li, Na, and K onset distances we minimized $E+PV$ on refined grids of volume and $c_4/a$ near the transition. We assigned band representations with the WIEN2k \texttt{irrep} program, using 50 intervals on each of the K--$\Gamma$ and $\Gamma$--M segments.

\subsection{Site charges and the geometric criterion}
\label{sec:criterion-test}
We computed the site populations from the electron number density $n(\mathbf r)$ integrated over equal-radius MT spheres. The WIEN2k \texttt{:NTO} population at site $\mathbf r_i$ and the charge difference are
\begin{equation}
 \begin{aligned}
 N_i^{\rm MT}&=\int_{|\mathbf r-\mathbf r_i|<R_{\rm MT}} n(\mathbf r)\,d^3r,\\
 |\Delta Q|&=e\,|N_1^{\rm MT}-N_2^{\rm MT}|.
 \end{aligned}
 \label{eq:mt-difference}
\end{equation}
For the orbital model we traced the occupied density matrix over the orbitals on $X_i$, summed over the BZ and both spin channels, and divided by two to obtain the projected charge per site in each WO (Supplementary Note~7).

For Eq.~\eqref{eq:criterion} we took $R$ from tabulated covalent radii~\cite{ref14,ref15} and evaluated $M$ by Ewald summation in a neutral cell~\cite{ref26,ref27}. We normalized the electrostatic energy per pair of opposite charge deviations, with one pair in the two-site BCC cell and two in the four-site hP4 cell.

For the five BCC alkali metals, the inset of Fig.~\ref{fig:bcc}(b) compares the calculated onset separation $r$ with $M_{\rm BCC}R$. Using the same energy threshold for all five elements, the fractional deviation $(r-M_{\rm BCC}R)/(M_{\rm BCC}R)$ ranges from $-5.8\%$ in K to $+4.0\%$ in Na (Supplementary Table~S1).

In hP4, the nearest $X_1$--$X_2$ separation is $r_{12}=(a^2/3+c_4^2/16)^{1/2}$. We evaluate $M_{\rm hP4}$ for opposite charge deviations on the two WOs represented by $X_1$ and $X_2$ using each element's axial ratio $c_4/a$ at the calculated transition. Relative to $M_{\rm hP4}R$, the onset separations are $5.4\%$ larger in Li, $2.8\%$ smaller in Na, and $7.1\%$ larger in K (Supplementary Tables~S2 and S3).

For Na we use the same charging radius, $R=1.55$~\AA, in both structures, so the difference between their predicted onset separations comes from the Madelung constant. The smaller Madelung constant of the hP4 charge pattern lowers the predicted onset separation from $2.73$~\AA\ in BCC to $2.22$~\AA\ in hP4. DFT gives $2.84$ and $2.15$~\AA, respectively (Supplementary Tables~S1--S3).

\subsection{Classification by symmetry generators}
We used the PyXtal subgroup catalogue and space-group operations~\cite{pyxtal2021}, following the subgroup and supergroup conventions of the Bilbao Crystallographic Server~\cite{wycksplit1998,supergroups2002,aroyo2006bilbao,delaflor2023symmdb}. For Type I, we classified the relations $\Gl=\langle G_{\rm dom},g\rangle$ by the broken generator $g$ and the resulting CDG $G_{\rm dom}$. For each embedding, we verified subgroup inclusion and checked that adjoining $g$ to $G_{\rm dom}$ regenerates $\Gl$ (Supplementary Note~4). We also extracted Type-I relations from charge patterns formed by assigning opposite charge deviations to one WO in a doubled primitive cell. We checked their symmetry independently with spglib 2.7.0~\cite{spglib2024}; the enumeration and WO-splitting checks are described in Supplementary Note~4 and Data~1.

For Type II, we closed an added spatial generator $g\notin\Gl$ with the parent operations to obtain $G=\langle\Gl,g\rangle$. For each embedding, we verified parent-group inclusion and determined the supergroup type and index (Supplementary Note~4). Tables~S5 and S6 organize these relations by the broken or added generator and the resulting group.

\section*{Data and code availability}
The manuscript and Supplementary Information contain the processed data for the figures and tables. Supplementary Data~1 and 2 provide the Type-I generator relations and Type-II generator extensions, respectively, together with their embedding and validation records. Supplementary Software~1 provides the classification programs, pinned crystallographic inputs, original charge-pattern records, and orbital-model checks. The underlying DFT input and output files are available from the corresponding author upon request.
\FloatBarrier
\section*{Acknowledgements}
We thank Wenlong Gao for insightful discussions. This work was supported by the National Key Research and Development Program of China (Grant No. 2024YFA1409800) and the National Natural Science Foundation of China (Grant No. 12688201).
\section*{Author contributions}
Qiu-Shi Huang and Su-Huai Wei conceived and designed the study. Qiu-Shi Huang carried out the theoretical derivations. Su-Huai Wei revised the manuscript. All authors discussed the results.
\section*{Competing interests}
The authors declare no competing interests.
\bibliography{references}
\end{document}


\raggedbottom
\title{Supplementary Information for Electronic symmetry beyond atomic space groups: A geometric criterion for breaking lattice and hidden symmetries}
\author{Qiu-Shi Huang}
\affiliation{Eastern Institute of Technology, Ningbo 315200, China}
\author{Xin-Gao Gong}
\affiliation{Key Laboratory for Computational Physical Sciences (MOE), State Key Laboratory of Surface Physics, Department of Physics, Fudan University, Shanghai 200433, China}
\author{Su-Huai Wei}
\email{suhuaiwei@eitech.edu.cn}
\affiliation{Eastern Institute of Technology, Ningbo 315200, China}
\maketitle
\begingroup
\footnotesize
\makeatletter\let\l@subsection\@gobbletwo\makeatother
\tableofcontents
\endgroup
\clearpage

\section{Supplementary Note 1. Definitions and symmetry setting}
\label{sec:notation}
The atomic space group $\Gl$ maps a site onto the other members of its Wyckoff orbit (WO). In Type I, the charge-domain group (CDG) $G_{\rm dom}$ preserves the site charges in one domain. In Type II, $G$ is the unitary electronic group generated by the representation of $\Gl$ and the hidden exchange $g$ in the same orbital Bloch space. With spinless time reversal $\mathcal T$, the extended group $\langle G,\mathcal T\rangle$ contains the antiunitary symmetry $\widetilde\Theta$ that protects the inherited band doublets.

We use real local orbitals for a nonmagnetic system without spin--orbit coupling. The matrices act in a single spin channel, and the spin multiplicity in the charge is $g_s=2$. Spinless time reversal $\mathcal T$ conjugates the coefficients and sends $\kk$ to $-\kk$. We write $\mathcal K$ for coefficient conjugation at fixed momentum. The classification treats scalar charge order, which is even under time reversal.

\subsection*{Notation used in the derivations}
\begin{description}
\item[$P$] Hydrostatic pressure.
\item[$\delta$, $\Delta Q$] Dimensionless transferred charge and signed site-charge difference, $\Delta Q=2e\delta$. Positive $\delta$ removes electrons from the first site.
\item[$F_{\rm pair}$, $N_p$] Free-energy change per oppositely charged pair and the number of such pairs in a cell.
\item[$\kappa$] Coulomb constant $(4\pi\varepsilon_0)^{-1}$. Bold $\kk$ always denotes momentum.
\item[$R$, $R_{\rm MT}$] Effective radius of the valence charge cloud and muffin-tin integration radius, respectively. The vector $\mathbf r_i$ denotes a site position and $\mathbf R$ a Bravais lattice vector.
\item[$r$, $M$] Nearest separation between opposite charge deviations and Madelung constant of their pattern. The geometric instability boundary is $r=MR$. Upright letters such as $\mathrm M$ denote points in the Brillouin zone.
\item[$\mathbf a_1,\mathbf a_2$] Hexagonal in-plane lattice vectors of length $a$ at $120^\circ$. A, B, and C label the three lateral layer positions in a close-packed stack.
\item[$h$, $c_2$, $c_4$] Layer spacing, AB repeat $c_2=2h$, and four-layer repeat $c_4=4h$. For hP4, $c/a$ means $c_4/a$.
\item[$t_c$, $T_{c_2}$, $\mathcal T$] BCC body-centering translation, translation by the two-layer AB period, and spinless time reversal. $T_\pm$ are the interlayer hopping matrices. In band labels $T$ denotes an irreducible representation on K--$\Gamma$.
\item[$\mathsf B(n)$] Orbital hopping matrix for a bond with vertical direction cosine $n$.
\item[$P_1,P_2$] Projectors onto the $X_1$ and $X_2$ site sectors. Each selects two sites and has rank $2N$ for $N$ orbitals per site.
\item[$\widetilde P_-$] Projector onto the ABAC boundary subspace inherited from the zone boundary of AB.
\item[$Q_i^{\rm proj}$, $Q_i^{\rm MT}$] Per-site charge from the model orbital projector and from a real-space muffin-tin integral, respectively.
\item[$\Delta\varepsilon$] Onsite energy difference between the two site sectors.
\item[$C_{\rm lay}$, $\Pi$, $J$] Cyclic permutation of the four layers, layer exchange under inversion, and orbital inversion-parity matrix, respectively.
\item[$L$, $N$] Number of layers in a periodic stack and number of retained orbitals per layer. The concrete four-layer $(s,p_z)$ model has $L=4$ and $N=2$.
\item[$\sigma_j,\chi_j$, $\theta_j$] Upward-bond chiralities of two stacking sequences and the phase on layer $j$ relating their Hamiltonians.
\item[$Y_\ell^0$] Axial spherical harmonic with angular momentum $\ell$ and azimuthal quantum number zero, measured about the common stacking axis.
\end{description}

\section{Supplementary Note 2. Computational definitions and protocols}
\label{sec:computation}
We used WIEN2k \cite{ref17,ref18} with the Perdew--Burke--Ernzerhof functional \cite{ref46}. The main-text Methods give the basis and convergence parameters. For BCC we compared two states at the same nuclear positions and electron count. The body-centering translation constrains the first state. In the second, we released this constraint and allowed an initially unequal site density to relax. Exact mesh folding gives the same physical wavevectors in both cells. We repeated the comparison with 24 subdivisions at representative onset points.

Let $E_{\rm sym}$ and $E_{\rm free}$ denote the respective energies per atom. For every BCC element, $r$ is the first sampled separation under compression at which
\begin{equation}
 E_{\rm sym}-E_{\rm free}>0.03~\mathrm{meV/atom}.
 \label{eq:onset}
\end{equation}
We compare these sampled onset distances with the theoretical boundary $A_2=0$ derived in Note~3.

We obtained the Na pressures from fits to the calculated energy--volume curve \cite{ref19,ref20}. For each hP4 element, we sampled a grid of volume and $c_4/a$, holding the nuclear positions fixed during each self-consistent electronic calculation. At a specified pressure $P$, we selected the structure by minimizing $E+PV$ over the converged grid and refined the grid near the switch between charge-equivalent and charge-transfer states. The site charges distinguish these two states. Note~10 lists the geometries used for the distance comparison and the representative structures used for the band spectra.

For site $i$, the WIEN2k \texttt{:NTO} electron population and the corresponding muffin-tin charge are
\begin{equation}
 N_i^{\rm MT}=\int_{|\mathbf r-\mathbf r_i|<R_{{\rm MT},i}}n(\mathbf r)\,d^3r,\qquad
 Q_i^{\rm MT}=Q_{\rm ref}^{\rm MT}-eN_i^{\rm MT}.
 \label{eq:mt-charge}
\end{equation}
The density $n$ includes both spin channels. We use equal sphere radii on the two sites, so the common reference $Q_{\rm ref}^{\rm MT}$ cancels in $|\Delta Q|=e|N_1^{\rm MT}-N_2^{\rm MT}|$. The sites are the corner and body center in BCC and $X_1=(0,0,0)$ and $X_2=(1/3,2/3,1/4)$ in ABAC. The integral counts all electrons inside the sphere, whereas the projected charge in Note~7 is defined from the selected model orbitals. We use the corresponding site-charge differences to follow the loss of charge equivalence in the DFT calculation and the orbital model, respectively.

\section{Supplementary Note 3. Landau free energy and the geometric instability criterion}
\label{sec:landau}
\subsection{Charge transfer and the free-energy minima}

Let $e>0$, and let $Q_0$ be the net site charge in the charge-equivalent state. Charge transfer preserves the total pair charge,
\begin{equation}
 Q_A=Q_0+e\delta,\qquad Q_B=Q_0-e\delta,\qquad Q_A+Q_B=2Q_0.
 \label{eq:delta}
\end{equation}
The exchange $g:A\leftrightarrow B$ reverses $\delta$. If $g$ is a symmetry of the reference Hamiltonian, $F_{\rm pair}(\delta,P)=F_{\rm pair}(-\delta,P)$, so only even powers of $\delta$ occur \cite{ref31,ref32}.

For a given pressure $P$, we vary $\delta$ at fixed atomic positions and relax the other electronic degrees of freedom at each step. The curvature of this constrained energy defines
\begin{equation}
 A_2(P)=\frac12\left.\frac{\partial^2F_{\rm pair}(\delta,P)}{\partial\delta^2}\right|_{\delta=0}.
 \label{eq:full-curvature}
\end{equation}
This curvature includes kinetic, exchange, and correlation contributions. To estimate $A_2$ electrostatically, we write the local charging cost and the intersite energy as
\begin{align}
 \Delta E_{\rm intra}&=A_{\rm intra}(P)\delta^2+B_{\rm intra}(P)\delta^4+O(\delta^6),\\
 \Delta E_{\rm inter}&=-A_{\rm inter}(P)\delta^2.
\end{align}
With $A_2=A_{\rm intra}-A_{\rm inter}$ in this estimate and $A_4=B_{\rm intra}>0$, the free-energy change is
\begin{equation}
 F_{\rm pair}=A_2\delta^2+A_4\delta^4,
 \label{eq:Fpair}
\end{equation}
to quartic order. The stationary equation is
\begin{equation}
 \frac{\partial F_{\rm pair}}{\partial\delta}=2\delta(A_2+2A_4\delta^2)=0.
 \label{eq:stationary}
\end{equation}
At $\delta=0$ the curvature is $2A_2$. The charge-equivalent state is a minimum for $A_2>0$ and loses its quadratic restoring force at $A_2=0$. For $A_2<0$, the two new minima satisfy
\begin{equation}
 \delta_\pm^2=-\frac{A_2}{2A_4},\qquad
 \left.\frac{\partial^2 F_{\rm pair}}{\partial\delta^2}\right|_{\delta_\pm}
 =2A_2+12A_4\delta_\pm^2=-4A_2>0.
 \label{eq:curvature}
\end{equation}
Substitution into Eq.~\eqref{eq:Fpair} gives
\begin{equation}
 F_{\rm pair}(\delta_\pm)
 =A_2\left(-\frac{A_2}{2A_4}\right)+A_4\frac{A_2^2}{4A_4^2}
 =-\frac{A_2^2}{4A_4}.
 \label{eq:energy-min}
\end{equation}
Along this continuous branch, $|\delta|$ grows as $(-A_2)^{1/2}$ and the energy gain as $A_2^2$. A cell with $N_p$ charge pairs gains $N_p$ times the energy per pair.

\subsection{Charging work and Madelung normalization}
\label{sec:electrostatic}

For a spherical shell of radius $R$ and charge $Q'$, the surface potential is $V(Q')=\kappa Q'/R$, where $\kappa=(4\pi\varepsilon_0)^{-1}$. Charging from zero to $Q$ costs
\begin{equation}
 E_{\rm self}(Q)=\int_0^Q V(Q')\,dQ'=\int_0^Q\frac{\kappa Q'}{R}\,dQ'
 =\frac{\kappa Q^2}{2R}.
 \label{eq:shell}
\end{equation}
The pair cost relative to two equivalent charges is
\begin{align}
 \Delta E_{\rm intra}
 &=\frac{\kappa}{2R}\left[(Q_0+e\delta)^2+(Q_0-e\delta)^2-2Q_0^2\right]\nonumber\\
 &=\frac{\kappa}{2R}\left[2e^2\delta^2\right]
 =\frac{\kappa e^2}{R}\delta^2.
 \label{eq:pair-self}
\end{align}
The linear terms cancel because the two sites have the same reference charge and charging environment. Each site costs $\kappa(e\delta)^2/(2R)$. We estimate the charging radius $R$ of the valence cloud from tabulated covalent radii \cite{ref14,ref15}.

For a periodic, charge-neutral pattern, write the deviation at site $i$ as $s_i e\delta$, with $s_i=\pm1$. If there are $N_p$ pairs per cell, its lattice contribution per pair is
\begin{equation}
 \Delta E_{\rm inter}=\frac{\kappa(e\delta)^2}{2N_p}
 \sum_{\alpha\in\mathrm{cell}}\sum_{j\ne\alpha}\frac{s_\alpha s_j}{d_{\alpha j}}.
 \label{eq:lattice}
\end{equation}
Here $d_{\alpha j}=|\mathbf r_j-\mathbf r_\alpha|$, and $j$ runs over the entire lattice except site $\alpha$. The factor $1/2$ counts each interaction once, and division by $N_p$ gives the energy per pair. Taking $r$ as the nearest separation between opposite charge deviations gives the Madelung constant
\begin{equation}
 M=-\frac{r}{2N_p}\sum_{\alpha\in\mathrm{cell}}\sum_{j\ne\alpha}
      \frac{s_\alpha s_j}{d_{\alpha j}}.
 \label{eq:M}
\end{equation}
$M$ is positive when the alternating lattice sum is attractive. Its value depends on the specified charge pattern, geometry, and distance normalization. With neutral-cell Ewald summation \cite{ref26,ref27}, Eq.~\eqref{eq:lattice} becomes
\begin{equation}
 \Delta E_{\rm inter}=-\frac{\kappa e^2M}{r}\delta^2.
\end{equation}
Combining the lattice energy with the shell charging cost in Eq.~\eqref{eq:pair-self} gives the electrostatic estimate
\begin{equation}
 A_2\simeq\kappa e^2\left(\frac1R-\frac Mr\right),\qquad
 r<MR\ \Longrightarrow\ A_2<0 .
 \label{eq:geometric}
\end{equation}

\subsection{Ewald evaluation}

For the neutral pattern with site-charge deviations $s_i e\delta$, the full-cell Ewald energy is \cite{ref26}
\begin{align}
 E_{\rm Ewald}={}&\kappa(e\delta)^2\Biggl[\frac{1}{2}\sum_{ij,\mathbf R}'s_i s_j
 \frac{\operatorname{erfc}(\eta_{\rm E}|\mathbf r_{ij}+\mathbf R|)}{|\mathbf r_{ij}+\mathbf R|}\nonumber\\
 &+\frac{2\pi}{\Omega}\sum_{\mathbf K\ne0}
 \frac{e^{-K^2/(4\eta_{\rm E}^2)}}{K^2}
 \left|\sum_j s_j e^{i\mathbf K\cdot\mathbf r_j}\right|^2
 -\frac{\eta_{\rm E}}{\sqrt\pi}\sum_j s_j^2\Biggr].
 \label{eq:ewald}
\end{align}
The Ewald parameter $\eta_{\rm E}$ controls convergence and $\Omega$ is the cell volume. The vectors $\mathbf R$ and $\mathbf K$ run over the direct and reciprocal lattices, respectively, and $\mathbf r_{ij}=\mathbf r_i-\mathbf r_j$. The prime omits the self-interaction at $i=j$, $\mathbf R=0$. Charge neutrality removes the $\mathbf K=0$ component.

Dividing the full-cell energy by $N_p$ gives the energy per pair and its Madelung constant,
\begin{equation}
 \Delta E_{\rm inter}=\frac{E_{\rm Ewald}}{N_p},\qquad
 M=-\frac{rE_{\rm Ewald}}{N_p\kappa(e\delta)^2}.
 \label{eq:ewald-normalization}
\end{equation}
The converged lattice sum is independent of $\eta_{\rm E}$. The two-site BCC cell has $N_p=1$, and the four-site hP4 cell has $N_p=2$, giving the same energy-per-pair normalization for both comparisons.

\section{Supplementary Note 4. Classification by symmetry generators}
\label{sec:classification}
\subsection{Type I: loss of a generator}

We describe a symmetry reduction by the relation $\Gl=\langle G_{\rm dom},g\rangle$, where $g\notin G_{\rm dom}$. The charge-domain group $G_{\rm dom}$ retains the operations of the ordered state, and adjoining the broken generator $g$ restores $\Gl$. For BCC, $g=t_c=\{\mathbb I_3\mid\tfrac12,\tfrac12,\tfrac12\}$ gives $Im\bar3m=\langle Pm\bar3m,t_c\rangle$.

Table~\ref{tab:classification-typeI} lists the parent group, broken generator, and resulting charge-domain group for 3746 representative classes covering all 230 parent groups.

\subsection{Type II: addition of a generator}
\label{sec:typeII-candidates}

For a parent group $\Gl$, we adjoin a spatial generator $g$ and close its products with the parent operations,
\begin{equation}
 G=\langle\Gl,g\rangle,\qquad g\notin\Gl.
 \label{eq:generator-extension}
\end{equation}
Table~\ref{tab:classification-typeII} lists the parent group, added generator, generated supergroup, and resulting index for 3721 representative classes covering all 230 parent groups.

\subsection{Group embeddings and closure checks}

We use all 8867 embeddings in the PyXtal subgroup catalogue \cite{pyxtal2021}, including both translationengleiche and klassengleiche relations. For each embedding, we reconstruct the operations from its exact cell transformation and check the conventional settings and group types with spglib \cite{spglib2024}. We verify subgroup inclusion and test closure by the transitive action of the subgroup and the displayed generator on the larger-group cosets. Every catalogue embedding passes this test.

For the charge-pattern data, let $T$ be the parent translation group. The subgroup $2T$, generated by twice each primitive translation, is normal in $\Gl$ and lies in every listed $G_{\rm dom}$. We therefore verify $\Gl=\langle G_{\rm dom},g\rangle$ in the finite quotient $\Gl/(2T)$.

We also include the single-generator relations extracted from the charge-pattern enumeration below and their inverses, with the coordinate transformations checked against the standard group operations. The spatial ABAC construction in Note~7 supplies an additional relation in both directions.

In both tables, records with the same pair of group types and the same generator type form one representative class. Each row gives one coordinate embedding and its index. Supplementary Data~1 and 2 retain the complete embeddings and validation records in table order.

\subsection{Charge-pattern realizations and validation}

We supplement the group catalogue with stabilizers of charge patterns at fixed atomic positions and chemical composition \cite{isodisplace2006,isodistort}. We vary the charges on one parent WO at a time. All other occupied WOs retain their parent-equivalent charges. Let $\mathbf a_i^{\rm p}$ be primitive translations and $\mathbf b_i^{\rm p}$ their reciprocal vectors, with $\mathbf a_i^{\rm p}\cdot\mathbf b_j^{\rm p}=2\pi\delta_{ij}$. A charge-order wavevector that doubles the primitive cell has the form
\begin{equation}
 \mathbf k_{\rm ord}=\frac12\sum_{i=1}^{3}\nu_i\mathbf b_i^{\rm p},
 \qquad \nu_i\in\{0,1\},\qquad (\nu_1,\nu_2,\nu_3)\ne(0,0,0).
 \label{eq:class-wavevector}
\end{equation}
The seven wavevectors fall into stars under the parent point group. For one representative of each star, we assign charge deviations on the active WO. If its reference charge is $Q_0$, the charge at site $i$ in cell $\mathbf R$ obeys
\begin{equation}
 Q_i(\mathbf R)-Q_0=[Q_i(0)-Q_0]e^{i\mathbf k_{\rm ord}\cdot\mathbf R},
 \qquad e^{i\mathbf k_{\rm ord}\cdot\mathbf R}\in\{+1,-1\}.
 \label{eq:class-charge}
\end{equation}
The lost translation reverses the sign of every deviation. It pairs opposite charges and makes the modulation neutral in the doubled cell. The operations that preserve a single assignment form its charge-domain group,
\begin{equation}
 G_{\rm dom}=\{g\in\Gl\,|\,Q_{g i}=Q_i\text{ for every site in the periodic pattern}\}.
 \label{eq:class-stabilizer}
\end{equation}
The action includes the cell translation attached to the mapped site. The electronic primitive volume is twice the parent volume, although additional lost point operations can make the group index exceed two.

For each star we take the quotient of its little space group by the doubled translation lattice. We enumerate the subgroups of this finite group that exclude the lost translation. On each parent WO we find the scalar charge patterns left invariant by each subgroup. We retain a subgroup if it is the full pointwise stabilizer of this space. This removes cases in which every allowed pattern has higher symmetry. A generic vector in a retained space realizes the stated stabilizer. We then test separately whether the space contains a pattern with just two opposite charge values.

We evaluated the PyXtal Wyckoff coordinates at generic parameter values \cite{pyxtal2021}, matched the conventional settings to Hall settings, and checked each CDG independently with spglib using the charges as site labels \cite{spglib2024}. At a special coordinate, the site belongs to its specialized Wyckoff position. As in conventional WO analysis \cite{wycksplit1998}, the splitting depends on the parent--child embedding. The space-group symbol can survive a change of primitive cell, and the same pair of group types can have different embeddings.

The enumeration covers 230 parent space groups, 1731 Wyckoff positions, and 981 inequivalent single-wavevector stars. Of the 57\,066 combinations of active WO and CDG embedding that admit an odd scalar modulation, 32\,720 allow $+e\delta$ or $-e\delta$ on every active site. In 9154 of these cases, the parent WO splits into exactly two WOs of $G_{\rm dom}$. In the others, several WOs can share a charge value. General scalar patterns may also have unequal magnitudes or leave some sites unmodulated.

We checked the CDG type and primitive-volume ratio independently for all 32\,720 two-charge patterns. A second set of generic coordinates tested all 57\,066 scalar stabilizers. For small WOs, exhaustive checks covered 36\,859 two-charge patterns on 5983 WO instances. We also independently enumerated the subgroups of 528 small finite groups. This resolved the two-charge existence test in every case.

Merging repeated WO realizations leaves 10\,099 distinct CDG embeddings in this charge-pattern data set. Of these, 5265 satisfy $\Gl=\langle G_{\rm dom},g\rangle$ for a single $g$ and enter Table~\ref{tab:classification-typeI}. The remaining 4834 require more than one additional generator to recover $\Gl$. Their complete generator sets remain in Supplementary Data~1; all 32\,720 charge-pattern branches are retained in Supplementary Software~1. The table also includes the catalogue relations described above, which are classified directly at the group level.

\subsection{Using the table in an electronic calculation}
For a Type-I charge-pattern realization, the embedding record in Supplementary Data~1 identifies the corresponding source record in Supplementary Software~1, which supplies the active WO, cell transformation, and charge assignment associated with a relation in Table~\ref{tab:classification-typeI}. These determine the doubled cell and initial charge density. Self-consistency can retain $G_{\rm dom}$ or allow further symmetry breaking.

The energy comparison uses the same atomic geometry, basis settings, and folded-equivalent $\kk$ mesh. The symmetry of the converged state is determined from the full charge density, including its orbital components.

\section{Supplementary Note 5. BCC symmetry breaking and the geometric criterion}
\label{sec:bcc}
\subsection{Body-centering translation and site charges}

In the BCC calculations, charge transfer between the corner and body-center sites of the $2a$ WO in $\Gl=Im\bar3m$ corresponds to the wavevector $(1,0,0)$, in units of $2\pi/a$ along the conventional cubic axes. The translation $t_c$ that exchanges these sites, written in conventional fractional coordinates, is
\begin{equation}
 t_c=\left\{E\,\middle|\,\tfrac12,\tfrac12,\tfrac12\right\}.
\end{equation}
The translation $t_c$ reverses $\Delta Q=Q_A-Q_B$. When the two charges differ, the cubic point group $m\bar3m$ survives and the charge-domain group is $G_{\rm dom}=Pm\bar3m$ (No.~221). The parent group splits into two cosets,
\begin{equation}
 \Gl=G_{\rm dom}\sqcup t_cG_{\rm dom},\qquad [\Gl:G_{\rm dom}]=2.
 \label{eq:coset}
\end{equation}
The square $t_c^2=\{E|1,1,1\}$ translates each charge sublattice into itself. Operations in $G_{\rm dom}$ preserve the sign of $\Delta Q$, and those in $t_cG_{\rm dom}$ reverse it.

For a direct charge test, let $P_A$ and $P_B$ project onto symmetry-related site regions and let $\rho$ include the occupied electronic states, normalized per two-site cell. Invariance under $t_c$ means $t_c\rho t_c^{-1}=\rho$. Since $P_B=t_cP_At_c^{-1}$,
\begin{equation}
 \Tr(P_B\rho)=\Tr(t_cP_At_c^{-1}\rho)
 =\Tr(P_At_c^{-1}\rho t_c)=\Tr(P_A\rho).
 \label{eq:bcc-trace}
\end{equation}
A charge difference between these two regions therefore breaks the electronic symmetry under $t_c$ even with both nuclei fixed.

\subsection{Geometric criterion and calculated onset}
The corner and body center are separated by $r=\sqrt3 a/2$, and the volume per atom is $a^3/2$. For their alternating charge pattern, $M_{\rm BCC}=1.762675$ \cite{ref16,ref26,ref27}. The condition $f<1$ follows from Eq.~\eqref{eq:geometric} with
\begin{equation}
 f=\frac{r}{M_{\rm BCC}R}.
\end{equation}
Table~\ref{tab:bcc} compares this boundary with the onset from Eq.~\eqref{eq:onset}. A positive $\Delta r$ means that charge transfer starts at a larger separation than predicted.

\begin{table}[htbp]
\centering
\caption{Onset of charge transfer in BCC alkali metals. The signed difference $\Delta r=r-M_{\rm BCC}R$ is given at the reported precision.}
\label{tab:bcc}
\begin{tabular}{lrrrr}
\toprule Element & $M_{\rm BCC}R$ (\AA) & $r$ (\AA) & $\Delta r$ (\AA) & $\Delta r/(M_{\rm BCC}R)$ (\%)\\\midrule
Li & 2.346 & 2.429 & $+0.082$ & $+3.5$\\
Na & 2.731 & 2.841 & $+0.110$ & $+4.0$\\
K  & 3.455 & 3.254 & $-0.202$ & $-5.8$\\
Rb & 3.703 & 3.529 & $-0.174$ & $-4.7$\\
Cs & 4.088 & 3.987 & $-0.101$ & $-2.5$\\
\bottomrule
\end{tabular}
\end{table}

\subsection{The other doubled charge pattern}

The corner--body-center pattern tested above is one of two symmetry-distinct charge patterns that double the primitive cell of the $2a$ WO. The other has wavevector $(\tfrac12,\tfrac12,0)$ and describes alternating charges on adjacent $(110)$ planes and has six symmetry-equivalent orientations. This pattern has $G_{\rm dom}=Cmmm$ (No.~65), with orthorhombic point group $mmm$ and index 12 in $\Gl$.

\section{Supplementary Note 6. Multiorbital gauge equivalence and its ABAB--ABAC realization}
\label{sec:multiorbital}
\subsection{Gauge-equivalence theorem}

Consider $L$ identical layers, each with $N$ orthonormal orbitals and the same Hermitian intralayer block $D(\kk)$. The upward hopping between adjacent layers is $T_+(\kk)$ or $T_-(\kk)$ according to the bond chirality. Hermiticity fixes the reverse block. We include cell-crossing phases in the same Bloch convention for both stacks. Suppose the two forward hopping matrices differ by a scalar phase,
\begin{equation}
  T_+(\kk)=e^{2i\phi(\kk)}T_-(\kk),
 \label{eq:general-phase}
\end{equation}
with $\phi$ real. The orbital matrices can be nondiagonal, and $D$ need not commute with $T_\pm$. Let $\sigma_j$ and $\chi_j$ be the $L$ upward-bond signs of the two stacks. If their net chiralities agree,
\begin{equation}
 \sum_{j=1}^{L}\sigma_j=\sum_{j=1}^{L}\chi_j,
 \label{eq:general-net}
\end{equation}
the Hamiltonians are unitarily equivalent at each momentum satisfying Eq.~\eqref{eq:general-phase}. This extends the scalar layer-phase construction of He \emph{et al.} \cite{ref10} to hopping matrices of any orbital dimension $N$.

\subsection{Interior bonds and periodic closure}
Let $|j\rangle$ denote the layer basis, and set
\begin{equation}
 \begin{aligned}
 W&=\sum_{j=1}^{L}e^{i\theta_j}|j\rangle\langle j|\otimes\I_N,\\
 \theta_1&=0,\qquad
 \theta_j=\phi\sum_{p=1}^{j-1}(\chi_p-\sigma_p).
 \end{aligned}
 \label{eq:general-W}
\end{equation}
Each factor has unit modulus, so $W$ is unitary. On layer $j$,
\begin{equation}
  e^{-i\theta_j}D e^{i\theta_j}=D,
\end{equation}
because the phase acts identically on all orbitals, including those coupled by $D$. On an interior bond, the same change of basis gives
\begin{equation}
 [W^\dagger H_\sigma W]_{j,j+1}
   =e^{i(\theta_{j+1}-\theta_j)}T_{\sigma_j}
   =e^{i(\chi_j-\sigma_j)\phi}T_{\sigma_j}
   =T_{\chi_j}.
 \label{eq:general-interior}
\end{equation}
For equal signs the multiplier is one. Reversing $+$ to $-$ multiplies the block by $e^{-2i\phi}$, and reversing $-$ to $+$ multiplies it by $e^{2i\phi}$, as required by Eq.~\eqref{eq:general-phase}. Taking the adjoint gives the reverse hopping.

The periodic gauge must use the same phase on layer 1 in the next cell. The recursion assigns
\begin{equation}
 \theta_{L+1}=\phi\sum_{j=1}^{L}(\chi_j-\sigma_j).
\end{equation}
Thus the closing bond is matched when
\begin{equation}
 e^{i(\theta_{L+1}-\theta_1)}=1.
 \label{eq:general-loop}
\end{equation}
Equal net chirality satisfies this condition for every $\phi(\kk)$. The complete Hamiltonian then obeys
\begin{equation}
 W^\dagger(\kk)H_\sigma(\kk)W(\kk)=H_\chi(\kk).
 \label{eq:general-H}
\end{equation}
The two stacks have the same eigenvalues and density of states. For unequal net chiralities, Eq.~\eqref{eq:general-loop} can still hold at individual momenta. In an open stack every one of the $L-1$ bonds can be matched by recursion because there is no closing bond.

For a short repeat, two periodic bonds can contribute to the same matrix entry. The transformation acts on each bond before their contributions are added.

\subsection{Axial orbitals and the common hopping phase}

Consider real orbitals centered at each atom,
\begin{equation}
 \varphi_\alpha(\mathbf r)=R_\alpha(r)Y_{\ell_\alpha}^{0}(\hat{\mathbf r}),
 \label{eq:axial-orbitals}
\end{equation}
where $\alpha=1,\ldots,N$. Each orbital has its own radial function $R_\alpha$ and angular momentum $\ell_\alpha$. Corresponding orbitals on different sites have the same parameters. All quantization axes lie along the layer normal, so rotation by $\widehat R_z(\varphi)$ about this axis leaves every orbital invariant.

In a rotationally covariant two-center model, let $t(\mathbf d)$ be the bond operator for displacement $\mathbf d$. Rotating the bond gives
\begin{equation}
 t[R_z(\varphi)\mathbf d]
  =\widehat R_z(\varphi)t(\mathbf d)\widehat R_z^{-1}(\varphi).
\end{equation}
Consequently, the matrix element between any two retained axial orbitals satisfies
\begin{equation}
 \begin{aligned}
 B_{\alpha\beta}[R_z(\varphi)\mathbf d]
 &=\langle\varphi_\alpha|\widehat R_z(\varphi)t(\mathbf d)
                   \widehat R_z^{-1}(\varphi)|\varphi_\beta\rangle\\
 &=\langle\varphi_\alpha|t(\mathbf d)|\varphi_\beta\rangle
  =B_{\alpha\beta}(\mathbf d).
 \end{aligned}
 \label{eq:axial-azimuth}
\end{equation}
Equation~\eqref{eq:axial-azimuth} also holds for $\ell_\alpha\ne\ell_\beta$. Thus the bond matrix, including its hybridization terms, is independent of azimuth. For a bond of length $d=\sqrt{a^2/3+h^2}$ and vertical direction cosine $n=h/d$, the $(s,p_z)$ example with real Slater--Koster parameters gives \cite{slater1954}
\begin{equation}
 \mathsf B(n)=\begin{pmatrix}
 V_{ss\sigma}&nV_{sp\sigma}\\
 -nV_{sp\sigma}&n^2V_{pp\sigma}+(1-n^2)V_{pp\pi}
 \end{pmatrix}.
 \label{eq:B}
\end{equation}
The off-diagonal signs follow from the phases of the real orbitals. The reverse hopping is $\mathsf B^\dagger$.

The three nearest upward bonds of a close-packed layer have lateral displacements
\begin{equation}
 \boldsymbol\rho_1=\frac{\mathbf a_1+2\mathbf a_2}{3},\quad
 \boldsymbol\rho_2=\frac{\mathbf a_1-\mathbf a_2}{3},\quad
  \boldsymbol\rho_3=\frac{-2\mathbf a_1-\mathbf a_2}{3}.
\end{equation}

Reversing the stacking chirality changes these displacements from $\boldsymbol\rho_j$ to $-\boldsymbol\rho_j$, while $d$ and $n$ stay fixed. Equation~\eqref{eq:axial-azimuth} therefore gives the same $\mathsf B(n)$ for both bond sets. Defining the in-plane structure factor by
\begin{equation}
 S(\kk_\parallel)=\sum_{\alpha=1}^3e^{i\kk_\parallel\cdot\boldsymbol\rho_\alpha}
  =|S|e^{i\phi}
\end{equation}
and summing the three bond phases gives
\begin{equation}
 \begin{aligned}
 T_+&=\sum_{j=1}^{3}e^{i\kk\cdot(\boldsymbol\rho_j+h\hat{\mathbf z})}\mathsf B(n)
      =e^{ik_zh}S\mathsf B(n),\\
 T_-&=\sum_{j=1}^{3}e^{i\kk\cdot(-\boldsymbol\rho_j+h\hat{\mathbf z})}\mathsf B(n)
      =e^{ik_zh}S^*\mathsf B(n).
 \end{aligned}
 \label{eq:axial-factorization}
\end{equation}
With $S=|S|e^{i\phi}$, these blocks obey Eq.~\eqref{eq:general-phase} for any finite set of $m=0$ orbitals, including $s$, $p_z$, $d_{z^2}$, and axial $f$ and higher-$\ell$ orbitals.

\subsection{Stacking geometry and Bloch convention}

Let $\mathbf b=(\mathbf a_1+2\mathbf a_2)/3$ be the lateral displacement from an A layer to a B layer. With the origin on layer 1 and the positive stacking axis along $\hat{\mathbf z}$, the four positions in the ABAB and ABAC cells are
\begin{equation}
 \begin{aligned}
 \boldsymbol\tau^{\rm ABAB}&=(0,\mathbf b+h\hat{\mathbf z},2h\hat{\mathbf z},\mathbf b+3h\hat{\mathbf z}),\\
 \boldsymbol\tau^{\rm ABAC}&=(0,\mathbf b+h\hat{\mathbf z},2h\hat{\mathbf z},-\mathbf b+3h\hat{\mathbf z}).
 \end{aligned}
 \label{eq:positions}
\end{equation}
Adding the lattice vector $\mathbf a_1+\mathbf a_2$ to the last ABAC position gives its conventional fractional coordinates $(2/3,1/3,3/4)$. This relabels its cell in the sum in Eq.~\eqref{eq:bloch}.

For either stacking, use real orthonormal local orbitals and the position-gauge Bloch basis
\begin{equation}
 |j\alpha,\kk\rangle=N_{\rm cell}^{-1/2}\sum_{\mathbf R}
 e^{i\kk\cdot(\mathbf R+\boldsymbol\tau_j)}|\mathbf Rj\alpha\rangle.
 \label{eq:bloch}
\end{equation}
Here $N_{\rm cell}$ is the number of cells in the Bloch sum, $j$ labels a layer, and $\alpha=1,\ldots,N$ labels its orbitals. Coefficients are ordered first by layer and then by orbital. For a reciprocal vector $\mathbf K$, the basis changes by
\begin{equation}
 V(\mathbf K)=\diag(e^{i\mathbf K\cdot\boldsymbol\tau_1},\ldots,
 e^{i\mathbf K\cdot\boldsymbol\tau_4})\otimes\I_N,\qquad
 H(\kk+\mathbf K)=V^\dagger H(\kk)V.
 \label{eq:sewing}
\end{equation}
We write $V_{\rm ABAB}$ and $V_{\rm ABAC}$ for the sewing matrices of the two stacks.

A spatial operation $a=\{\mathcal R|\mathbf t\}$ maps orbital centers according to
\begin{equation}
 \mathcal R\boldsymbol\tau_j+\mathbf t
 =\boldsymbol\tau_{p(j)}+\mathbf L_j,
 \label{eq:spatial-embedding}
\end{equation}
where $\mathbf L_j$ is a lattice vector. Its action on Eq.~\eqref{eq:bloch} gives

\begin{equation}
 D_a(\kk)=e^{-i(\mathcal R\kk)\cdot\mathbf t}\Pi_a\otimes d_{\rm orb}(\mathcal R).
 \label{eq:representation}
\end{equation}
To obtain the phase, set $\mathbf R'=\mathcal R\mathbf R+\mathbf L_j$ in the transformed sum. Then $\kk\cdot(\mathbf R+\boldsymbol\tau_j)=(\mathcal R\kk)\cdot(\mathbf R'+\boldsymbol\tau_{p(j)}-\mathbf t)$. The common phase in Eq.~\eqref{eq:representation} already includes the cell crossing. The matrix $\Pi_a$ permutes the sites, and $d_{\rm orb}$ rotates the local orbitals.

\subsection{The four-layer Hamiltonian and layer gauge}
\label{sec:gauge}

Both stacks use the same orbital parameters and intralayer block $D(\kk)$. The retained hoppings preserve the atomic space group of each stacking and spinless time reversal. In the $(s,p_z)$ example, horizontal two-center bonds give an even, inversion-compatible $D(\kk)$ with vanishing $s$--$p_z$ hybridization.

Denote upward layer chiralities by $\sigma_j=\pm1$ \cite{ref28}. For $N$ orbitals on each of the four layers, the Hamiltonian is
\begin{equation}
 H_\sigma(\kk)=\begin{pmatrix}
 D&T_{\sigma_1}&0&T_{\sigma_4}^\dagger\\
 T_{\sigma_1}^\dagger&D&T_{\sigma_2}&0\\
 0&T_{\sigma_2}^\dagger&D&T_{\sigma_3}\\
 T_{\sigma_4}&0&T_{\sigma_3}^\dagger&D
 \end{pmatrix}.
 \label{eq:Hcomplete}
\end{equation}
Each entry is an $N\times N$ block. The Bloch phase of the bond from layer 4 to layer 1 in the next cell is already in $T_{\sigma_4}$ through Eq.~\eqref{eq:axial-factorization}.

The upward chiralities are $(+,-,+,-)$ for ABAB and $(+,-,-,+)$ for ABAC. Only the last two bonds change chirality. Equation~\eqref{eq:general-phase} shows that multiplying the layer-4 basis by $e^{-2i\phi}$ reverses the third hopping. The closing bond then picks up the conjugate phase. For these sequences the general gauge $W$ becomes the four-layer gauge $U_{\rm L}$,
\begin{equation}
 \theta=(0,0,0,-2\phi),\qquad
 U_{\rm L}=\diag(1,1,1,\eta)\otimes\I_N,\qquad \eta=e^{-2i\phi}=S^*/S.
 \label{eq:Ug}
\end{equation}
For $S\ne0$, $|\eta|=1$. At $S=0$ both hopping blocks vanish, so we choose a phase of unit modulus on a fundamental momentum domain and extend it by inversion conjugation and reciprocal sewing. Acting on the four upward bonds gives
\begin{equation}
 (T_+,T_-,T_+,T_-)
 \mapsto (T_+,T_-,\eta T_+,\eta^*T_-)
 =(T_+,T_-,T_-,T_+).
\end{equation}
The transformed hoppings are precisely those of ABAC, giving
\begin{equation}
 H_{\rm ABAC}(\kk)=U_{\rm L}^\dagger(\kk)H_{\rm ABAB}(\kk)U_{\rm L}(\kk),\qquad
 |u_{n\kk}^{\rm ABAC}\rangle=U_{\rm L}^\dagger(\kk)|u_{n\kk}^{\rm ABAB}\rangle.
 \label{eq:gauge}
\end{equation}
For an in-plane reciprocal vector $\mathbf K$, the three lateral bonds differ by lattice vectors, so $S(\kk+\mathbf K)=e^{i\mathbf K\cdot\mathbf b}S(\kk)$ and $\eta(\kk+\mathbf K)=e^{-2i\mathbf K\cdot\mathbf b}\eta(\kk)$. Since $\boldsymbol\tau_4^{\rm ABAC}-\boldsymbol\tau_4^{\rm ABAB}=-2\mathbf b$ modulo a lattice vector, the gauge satisfies
\begin{equation}
 U_{\rm L}(\kk+\mathbf K)=V_{\rm ABAB}^\dagger(\mathbf K)U_{\rm L}(\kk)V_{\rm ABAC}(\mathbf K).
 \label{eq:Ug-sewing}
\end{equation}
Equation~\eqref{eq:Ug-sewing} also holds for a reciprocal vector along $z$ because the layer heights coincide and $U_{\rm L}$ is independent of $k_z$.

\section{Supplementary Note 7. Hidden exchange, charge equivalence, and the electronic supergroup}
\label{sec:exchange}
\subsection{Transporting the ABAB inversion}

Let $\mathcal I_{12}$ denote the ABAB inversion about the midpoint of neighboring sites on layers 1 and 2. In the chosen coordinates $\mathcal I_{12}$ is
\begin{equation}
 \mathcal I_{12}=\{-\I_3\mid\mathbf b+h\hat{\mathbf z}\}.
\end{equation}
The point matrix $-\I_3$ reverses all three components of a displacement from the inversion center. The operation $\mathcal I_{12}$ exchanges layers $1\leftrightarrow2$ and $3\leftrightarrow4$ modulo translations. Its layer permutation and orbital parity matrix are
\begin{equation}
 \Pi=\begin{pmatrix}0&1&0&0\\1&0&0&0\\0&0&0&1\\0&0&1&0\end{pmatrix},\qquad
 J=\diag[(-1)^{\ell_1},\ldots,(-1)^{\ell_N}].
\end{equation}
The matrix $J$ contains the inversion parities of the retained axial orbitals. For $(s,p_z)$ it is $\diag(1,-1)$. Equation~\eqref{eq:representation} gives the Bloch matrix $D_{12}(\kk)$ of $\mathcal I_{12}$, denoted by $\mathcal I_{12}(\kk)$ in the main text,
\begin{equation}
 D_{12}(\kk)=e^{i\kk\cdot(\mathbf b+h\hat{\mathbf z})}\Pi\otimes J,\qquad
 D_{12}(-\kk)D_{12}(\kk)=\I_{4N}.
 \label{eq:Dg}
\end{equation}
Inversion relates two momenta through $D_{12}(\kk)H_{\rm ABAB}(\kk)D_{12}^\dagger(\kk)=H_{\rm ABAB}(-\kk)$. Moving layer 4 from B to C produces ABAC, whose atomic space group $\Gl$ lacks $\mathcal I_{12}$.

The layer gauge $U_{\rm L}$ of Eq.~\eqref{eq:Ug} maps $D_{12}(\kk)$ to the ABAC exchange $g(\kk)$,
\begin{equation}
 g(\kk)=U_{\rm L}^\dagger(-\kk)D_{12}(\kk)U_{\rm L}(\kk).
 \label{eq:gtilde}
\end{equation}
Substituting Eq.~\eqref{eq:gauge} gives
\begin{align}
 g H_{\rm ABAC}(\kk)g^\dagger
 &=U_{\rm L}^\dagger(-\kk)D_{12}(\kk)H_{\rm ABAB}(\kk)D_{12}^\dagger(\kk)U_{\rm L}(-\kk)\nonumber\\
 &=U_{\rm L}^\dagger(-\kk)H_{\rm ABAB}(-\kk)U_{\rm L}(-\kk)=H_{\rm ABAC}(-\kk).
 \label{eq:covariance}
\end{align}
Since $U_{\rm L}(-\kk)=U_{\rm L}(\kk)^*$, the hidden exchange $g(\kk)$ has the explicit form
\begin{equation}
 g(\kk)=e^{i\kk\cdot(\mathbf b+h\hat{\mathbf z})}
 \begin{pmatrix}0&1&0&0\\1&0&0&0\\0&0&0&\eta\\0&0&\eta&0\end{pmatrix}\otimes J.
 \label{eq:gtilde-matrix}
\end{equation}
The operator is unitary, with $g(-\kk)g(\kk)=\I_{4N}$.

\subsection{Projector exchange and the charge trace}
Let $P_1$ select the orbitals on layers 1 and 3, the WO containing $X_1$, and let $P_2$ select those on layers 2 and 4, the WO containing $X_2$. In the layer basis,
\begin{equation}
 P_1=\diag(1,0,1,0)\otimes\I_N,\qquad
 P_2=\diag(0,1,0,1)\otimes\I_N=\I_{4N}-P_1.
\end{equation}
Each WO contains two sites with $N$ orbitals per site. Since $U_{\rm L}$ commutes with both projectors and $D_{12}$ interchanges them,
\begin{equation}
 g P_1 g^\dagger=P_2,\qquad
 g P_2 g^\dagger=P_1.
 \label{eq:projector-exchange}
\end{equation}
The one-spin equilibrium density matrix is $\rho(\kk)=f_{\rm FD}[H_{\rm ABAC}(\kk)-\mu]$, where $f_{\rm FD}$ is the Fermi--Dirac distribution and $\mu$ is the chemical potential. Its occupations depend only on energy and are equal within each degenerate subspace. The Hamiltonian covariance in Eq.~\eqref{eq:covariance} therefore gives
\begin{equation}
 \rho(-\kk)=g(\kk)\rho(\kk)g^\dagger(\kk).
 \label{eq:rho}
\end{equation}
With normalized Brillouin-zone weights $\sum_{\kk}w_{\kk}=1$, the projected charge per site is
\begin{equation}
 Q_i^{\rm proj}=Q_{\rm ref}^{\rm proj}-\frac{eg_s}{2}
       \sum_{\kk}w_{\kk}\Tr[P_i\rho(\kk)],\qquad g_s=2.
 \label{eq:Qproj}
\end{equation}
The reference is identical on the two elemental sectors, and $1/2$ divides by their two sites. Relabelling the integral or symmetry-complete sum by $\kk\mapsto-\kk$ yields
\begin{align}
 Q_2^{\rm proj}-Q_{\rm ref}^{\rm proj}
 &=-\frac{eg_s}{2}\sum_{\kk}w_{\kk}\Tr[P_2\rho(-\kk)]\nonumber\\
 &=-\frac{eg_s}{2}\sum_{\kk}w_{\kk}\Tr[
   g P_1 g^\dagger g\rho(\kk)g^\dagger]\nonumber\\
 &=-\frac{eg_s}{2}\sum_{\kk}w_{\kk}\Tr[P_1\rho(\kk)]
 =Q_1^{\rm proj}-Q_{\rm ref}^{\rm proj}.
 \label{eq:Qequality}
\end{align}
The second line uses Eqs.~\eqref{eq:projector-exchange} and \eqref{eq:rho}. In the third, cyclic invariance of the trace cancels $g^\dagger g$. Within each WO the sites are related by $\Gl$, so the equality holds for the individual sites $X_1$ and $X_2$.

\subsection{The generated electronic group}
The ABAC lattice inversion $\mathcal I_0=\{-\I_3\mid\mathbf 0\}$ interchanges layers 2 and 4 and fixes layers 1 and 3. Like the other lattice operations in the Bloch representation $D_{\rm ABAC}(\Gl)$, it preserves each WO, while $g$ exchanges the two site projectors [Eq.~\eqref{eq:projector-exchange}]. Together the lattice operations and $g$ generate the unitary electronic group
\begin{equation}
 G=\langle D_{\rm ABAC}(\Gl),g\rangle\supset D_{\rm ABAC}(\Gl).
 \label{eq:generated-group}
\end{equation}
The brackets include products and inverses with their momentum maps. Each product inherits Hamiltonian covariance from its factors. The inclusion is strict because $g$ exchanges the two WOs.

\subsection{Spatial extension used in the ABAC classification entry}

For the ABAC row of Table~\ref{tab:classification-typeII}, the added spatial generator is $\mathcal I_{12}=\{-\mathbb I_3\mid\tfrac13,\tfrac23,\tfrac14\}$. Ordinary translation phases remain part of the space-group representation. Closing $P6_3/mmc$ with $\mathcal I_{12}$ gives 288 operations modulo the original lattice translations, compared with 24 for $P6_3/mmc$. The resulting spatial group is $P6/mmm$, with index 12. A primitive basis of this supergroup is
\[
 \mathbf a'_1=\frac{2\mathbf a_1+\mathbf a_2}{3},\qquad
 \mathbf a'_2=\frac{-\mathbf a_1+\mathbf a_2}{3},\qquad
 \mathbf c'=\frac{c_4}{4}\hat{\mathbf z}.
\]
Its cell volume is $1/12$ of the original cell volume. Table~\ref{tab:classification-typeII} reports this phase-reduced spatial group. The full hidden operator $g(\kk)$ in Eq.~\eqref{eq:gtilde} retains the layer-gauge phase $\eta(\kk)$. This phase generally differs from the Bloch phase of a fixed translation, so the electronic exchange can act nonlocally in real space.

\section{Supplementary Note 8. AB boundary doublets and their protection in ABAC}
\label{sec:boundary}
\subsection{Screw symmetry and the AB boundary}

The AB parent has two layers per period $c_2=2h$. With the origin on the first atom, the second lies at $\mathbf b+h\hat{\mathbf z}$, where $\mathbf b=(\mathbf a_1+2\mathbf a_2)/3$. The screw operation that exchanges the layers is
\begin{equation}
 S_{2z}=\{C_{2z}|\mathbf b+h\hat{\mathbf z}\}.
 \label{eq:screw}
\end{equation}
It sends the first layer to the second and the second to the first in the next AB cell. Because $C_{2z}\mathbf b=-\mathbf b$, the two lateral translations cancel when it is applied twice,
\begin{equation}
 S_{2z}^2=\{E|2h\hat{\mathbf z}\}=T_{c_2}.
 \label{eq:screw-square}
\end{equation}
For real spinless orbitals, $\mathcal T^2=+1$ and time reversal commutes with spatial operations. Define $\Theta=S_{2z}\mathcal T$. Its momentum action and square are
\begin{equation}
 \theta\kk=(k_x,k_y,-k_z),\qquad
 \Theta^2=S_{2z}^2\mathcal T^2=T_{c_2}.
 \label{eq:Theta}
\end{equation}
On $k_z=\pi/c_2$, the initial and final momenta differ by a reciprocal vector of the AB cell. Translation acts on a Bloch state as $T_{c_2}\mapsto e^{-ik_zc_2}=-1$. Thus $\Theta$ leaves the boundary momentum invariant and squares to $-1$ there \cite{wu2018nodalsurface}.

For a boundary eigenstate $H|u\rangle=E|u\rangle$, symmetry gives $H\Theta|u\rangle=E\Theta|u\rangle$. Antiunitarity gives $\langle\Theta v|\Theta w\rangle=\langle w|v\rangle$. Taking $v=u$ and $w=\Theta u$ yields
\begin{equation}
 \langle\Theta u|\Theta^2u\rangle=\langle\Theta u|u\rangle.
 \label{eq:orthogonal-partners}
\end{equation}
The left-hand side equals $-\langle\Theta u|u\rangle$, so the overlap vanishes. The screw and time reversal thus give two orthogonal states of the same energy within a single spin channel.

The boundary loop contains the points
\begin{equation}
 \mathrm A=(0,0,1/2),\qquad \mathrm L=(1/2,0,1/2),\qquad \mathrm H=(1/3,1/3,1/2),
\end{equation}
in fractional reciprocal coordinates of the AB cell. The degeneracy holds along A--L--H--A and throughout the boundary plane.

\subsection{Four-layer folding and reciprocal sewing}
\label{sec:folding}

The ABAB cell doubles the AB repeat, $c_4=2c_2=4h$. Its reciprocal repeat is $\mathbf G_z=(2\pi/c_4)\hat{\mathbf z}$. A parent momentum with $k_z=\pi/c_2=2\pi/c_4$ is therefore folded to $k_z=0$. The high-symmetry points map as
\begin{equation}
 \mathrm A\mapsto(0,0,1)\equiv\Gamma,\qquad
 \mathrm L\mapsto(1/2,0,1)\equiv \mathrm M,\qquad
 \mathrm H\mapsto(1/3,1/3,1)\equiv \mathrm K.
 \label{eq:point-folding}
\end{equation}
The coordinates after the arrows use the four-layer reciprocal basis.

In ABAB, the screw $S_{2z}$ cyclically permutes the four layers. In their order along the stacking axis its layer matrix is
\begin{equation}
 C_{\rm lay}=\begin{pmatrix}0&0&0&1\\1&0&0&0\\0&1&0&0\\0&0&1&0\end{pmatrix}.
 \label{eq:C}
\end{equation}
Every axial orbital has eigenvalue $+1$ under $C_{2z}$. Combining the cyclic permutation with the Bloch phase in Eqs.~\eqref{eq:screw} and \eqref{eq:representation} gives
\begin{equation}
 D_{S_{2z}}(\kk)=e^{i\kk_\parallel\cdot\mathbf b-ik_zh}C_{\rm lay}\otimes\I_N,\qquad
 D_{T_{c_2}}(\kk)=e^{-ik_zc_2}C_{\rm lay}^2\otimes\I_N.
 \label{eq:screw-matrix}
\end{equation}
At basal momentum, the matrix $C_{\rm lay}^2$ has eigenvalues $+1$ and $-1$ in the two folded sectors. Their projectors are
\begin{equation}
 P_\pm^{\rm fold}=\frac{\I_4\pm C_{\rm lay}^2}{2}\otimes\I_N.
 \label{eq:fold-projectors}
\end{equation}
The minus subspace inherits the AB boundary doublets, with both partners in the same subspace. The plus subspace inherits the AB basal bands.

Combining $S_{2z}$ with time reversal $\mathcal T$ gives $\Theta=A(\kk)\mathcal K$, with momentum map $\theta\kk$ from Eq.~\eqref{eq:Theta}, and
\begin{equation}
 A(\kk)=e^{-i\kk_\parallel\cdot\mathbf b+ik_zh}C_{\rm lay}\otimes\I_N,\qquad
 A(\theta\kk)A(\kk)^*=D_{T_{c_2}}(\kk).
 \label{eq:A}
\end{equation}
On the folded basal plane write $\Theta_{\rm fold}=A_{\rm fold}\mathcal K$, where
\begin{equation}
 A_{\rm fold}=e^{-i\kk_\parallel\cdot\mathbf b}C_{\rm lay}\otimes\I_N,\qquad
 A_{\rm fold}A_{\rm fold}^*=C_{\rm lay}^2\otimes\I_N.
 \label{eq:Afold}
\end{equation}
In the layer basis $(e_1-e_3,e_2-e_4)/\sqrt2$ of the minus subspace, $C_{\rm lay}$ is $\bigl(\begin{smallmatrix}0&-1\\1&0\end{smallmatrix}\bigr)$. Thus $C_{\rm lay}^2\otimes\I_N=-\I_{2N}$ in the minus subspace.

To follow the Bloch phases through the fold, take $\mathbf q=\kk_0+\mathbf G_z$ with $k_{0z}=0$. Then $\theta\mathbf q=\mathbf q-2\mathbf G_z$ and
\begin{equation}
 V_{\rm ABAB}(\mathbf G_z)=\diag(1,i,-1,-i)\otimes\I_N.
\end{equation}
Identifying the output fibre at $\theta\mathbf q$ with that at $\mathbf q$, and then moving to the basal fibre, gives
\begin{equation}
 A_{\rm fold}(\kk_0)=V_{\rm ABAB}(\mathbf G_z)
       [V_{\rm ABAB}(-2\mathbf G_z)A(\mathbf q)]V_{\rm ABAB}(\mathbf G_z)^T.
 \label{eq:explicit-sewing}
\end{equation}
Antiunitarity gives the transpose on the right of Eq.~\eqref{eq:explicit-sewing}. Substituting $V_{\rm ABAB}$ recovers the basal-plane matrix in Eq.~\eqref{eq:Afold}.

\subsection{Hidden antiunitary symmetry in ABAC}

The horizontal mirror $m_h$ through layer~2 sends $z$ to $2h-z$. Modulo four-layer translations, it interchanges layers~1 and~3 and fixes layers~2 and~4. It also leaves $\kk_\parallel$ unchanged, so the gauge phase on layer~4 is invariant under $m_h$. The layer gauge therefore preserves this mirror. The identity $U_{\rm L}(-\kk)=U_{\rm L}(\kk)^*$ likewise preserves spinless time reversal. Since $S_{2z}=m_h\mathcal I_{12}$ in ABAB, the transported operations are
\[
 \begin{aligned}
 \widetilde S_{2z}
 &=U_{\rm L}^\dagger S_{2z}U_{\rm L}
 =m_hU_{\rm L}^\dagger\mathcal I_{12}U_{\rm L}=m_hg,\\
 \widetilde\Theta
 &=U_{\rm L}^\dagger\Theta_{\rm fold}U_{\rm L}
 =m_hU_{\rm L}^\dagger\mathcal I_{12}U_{\rm L}\mathcal T=m_hg\mathcal T\in\langle G,\mathcal T\rangle.
 \end{aligned}
\]
With $T_{c_2}$ acting in the common four-layer cell, squaring $\widetilde\Theta$ gives the transported translation,
\[
 \widetilde\Theta^2
 =U_{\rm L}^\dagger\Theta_{\rm fold}^2U_{\rm L}
 =U_{\rm L}^\dagger T_{c_2}U_{\rm L}
 =\widetilde T_{c_2}.
\]

Both $g$ and $\mathcal T$ reverse $\kk$, so their product $g\mathcal T$ leaves momentum unchanged. The remaining mirror $m_h$ reverses only $k_z$, giving
\[
 \widetilde\Theta:(k_x,k_y,k_z)\longmapsto(k_x,k_y,-k_z).
\]
Thus every wavevector on the four-layer basal plane $k_z=0$ is fixed by $\widetilde\Theta$.

To obtain its matrix on the basal plane, write
\begin{equation}
 \widetilde\Theta=\widetilde A\mathcal K,\qquad
 \widetilde A=U_{\rm L}^\dagger A_{\rm fold}U_{\rm L}^*.
 \label{eq:antiunitary-transport}
\end{equation}
Complex conjugation in $\Theta_{\rm fold}$ gives the factor $U_{\rm L}^*$. The ABAB antiunitary covariance then gives $\widetilde A H_{\rm ABAC}^*\widetilde A^\dagger=H_{\rm ABAC}$, or equivalently $H_{\rm ABAC}(\kk)\widetilde\Theta=\widetilde\Theta H_{\rm ABAC}(\kk)$. For an eigenstate $|u_-\rangle$ in the inherited boundary subspace, with $H_{\rm ABAC}(\kk)|u_-\rangle=E|u_-\rangle$, this relation implies
\[
 \begin{aligned}
 H_{\rm ABAC}(\kk)\widetilde\Theta|u_-\rangle
 &=\widetilde\Theta H_{\rm ABAC}(\kk)|u_-\rangle\\
 &=E\widetilde\Theta|u_-\rangle,
 \end{aligned}
\]
where the energy $E$ is real. The transformed state therefore has the same momentum and energy as the original state. Their orthogonality follows from the square of $\widetilde\Theta$ on the inherited boundary subspace. To find this square, multiply the two matrices,
\begin{align}
 \widetilde A\widetilde A^*
 &=U_{\rm L}^\dagger A_{\rm fold}U_{\rm L}^*U_{\rm L}^T A_{\rm fold}^*U_{\rm L}\nonumber\\
 &=U_{\rm L}^\dagger(A_{\rm fold}A_{\rm fold}^*)U_{\rm L}
 =U_{\rm L}^\dagger(C_{\rm lay}^2\otimes\I_N)U_{\rm L}.
 \label{eq:antiunitary-square}
\end{align}
The identity $U_{\rm L}^*U_{\rm L}^T=\I_{4N}$ follows from unitarity. Define the transported boundary projector
\begin{equation}
 \widetilde P_-=U_{\rm L}^\dagger P_-^{\rm fold}U_{\rm L}.
\end{equation}
It commutes with $H_{\rm ABAC}$ because its ABAB counterpart commutes with $H_{\rm ABAB}$. The antiunitary symmetry $\widetilde\Theta$ preserves this boundary subspace and
\begin{equation}
 \widetilde\Theta^2\widetilde P_-=-\widetilde P_-.
 \label{eq:minus-sector}
\end{equation}
By the orthogonality argument in Eq.~\eqref{eq:orthogonal-partners}, eigenstates in $\widetilde P_-$ occur in degenerate pairs. In the plus subspace $\widetilde\Theta^2=+1$, which permits nondegenerate states.

\section{Supplementary Note 9. Charge transfer and doublet splitting}
\label{sec:breakdown}
\subsection{Onsite response and broken symmetries}

When $r<MR$, charge transfer between $X_1$ and $X_2$ changes their local effective potentials. Let $\varepsilon_1$ and $\varepsilon_2$ be their orbital-independent onsite energies. To first order in the projected charge difference,
\begin{equation}
 \Delta\varepsilon=\varepsilon_1-\varepsilon_2
 \simeq\lambda\left(Q_1^{\rm proj}-Q_2^{\rm proj}\right),
 \label{eq:charge-response}
\end{equation}
where $\lambda$ is the linear response coefficient of the onsite energy difference. Absorbing the mean $(\varepsilon_1+\varepsilon_2)/2$ into the charge-equivalent Hamiltonian $H_0$ gives
\begin{equation}
 \Delta H_Q=\frac{\Delta\varepsilon}{2}(P_1-P_2),\qquad
 g\Delta H_Q g^\dagger=-\Delta H_Q.
 \label{eq:DeltaH}
\end{equation}
The cyclic permutation $C_{\rm lay}$ exchanges the odd and even layers. Complex conjugation preserves the real diagonal projectors, and the layer gauge $U_{\rm L}$ commutes with $P_1$ and $P_2$. Consequently,
\begin{equation}
 \widetilde\Theta\Delta H_Q\widetilde\Theta^{-1}=-\Delta H_Q.
\end{equation}
A nonzero $\Delta\varepsilon$ therefore breaks both $g$ and $\widetilde\Theta$.

\subsection{First-order splitting and selection rules}

At a fixed basal momentum choose an isolated protected doublet with normalized partners $|u\rangle$ and $|v\rangle=\widetilde\Theta|u\rangle$. They are orthogonal because $\widetilde\Theta^2=-1$ in this subspace. Consider a small Hermitian perturbation $V$ that is odd under $\widetilde\Theta$,
\begin{equation}
 \widetilde\Theta V\widetilde\Theta^{-1}=-V.
 \label{eq:odd-perturbation}
\end{equation}
The onsite-imbalance term $V=\Delta H_Q$ satisfies this condition by Eq.~\eqref{eq:DeltaH} and its antiunitary counterpart. Since the diagonal expectation of a Hermitian operator is real, antiunitarity gives
\begin{equation}
 \langle v|V|v\rangle=-\langle u|V|u\rangle.
\end{equation}
Write $V_{11}=\langle u|V|u\rangle$ and $V_{12}=\langle u|V|v\rangle$ in this two-state basis, with real $V_{11}$. The first-order matrix and its level separation are
\begin{equation}
 V_{\rm pair}=\begin{pmatrix}V_{11}&V_{12}\\V_{12}^*&-V_{11}\end{pmatrix},\qquad
 \Delta=2\sqrt{V_{11}^2+|V_{12}|^2}.
 \label{eq:doublet-response}
\end{equation}

If the partners $|u_1\rangle=|u\rangle$ and $|u_2\rangle=|v\rangle$ carry distinct one-dimensional irreducible representations of the little group, each operation $\mathcal R$ acts as $D_{\mathcal R}|u_i\rangle=d_i(\mathcal R)|u_i\rangle$. Here $D_{\mathcal R}$ is its matrix in the electronic basis and $d_i(\mathcal R)$ is its eigenvalue on $|u_i\rangle$. A perturbation $V$ preserving this group satisfies $D_{\mathcal R}^\dagger V D_{\mathcal R}=V$, so
\begin{equation}
 V_{12}=\langle u_1|D_{\mathcal R}^\dagger V D_{\mathcal R}|u_2\rangle
       =d_1^*(\mathcal R)d_2(\mathcal R)V_{12}.
\end{equation}
There is an operation for which $d_1^*d_2\ne1$, so $V_{12}=0$. The splitting in Eq.~\eqref{eq:doublet-response} then reduces to $2|V_{11}|$.

At $\Gamma$, the ABAC states $\Gamma_3^-$ and $\Gamma_4^+$ have opposite inversion parities. The charge-transfer potential preserves $\mathcal I_0$, so these states shift without mixing. Substituting $V=\Delta H_Q$ gives
\begin{equation}
 V_{11}=\frac{\Delta\varepsilon}{2}\langle u|P_1-P_2|u\rangle,\qquad
 \Delta\simeq|\Delta\varepsilon|\,|\langle u|P_1-P_2|u\rangle|.
 \label{eq:site-weight-splitting}
\end{equation}
The factor $\langle u|P_1-P_2|u\rangle$ is the difference between the state's weights on the two WOs. Together with Eq.~\eqref{eq:charge-response}, this connects charge imbalance to the first-order doublet splitting.

\subsection{Other orbital and hopping contributions}

The full projected Hamiltonian can also contain orbital and longer-range hopping terms,
\begin{equation}
 H(\kk,P)=H_0(\kk,P)+\Delta H_{\rm orb}(\kk,P)+\Delta H_{\rm lr}(\kk,P)+\Delta H_Q(P).
 \label{eq:corrections}
\end{equation}
The reference Hamiltonian $H_0$ obeys the phase relation in Eq.~\eqref{eq:general-phase}. The correction $\Delta H_{\rm orb}$ includes orbital terms that break this relation. For example, $p_x$ and $p_y$ can give orbital-dependent angular factors, whereas axial orbitals share the common phase derived in Note~6. For orbital-dependent onsite shifts, the difference is an $N\times N$ matrix, whose exchange-odd component breaks $g$ and $\widetilde\Theta$.

The term $\Delta H_{\rm lr}$ contains longer-range interlayer hoppings. A block from layer $j$ to $j+p$ preserves the gauge equivalence if multiplication by $e^{i(\theta_{j+p}-\theta_j)}$ gives the corresponding block in the other stack. Hoppings that violate this condition can split the doublet even when the site charges are equal. Spin--orbit coupling and environment-dependent many-center hoppings likewise require testing Eq.~\eqref{eq:general-phase} in the corresponding orbital basis.

\section{Supplementary Note 10. hP4 geometries, charge-transfer onset, and band spectra}
\label{sec:hp4-validation}
\subsection{Geometry and the geometric criterion}

For hP4 coordinates $(0,0,0)$ and $(1/3,2/3,1/4)$, the hexagonal metric gives
\begin{equation}
 |x\mathbf a_1+y\mathbf a_2|^2=a^2(x^2+y^2-xy),\qquad
 r_{12}=\sqrt{\frac{a^2}{3}+\frac{c_4^2}{16}}.
 \label{eq:r12}
\end{equation}
The vertical separation is $c_4/4=h$. For each element we used $c_4/a$ at the transition to evaluate $r_{12}$ and $M_{\rm hP4}$ (Tables~\ref{tab:hp4} and \ref{tab:hp4-parameters}).

In the layer order of Eq.~\eqref{eq:positions}, the charge deviations are $(+e\delta,-e\delta,+e\delta,-e\delta)$. The four-site cell contains two charge pairs, so the Ewald normalization of Note~3 gives
\begin{equation}
 \Delta E_{\rm inter}=\frac{E_{\rm Ewald}}{2},\qquad
 M_{\rm hP4}=-\frac{r_{12}E_{\rm Ewald}}{2\kappa(e\delta)^2}.
\end{equation}

\begin{table}[htbp]
\centering
\caption{Onset of charge transfer in hP4 alkali metals. The predicted separation is $M_{\rm hP4}R$, and $r$ is the nearest $X_1$--$X_2$ separation at the calculated transition. The signed difference is $\Delta r=r-M_{\rm hP4}R$.}
\label{tab:hp4}
\begin{tabular}{lrrrr}
\toprule Element & $M_{\rm hP4}R$ (\AA) & $r$ (\AA) & $\Delta r$ (\AA) & $\Delta r/(M_{\rm hP4}R)$ (\%)\\\midrule
Li & 1.852 & 1.953 & $+0.101$ & $+5.4$\\
Na & 2.216 & 2.154 & $-0.062$ & $-2.8$\\
K  & 2.745 & 2.939 & $+0.194$ & $+7.1$\\
\bottomrule
\end{tabular}
\end{table}
\begin{table}[htbp]
\centering
\caption{hP4 parameters used in Table~\ref{tab:hp4}. $M_{\rm hP4}$ is the Madelung constant for opposite charge deviations on $X_1$ and $X_2$, and $R$ is the charging radius.}
\label{tab:hp4-parameters}
\begin{tabular}{lrr}
\toprule Element & $M_{\rm hP4}$ & $R$ (\AA)\\\midrule
Li & 1.393 & 1.33\\
Na & 1.4298 & 1.55\\
K  & 1.400 & 1.96\\
\bottomrule
\end{tabular}
\end{table}

\subsection{Band representations and splittings}
\label{sec:irreps}

We use EQ for the charge-equivalent state and CT for the charge-transfer state. We assigned their band irreducible representations with the WIEN2k \texttt{irrep} program \cite{ref18}. The symbol $T$ labels the open K--$\Gamma$ line and $\Sigma$ the open $\Gamma$--M line. We used 50 intervals on each segment. Table~\ref{tab:irreps} gives the assignments at
\begin{equation}
 \kk_T=(1/6,1/6,0),\qquad \kk_\Gamma=(0,0,0),\qquad \kk_\Sigma=(1/4,0,0)
\end{equation}
in fractional reciprocal coordinates of the four-layer cell. A slash orders the two bands by increasing energy, and a direct sum labels the two representations in the EQ pair. The exact degeneracy in Note~8 is a property of the axial orbital Hamiltonian. For the DFT bands, we track the corresponding pair and report its actual energy separation, including residual splittings in the EQ state.

\begin{table}[htbp]
\centering\small
\setlength{\tabcolsep}{3.2pt}
\caption{Irreducible representations and splittings $\Delta$ (meV) of the two hP4 bands in Li, Na, and K. For each element the rows give representative EQ and CT structures.}
\label{tab:irreps}
\begin{tabular}{lccrcrcr}
\toprule
Element & Branch & \multicolumn{2}{c}{K--$\Gamma$} & \multicolumn{2}{c}{$\Gamma$} & \multicolumn{2}{c}{$\Gamma$--M}\\
 & & Irreps & $\Delta$ & Irreps & $\Delta$ & Irreps & $\Delta$\\\midrule
Li & EQ & $T_3\oplus T_4$ & 5.5 & $\Gamma_4^+\oplus\Gamma_3^-$ & 0.1 & $\Sigma_2\oplus\Sigma_1$ & 2.5\\
Li & CT & $T_3/T_4$ & 93.2 & $\Gamma_4^+/\Gamma_3^-$ & 23.1 & $\Sigma_2/\Sigma_1$ & 57.9\\
Na & EQ & $T_3\oplus T_4$ & 6.6 & $\Gamma_3^-\oplus\Gamma_4^+$ & 0.0 & $\Sigma_2\oplus\Sigma_1$ & 3.8\\
Na & CT & $T_3/T_4$ & 757.5 & $\Gamma_4^+/\Gamma_3^-$ & 258.5 & $\Sigma_2/\Sigma_1$ & 590.2\\
K & EQ & $T_3\oplus T_4$ & 18.7 & $\Gamma_4^+\oplus\Gamma_3^-$ & 4.1 & $\Sigma_2\oplus\Sigma_1$ & 12.5\\
K & CT & $T_3/T_4$ & 2016.9 & $\Gamma_4^+/\Gamma_3^-$ & 950.2 & $\Sigma_2/\Sigma_1$ & 1895.9\\
\bottomrule
\end{tabular}
\end{table}

\begin{figure}[!ht]
\centering
\includegraphics[width=0.96\linewidth,height=0.80\textheight,keepaspectratio]{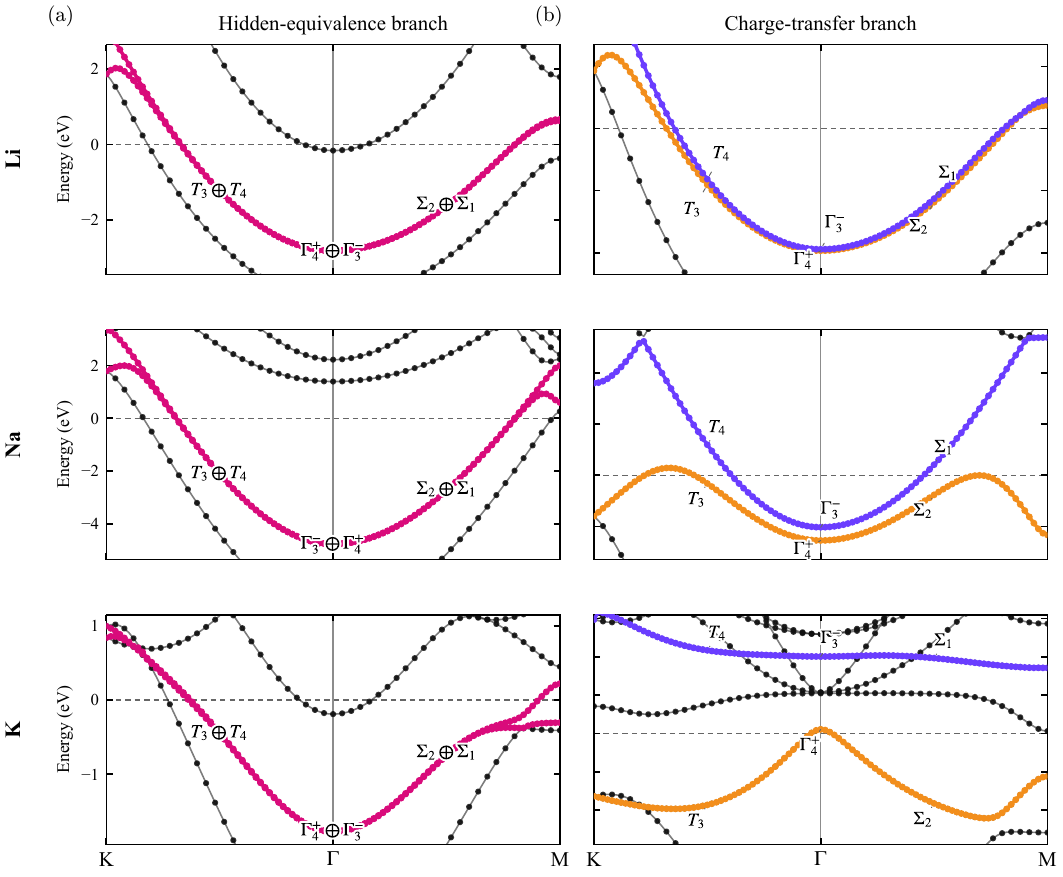}
\caption{hP4 bands of Li, Na, and K for the EQ (a) and CT (b) structures in Table~\ref{tab:irreps}. Each panel uses the Fermi energy of its structure as the energy zero. Magenta marks the tracked pair of bands corresponding to the model doublet. Purple and orange distinguish the two bands in the CT state. In K their representations follow $T_3/T_4\rightarrow\Gamma_4^+/\Gamma_3^-\rightarrow\Sigma_2/\Sigma_1$ through the crossings.}
\label{fig:bands}
\end{figure}

Figure~\ref{fig:bands} follows the pair of DFT bands corresponding to the model doublet from the EQ to the CT state. For K we used the first converged CT structure beyond the transition. At allowed crossings the energy ordering changes, so we tracked the two states by the compatibility of their irreducible representations.

\clearpage
\subsection{Pressure evolution of the stacking spectra}
\begin{figure}[!htbp]
\centering
\includegraphics[width=0.98\linewidth]{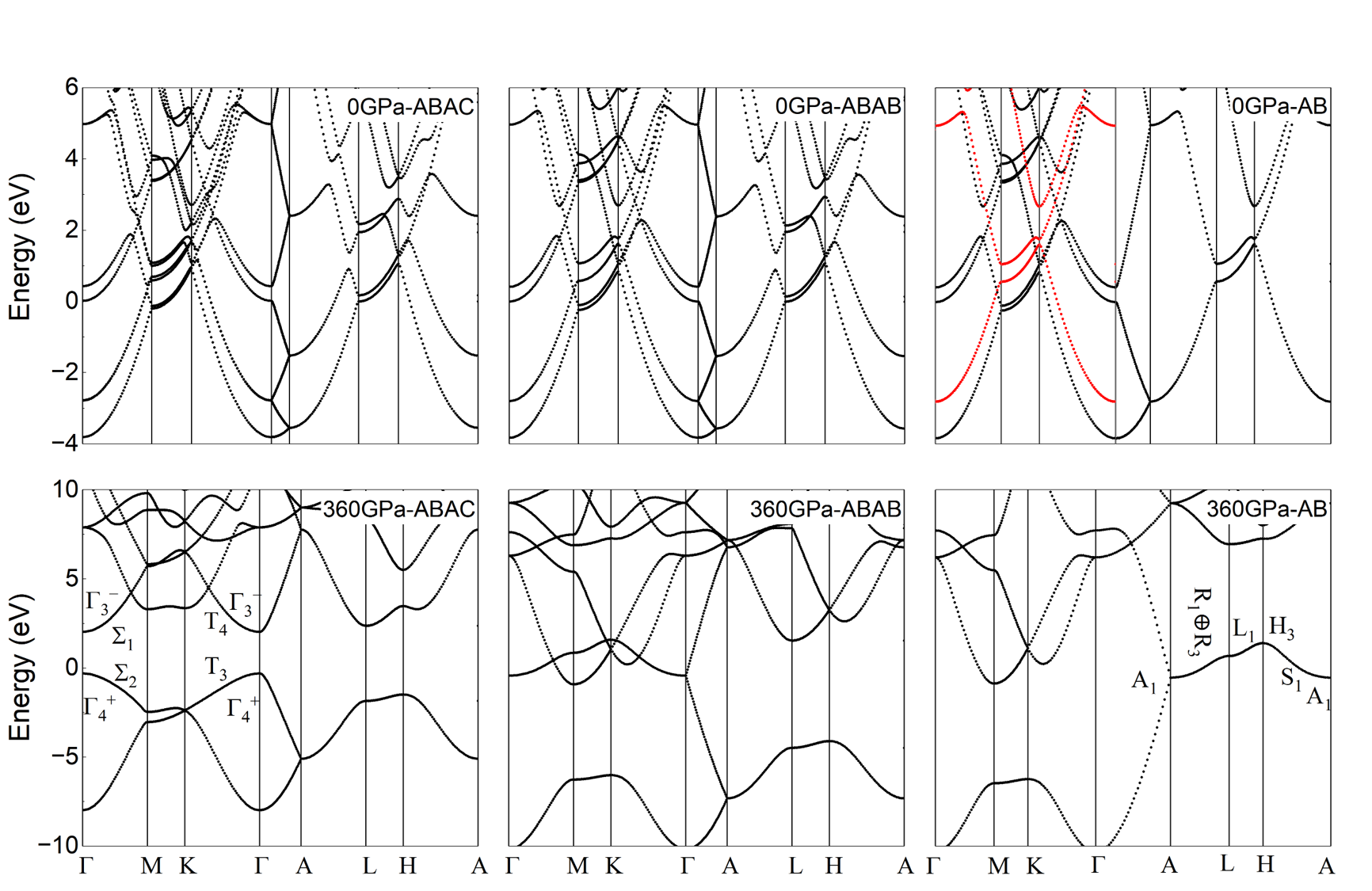}
\caption{Na bands at 0~GPa (top) and 360~GPa (bottom) for ABAC, ABAB, and the two-layer AB parent, from left to right. Each column uses its own crystallographic Brillouin zone. In the upper right panel, red shows the AB boundary bands on A--L--H--A at the in-plane coordinates where they fold onto $\Gamma$--M--K--$\Gamma$ in ABAB. Main-text Fig.~3(b) shows the original AB boundary path beside the two four-layer basal paths.}
\label{fig:stackings}
\end{figure}

The gauge equivalence in Note~6 relates the spectra of ABAC and ABAB. Figure~\ref{fig:stackings} compares the DFT bands of these stackings with those of the two-layer AB parent at 0 and 360~GPa. At 0~GPa the ABAC bands closely follow the folded ABAB spectrum; at 360~GPa the two four-layer spectra differ near the Fermi level.

Within ABAC, the pressure-driven separation of the $\Gamma_3^-$ and $\Gamma_4^+$ branches tracked in main-text Fig.~4 reduces valence--conduction overlap, as described by the symmetry-breaking mechanism in Note~9. Let $E_c(\kk)$ and $E_v(\kk)$ denote the lowest conduction-band and highest valence-band energies at each $\kk$ for the fixed electron filling. When the overlap is removed throughout the Brillouin zone, the system has a positive indirect gap,
\[
 E_g=\min_{\kk}E_c(\kk)-\max_{\kk}E_v(\kk)>0.
\]

\clearpage
\bibliography{references}
\clearpage\thispagestyle{plain}
\refstepcounter{table}\label{tab:classification-typeI}
\addcontentsline{toc}{section}{Table S5. Type-I generator relations}
\pdfbookmark[2]{P1 (1)}{table-typeI-sg1}
\pdfbookmark[2]{P-1 (2)}{table-typeI-sg2}
\pdfbookmark[2]{P2 (3)}{table-typeI-sg3}
\pdfbookmark[2]{P21 (4)}{table-typeI-sg4}
\pdfbookmark[2]{C2 (5)}{table-typeI-sg5}
\pdfbookmark[2]{Pm (6)}{table-typeI-sg6}
\pdfbookmark[2]{Pc (7)}{table-typeI-sg7}
\pdfbookmark[2]{Cm (8)}{table-typeI-sg8}
\noindent\vbox to\dimexpr\textheight-4pt\relax{\vfil\hbox to\textwidth{\hfil\includegraphics[page=1,trim=18mm 15mm 18mm 17mm,clip,width=\textwidth,height=\dimexpr\textheight-4pt\relax,keepaspectratio]{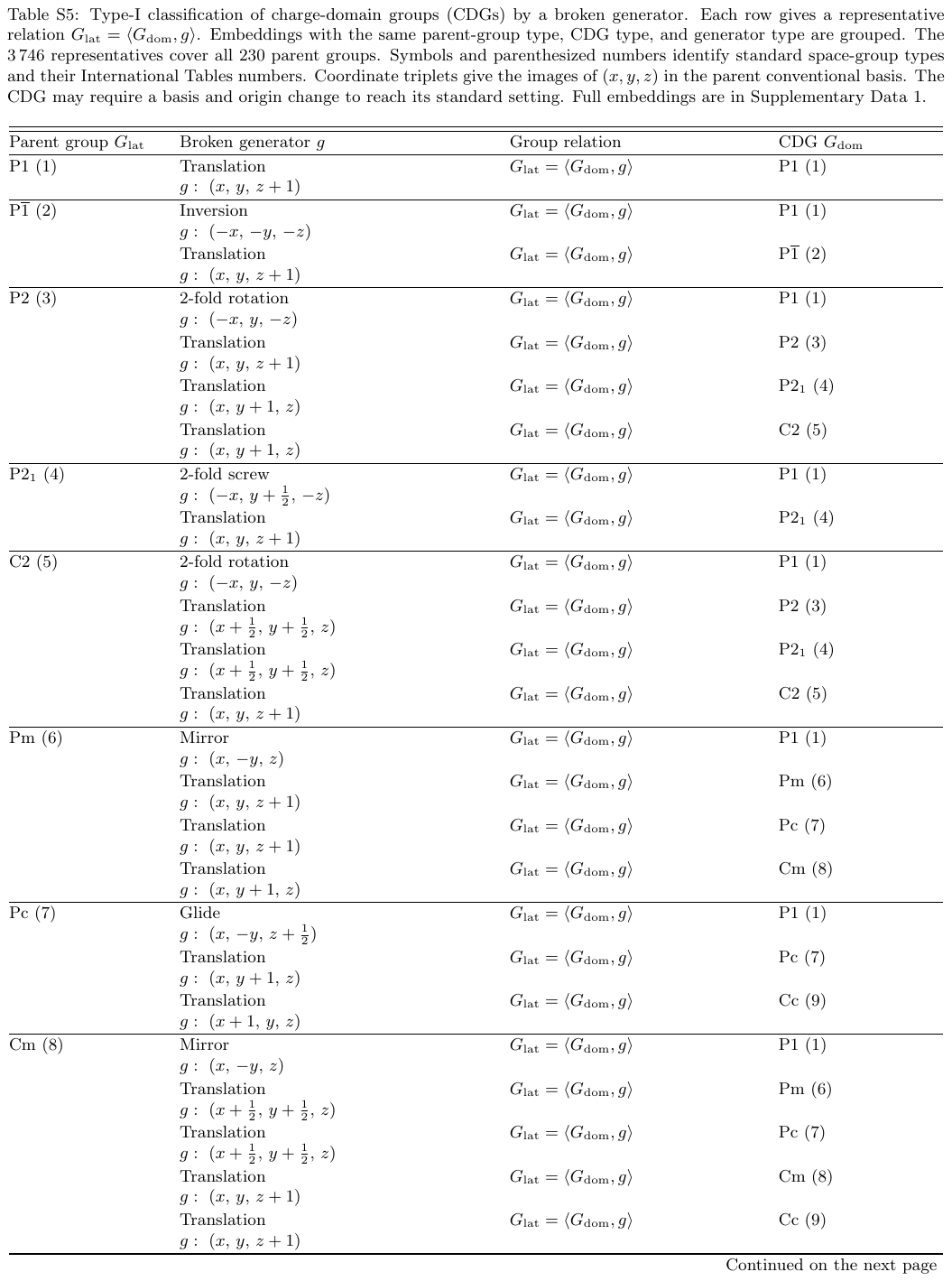}\hfil}\vfil}\par
\clearpage\thispagestyle{plain}
\pdfbookmark[2]{Cc (9)}{table-typeI-sg9}
\pdfbookmark[2]{P2/m (10)}{table-typeI-sg10}
\pdfbookmark[2]{P21/m (11)}{table-typeI-sg11}
\pdfbookmark[2]{C2/m (12)}{table-typeI-sg12}
\pdfbookmark[2]{P2/c (13)}{table-typeI-sg13}
\noindent\vbox to\dimexpr\textheight-4pt\relax{\vfil\hbox to\textwidth{\hfil\includegraphics[page=2,trim=18mm 15mm 18mm 17mm,clip,width=\textwidth,height=\dimexpr\textheight-4pt\relax,keepaspectratio]{classification_typeI_embed.pdf}\hfil}\vfil}\par
\clearpage\thispagestyle{plain}
\pdfbookmark[2]{P21/c (14)}{table-typeI-sg14}
\pdfbookmark[2]{C2/c (15)}{table-typeI-sg15}
\pdfbookmark[2]{P222 (16)}{table-typeI-sg16}
\pdfbookmark[2]{P2221 (17)}{table-typeI-sg17}
\noindent\vbox to\dimexpr\textheight-4pt\relax{\vfil\hbox to\textwidth{\hfil\includegraphics[page=3,trim=18mm 15mm 18mm 17mm,clip,width=\textwidth,height=\dimexpr\textheight-4pt\relax,keepaspectratio]{classification_typeI_embed.pdf}\hfil}\vfil}\par
\clearpage\thispagestyle{plain}
\pdfbookmark[2]{P21212 (18)}{table-typeI-sg18}
\pdfbookmark[2]{P212121 (19)}{table-typeI-sg19}
\pdfbookmark[2]{C2221 (20)}{table-typeI-sg20}
\pdfbookmark[2]{C222 (21)}{table-typeI-sg21}
\pdfbookmark[2]{F222 (22)}{table-typeI-sg22}
\noindent\vbox to\dimexpr\textheight-4pt\relax{\vfil\hbox to\textwidth{\hfil\includegraphics[page=4,trim=18mm 15mm 18mm 17mm,clip,width=\textwidth,height=\dimexpr\textheight-4pt\relax,keepaspectratio]{classification_typeI_embed.pdf}\hfil}\vfil}\par
\clearpage\thispagestyle{plain}
\pdfbookmark[2]{I222 (23)}{table-typeI-sg23}
\pdfbookmark[2]{I212121 (24)}{table-typeI-sg24}
\pdfbookmark[2]{Pmm2 (25)}{table-typeI-sg25}
\pdfbookmark[2]{Pmc21 (26)}{table-typeI-sg26}
\noindent\vbox to\dimexpr\textheight-4pt\relax{\vfil\hbox to\textwidth{\hfil\includegraphics[page=5,trim=18mm 15mm 18mm 17mm,clip,width=\textwidth,height=\dimexpr\textheight-4pt\relax,keepaspectratio]{classification_typeI_embed.pdf}\hfil}\vfil}\par
\clearpage\thispagestyle{plain}
\pdfbookmark[2]{Pcc2 (27)}{table-typeI-sg27}
\pdfbookmark[2]{Pma2 (28)}{table-typeI-sg28}
\pdfbookmark[2]{Pca21 (29)}{table-typeI-sg29}
\noindent\vbox to\dimexpr\textheight-4pt\relax{\vfil\hbox to\textwidth{\hfil\includegraphics[page=6,trim=18mm 15mm 18mm 17mm,clip,width=\textwidth,height=\dimexpr\textheight-4pt\relax,keepaspectratio]{classification_typeI_embed.pdf}\hfil}\vfil}\par
\clearpage\thispagestyle{plain}
\pdfbookmark[2]{Pnc2 (30)}{table-typeI-sg30}
\pdfbookmark[2]{Pmn21 (31)}{table-typeI-sg31}
\pdfbookmark[2]{Pba2 (32)}{table-typeI-sg32}
\pdfbookmark[2]{Pna21 (33)}{table-typeI-sg33}
\noindent\vbox to\dimexpr\textheight-4pt\relax{\vfil\hbox to\textwidth{\hfil\includegraphics[page=7,trim=18mm 15mm 18mm 17mm,clip,width=\textwidth,height=\dimexpr\textheight-4pt\relax,keepaspectratio]{classification_typeI_embed.pdf}\hfil}\vfil}\par
\clearpage\thispagestyle{plain}
\pdfbookmark[2]{Pnn2 (34)}{table-typeI-sg34}
\pdfbookmark[2]{Cmm2 (35)}{table-typeI-sg35}
\pdfbookmark[2]{Cmc21 (36)}{table-typeI-sg36}
\noindent\vbox to\dimexpr\textheight-4pt\relax{\vfil\hbox to\textwidth{\hfil\includegraphics[page=8,trim=18mm 15mm 18mm 17mm,clip,width=\textwidth,height=\dimexpr\textheight-4pt\relax,keepaspectratio]{classification_typeI_embed.pdf}\hfil}\vfil}\par
\clearpage\thispagestyle{plain}
\pdfbookmark[2]{Ccc2 (37)}{table-typeI-sg37}
\pdfbookmark[2]{Amm2 (38)}{table-typeI-sg38}
\pdfbookmark[2]{Aem2 (39)}{table-typeI-sg39}
\noindent\vbox to\dimexpr\textheight-4pt\relax{\vfil\hbox to\textwidth{\hfil\includegraphics[page=9,trim=18mm 15mm 18mm 17mm,clip,width=\textwidth,height=\dimexpr\textheight-4pt\relax,keepaspectratio]{classification_typeI_embed.pdf}\hfil}\vfil}\par
\clearpage\thispagestyle{plain}
\pdfbookmark[2]{Ama2 (40)}{table-typeI-sg40}
\pdfbookmark[2]{Aea2 (41)}{table-typeI-sg41}
\pdfbookmark[2]{Fmm2 (42)}{table-typeI-sg42}
\noindent\vbox to\dimexpr\textheight-4pt\relax{\vfil\hbox to\textwidth{\hfil\includegraphics[page=10,trim=18mm 15mm 18mm 17mm,clip,width=\textwidth,height=\dimexpr\textheight-4pt\relax,keepaspectratio]{classification_typeI_embed.pdf}\hfil}\vfil}\par
\clearpage\thispagestyle{plain}
\pdfbookmark[2]{Fdd2 (43)}{table-typeI-sg43}
\pdfbookmark[2]{Imm2 (44)}{table-typeI-sg44}
\pdfbookmark[2]{Iba2 (45)}{table-typeI-sg45}
\pdfbookmark[2]{Ima2 (46)}{table-typeI-sg46}
\noindent\vbox to\dimexpr\textheight-4pt\relax{\vfil\hbox to\textwidth{\hfil\includegraphics[page=11,trim=18mm 15mm 18mm 17mm,clip,width=\textwidth,height=\dimexpr\textheight-4pt\relax,keepaspectratio]{classification_typeI_embed.pdf}\hfil}\vfil}\par
\clearpage\thispagestyle{plain}
\pdfbookmark[2]{Pmmm (47)}{table-typeI-sg47}
\pdfbookmark[2]{Pnnn (48)}{table-typeI-sg48}
\pdfbookmark[2]{Pccm (49)}{table-typeI-sg49}
\noindent\vbox to\dimexpr\textheight-4pt\relax{\vfil\hbox to\textwidth{\hfil\includegraphics[page=12,trim=18mm 15mm 18mm 17mm,clip,width=\textwidth,height=\dimexpr\textheight-4pt\relax,keepaspectratio]{classification_typeI_embed.pdf}\hfil}\vfil}\par
\clearpage\thispagestyle{plain}
\pdfbookmark[2]{Pban (50)}{table-typeI-sg50}
\pdfbookmark[2]{Pmma (51)}{table-typeI-sg51}
\noindent\vbox to\dimexpr\textheight-4pt\relax{\vfil\hbox to\textwidth{\hfil\includegraphics[page=13,trim=18mm 15mm 18mm 17mm,clip,width=\textwidth,height=\dimexpr\textheight-4pt\relax,keepaspectratio]{classification_typeI_embed.pdf}\hfil}\vfil}\par
\clearpage\thispagestyle{plain}
\pdfbookmark[2]{Pnna (52)}{table-typeI-sg52}
\noindent\vbox to\dimexpr\textheight-4pt\relax{\vfil\hbox to\textwidth{\hfil\includegraphics[page=14,trim=18mm 15mm 18mm 17mm,clip,width=\textwidth,height=\dimexpr\textheight-4pt\relax,keepaspectratio]{classification_typeI_embed.pdf}\hfil}\vfil}\par
\clearpage\thispagestyle{plain}
\pdfbookmark[2]{Pmna (53)}{table-typeI-sg53}
\pdfbookmark[2]{Pcca (54)}{table-typeI-sg54}
\noindent\vbox to\dimexpr\textheight-4pt\relax{\vfil\hbox to\textwidth{\hfil\includegraphics[page=15,trim=18mm 15mm 18mm 17mm,clip,width=\textwidth,height=\dimexpr\textheight-4pt\relax,keepaspectratio]{classification_typeI_embed.pdf}\hfil}\vfil}\par
\clearpage\thispagestyle{plain}
\pdfbookmark[2]{Pbam (55)}{table-typeI-sg55}
\pdfbookmark[2]{Pccn (56)}{table-typeI-sg56}
\noindent\vbox to\dimexpr\textheight-4pt\relax{\vfil\hbox to\textwidth{\hfil\includegraphics[page=16,trim=18mm 15mm 18mm 17mm,clip,width=\textwidth,height=\dimexpr\textheight-4pt\relax,keepaspectratio]{classification_typeI_embed.pdf}\hfil}\vfil}\par
\clearpage\thispagestyle{plain}
\pdfbookmark[2]{Pbcm (57)}{table-typeI-sg57}
\pdfbookmark[2]{Pnnm (58)}{table-typeI-sg58}
\noindent\vbox to\dimexpr\textheight-4pt\relax{\vfil\hbox to\textwidth{\hfil\includegraphics[page=17,trim=18mm 15mm 18mm 17mm,clip,width=\textwidth,height=\dimexpr\textheight-4pt\relax,keepaspectratio]{classification_typeI_embed.pdf}\hfil}\vfil}\par
\clearpage\thispagestyle{plain}
\pdfbookmark[2]{Pmmn (59)}{table-typeI-sg59}
\pdfbookmark[2]{Pbcn (60)}{table-typeI-sg60}
\noindent\vbox to\dimexpr\textheight-4pt\relax{\vfil\hbox to\textwidth{\hfil\includegraphics[page=18,trim=18mm 15mm 18mm 17mm,clip,width=\textwidth,height=\dimexpr\textheight-4pt\relax,keepaspectratio]{classification_typeI_embed.pdf}\hfil}\vfil}\par
\clearpage\thispagestyle{plain}
\pdfbookmark[2]{Pbca (61)}{table-typeI-sg61}
\pdfbookmark[2]{Pnma (62)}{table-typeI-sg62}
\pdfbookmark[2]{Cmcm (63)}{table-typeI-sg63}
\noindent\vbox to\dimexpr\textheight-4pt\relax{\vfil\hbox to\textwidth{\hfil\includegraphics[page=19,trim=18mm 15mm 18mm 17mm,clip,width=\textwidth,height=\dimexpr\textheight-4pt\relax,keepaspectratio]{classification_typeI_embed.pdf}\hfil}\vfil}\par
\clearpage\thispagestyle{plain}
\pdfbookmark[2]{Cmce (64)}{table-typeI-sg64}
\pdfbookmark[2]{Cmmm (65)}{table-typeI-sg65}
\noindent\vbox to\dimexpr\textheight-4pt\relax{\vfil\hbox to\textwidth{\hfil\includegraphics[page=20,trim=18mm 15mm 18mm 17mm,clip,width=\textwidth,height=\dimexpr\textheight-4pt\relax,keepaspectratio]{classification_typeI_embed.pdf}\hfil}\vfil}\par
\clearpage\thispagestyle{plain}
\pdfbookmark[2]{Cccm (66)}{table-typeI-sg66}
\noindent\vbox to\dimexpr\textheight-4pt\relax{\vfil\hbox to\textwidth{\hfil\includegraphics[page=21,trim=18mm 15mm 18mm 17mm,clip,width=\textwidth,height=\dimexpr\textheight-4pt\relax,keepaspectratio]{classification_typeI_embed.pdf}\hfil}\vfil}\par
\clearpage\thispagestyle{plain}
\pdfbookmark[2]{Cmme (67)}{table-typeI-sg67}
\pdfbookmark[2]{Ccce (68)}{table-typeI-sg68}
\noindent\vbox to\dimexpr\textheight-4pt\relax{\vfil\hbox to\textwidth{\hfil\includegraphics[page=22,trim=18mm 15mm 18mm 17mm,clip,width=\textwidth,height=\dimexpr\textheight-4pt\relax,keepaspectratio]{classification_typeI_embed.pdf}\hfil}\vfil}\par
\clearpage\thispagestyle{plain}
\pdfbookmark[2]{Fmmm (69)}{table-typeI-sg69}
\pdfbookmark[2]{Fddd (70)}{table-typeI-sg70}
\pdfbookmark[2]{Immm (71)}{table-typeI-sg71}
\noindent\vbox to\dimexpr\textheight-4pt\relax{\vfil\hbox to\textwidth{\hfil\includegraphics[page=23,trim=18mm 15mm 18mm 17mm,clip,width=\textwidth,height=\dimexpr\textheight-4pt\relax,keepaspectratio]{classification_typeI_embed.pdf}\hfil}\vfil}\par
\clearpage\thispagestyle{plain}
\pdfbookmark[2]{Ibam (72)}{table-typeI-sg72}
\pdfbookmark[2]{Ibca (73)}{table-typeI-sg73}
\pdfbookmark[2]{Imma (74)}{table-typeI-sg74}
\noindent\vbox to\dimexpr\textheight-4pt\relax{\vfil\hbox to\textwidth{\hfil\includegraphics[page=24,trim=18mm 15mm 18mm 17mm,clip,width=\textwidth,height=\dimexpr\textheight-4pt\relax,keepaspectratio]{classification_typeI_embed.pdf}\hfil}\vfil}\par
\clearpage\thispagestyle{plain}
\pdfbookmark[2]{P4 (75)}{table-typeI-sg75}
\pdfbookmark[2]{P41 (76)}{table-typeI-sg76}
\pdfbookmark[2]{P42 (77)}{table-typeI-sg77}
\pdfbookmark[2]{P43 (78)}{table-typeI-sg78}
\noindent\vbox to\dimexpr\textheight-4pt\relax{\vfil\hbox to\textwidth{\hfil\includegraphics[page=25,trim=18mm 15mm 18mm 17mm,clip,width=\textwidth,height=\dimexpr\textheight-4pt\relax,keepaspectratio]{classification_typeI_embed.pdf}\hfil}\vfil}\par
\clearpage\thispagestyle{plain}
\pdfbookmark[2]{I4 (79)}{table-typeI-sg79}
\pdfbookmark[2]{I41 (80)}{table-typeI-sg80}
\pdfbookmark[2]{P-4 (81)}{table-typeI-sg81}
\pdfbookmark[2]{I-4 (82)}{table-typeI-sg82}
\pdfbookmark[2]{P4/m (83)}{table-typeI-sg83}
\noindent\vbox to\dimexpr\textheight-4pt\relax{\vfil\hbox to\textwidth{\hfil\includegraphics[page=26,trim=18mm 15mm 18mm 17mm,clip,width=\textwidth,height=\dimexpr\textheight-4pt\relax,keepaspectratio]{classification_typeI_embed.pdf}\hfil}\vfil}\par
\clearpage\thispagestyle{plain}
\pdfbookmark[2]{P42/m (84)}{table-typeI-sg84}
\pdfbookmark[2]{P4/n (85)}{table-typeI-sg85}
\noindent\vbox to\dimexpr\textheight-4pt\relax{\vfil\hbox to\textwidth{\hfil\includegraphics[page=27,trim=18mm 15mm 18mm 17mm,clip,width=\textwidth,height=\dimexpr\textheight-4pt\relax,keepaspectratio]{classification_typeI_embed.pdf}\hfil}\vfil}\par
\clearpage\thispagestyle{plain}
\pdfbookmark[2]{P42/n (86)}{table-typeI-sg86}
\pdfbookmark[2]{I4/m (87)}{table-typeI-sg87}
\noindent\vbox to\dimexpr\textheight-4pt\relax{\vfil\hbox to\textwidth{\hfil\includegraphics[page=28,trim=18mm 15mm 18mm 17mm,clip,width=\textwidth,height=\dimexpr\textheight-4pt\relax,keepaspectratio]{classification_typeI_embed.pdf}\hfil}\vfil}\par
\clearpage\thispagestyle{plain}
\pdfbookmark[2]{I41/a (88)}{table-typeI-sg88}
\pdfbookmark[2]{P422 (89)}{table-typeI-sg89}
\noindent\vbox to\dimexpr\textheight-4pt\relax{\vfil\hbox to\textwidth{\hfil\includegraphics[page=29,trim=18mm 15mm 18mm 17mm,clip,width=\textwidth,height=\dimexpr\textheight-4pt\relax,keepaspectratio]{classification_typeI_embed.pdf}\hfil}\vfil}\par
\clearpage\thispagestyle{plain}
\pdfbookmark[2]{P4212 (90)}{table-typeI-sg90}
\pdfbookmark[2]{P4122 (91)}{table-typeI-sg91}
\noindent\vbox to\dimexpr\textheight-4pt\relax{\vfil\hbox to\textwidth{\hfil\includegraphics[page=30,trim=18mm 15mm 18mm 17mm,clip,width=\textwidth,height=\dimexpr\textheight-4pt\relax,keepaspectratio]{classification_typeI_embed.pdf}\hfil}\vfil}\par
\clearpage\thispagestyle{plain}
\pdfbookmark[2]{P41212 (92)}{table-typeI-sg92}
\pdfbookmark[2]{P4222 (93)}{table-typeI-sg93}
\noindent\vbox to\dimexpr\textheight-4pt\relax{\vfil\hbox to\textwidth{\hfil\includegraphics[page=31,trim=18mm 15mm 18mm 17mm,clip,width=\textwidth,height=\dimexpr\textheight-4pt\relax,keepaspectratio]{classification_typeI_embed.pdf}\hfil}\vfil}\par
\clearpage\thispagestyle{plain}
\pdfbookmark[2]{P42212 (94)}{table-typeI-sg94}
\pdfbookmark[2]{P4322 (95)}{table-typeI-sg95}
\noindent\vbox to\dimexpr\textheight-4pt\relax{\vfil\hbox to\textwidth{\hfil\includegraphics[page=32,trim=18mm 15mm 18mm 17mm,clip,width=\textwidth,height=\dimexpr\textheight-4pt\relax,keepaspectratio]{classification_typeI_embed.pdf}\hfil}\vfil}\par
\clearpage\thispagestyle{plain}
\pdfbookmark[2]{P43212 (96)}{table-typeI-sg96}
\pdfbookmark[2]{I422 (97)}{table-typeI-sg97}
\noindent\vbox to\dimexpr\textheight-4pt\relax{\vfil\hbox to\textwidth{\hfil\includegraphics[page=33,trim=18mm 15mm 18mm 17mm,clip,width=\textwidth,height=\dimexpr\textheight-4pt\relax,keepaspectratio]{classification_typeI_embed.pdf}\hfil}\vfil}\par
\clearpage\thispagestyle{plain}
\pdfbookmark[2]{I4122 (98)}{table-typeI-sg98}
\pdfbookmark[2]{P4mm (99)}{table-typeI-sg99}
\noindent\vbox to\dimexpr\textheight-4pt\relax{\vfil\hbox to\textwidth{\hfil\includegraphics[page=34,trim=18mm 15mm 18mm 17mm,clip,width=\textwidth,height=\dimexpr\textheight-4pt\relax,keepaspectratio]{classification_typeI_embed.pdf}\hfil}\vfil}\par
\clearpage\thispagestyle{plain}
\pdfbookmark[2]{P4bm (100)}{table-typeI-sg100}
\noindent\vbox to\dimexpr\textheight-4pt\relax{\vfil\hbox to\textwidth{\hfil\includegraphics[page=35,trim=18mm 15mm 18mm 17mm,clip,width=\textwidth,height=\dimexpr\textheight-4pt\relax,keepaspectratio]{classification_typeI_embed.pdf}\hfil}\vfil}\par
\clearpage\thispagestyle{plain}
\pdfbookmark[2]{P42cm (101)}{table-typeI-sg101}
\pdfbookmark[2]{P42nm (102)}{table-typeI-sg102}
\noindent\vbox to\dimexpr\textheight-4pt\relax{\vfil\hbox to\textwidth{\hfil\includegraphics[page=36,trim=18mm 15mm 18mm 17mm,clip,width=\textwidth,height=\dimexpr\textheight-4pt\relax,keepaspectratio]{classification_typeI_embed.pdf}\hfil}\vfil}\par
\clearpage\thispagestyle{plain}
\pdfbookmark[2]{P4cc (103)}{table-typeI-sg103}
\pdfbookmark[2]{P4nc (104)}{table-typeI-sg104}
\noindent\vbox to\dimexpr\textheight-4pt\relax{\vfil\hbox to\textwidth{\hfil\includegraphics[page=37,trim=18mm 15mm 18mm 17mm,clip,width=\textwidth,height=\dimexpr\textheight-4pt\relax,keepaspectratio]{classification_typeI_embed.pdf}\hfil}\vfil}\par
\clearpage\thispagestyle{plain}
\pdfbookmark[2]{P42mc (105)}{table-typeI-sg105}
\pdfbookmark[2]{P42bc (106)}{table-typeI-sg106}
\noindent\vbox to\dimexpr\textheight-4pt\relax{\vfil\hbox to\textwidth{\hfil\includegraphics[page=38,trim=18mm 15mm 18mm 17mm,clip,width=\textwidth,height=\dimexpr\textheight-4pt\relax,keepaspectratio]{classification_typeI_embed.pdf}\hfil}\vfil}\par
\clearpage\thispagestyle{plain}
\pdfbookmark[2]{I4mm (107)}{table-typeI-sg107}
\pdfbookmark[2]{I4cm (108)}{table-typeI-sg108}
\noindent\vbox to\dimexpr\textheight-4pt\relax{\vfil\hbox to\textwidth{\hfil\includegraphics[page=39,trim=18mm 15mm 18mm 17mm,clip,width=\textwidth,height=\dimexpr\textheight-4pt\relax,keepaspectratio]{classification_typeI_embed.pdf}\hfil}\vfil}\par
\clearpage\thispagestyle{plain}
\pdfbookmark[2]{I41md (109)}{table-typeI-sg109}
\noindent\vbox to\dimexpr\textheight-4pt\relax{\vfil\hbox to\textwidth{\hfil\includegraphics[page=40,trim=18mm 15mm 18mm 17mm,clip,width=\textwidth,height=\dimexpr\textheight-4pt\relax,keepaspectratio]{classification_typeI_embed.pdf}\hfil}\vfil}\par
\clearpage\thispagestyle{plain}
\pdfbookmark[2]{I41cd (110)}{table-typeI-sg110}
\pdfbookmark[2]{P-42m (111)}{table-typeI-sg111}
\noindent\vbox to\dimexpr\textheight-4pt\relax{\vfil\hbox to\textwidth{\hfil\includegraphics[page=41,trim=18mm 15mm 18mm 17mm,clip,width=\textwidth,height=\dimexpr\textheight-4pt\relax,keepaspectratio]{classification_typeI_embed.pdf}\hfil}\vfil}\par
\clearpage\thispagestyle{plain}
\pdfbookmark[2]{P-42c (112)}{table-typeI-sg112}
\pdfbookmark[2]{P-421m (113)}{table-typeI-sg113}
\noindent\vbox to\dimexpr\textheight-4pt\relax{\vfil\hbox to\textwidth{\hfil\includegraphics[page=42,trim=18mm 15mm 18mm 17mm,clip,width=\textwidth,height=\dimexpr\textheight-4pt\relax,keepaspectratio]{classification_typeI_embed.pdf}\hfil}\vfil}\par
\clearpage\thispagestyle{plain}
\pdfbookmark[2]{P-421c (114)}{table-typeI-sg114}
\pdfbookmark[2]{P-4m2 (115)}{table-typeI-sg115}
\noindent\vbox to\dimexpr\textheight-4pt\relax{\vfil\hbox to\textwidth{\hfil\includegraphics[page=43,trim=18mm 15mm 18mm 17mm,clip,width=\textwidth,height=\dimexpr\textheight-4pt\relax,keepaspectratio]{classification_typeI_embed.pdf}\hfil}\vfil}\par
\clearpage\thispagestyle{plain}
\pdfbookmark[2]{P-4c2 (116)}{table-typeI-sg116}
\pdfbookmark[2]{P-4b2 (117)}{table-typeI-sg117}
\noindent\vbox to\dimexpr\textheight-4pt\relax{\vfil\hbox to\textwidth{\hfil\includegraphics[page=44,trim=18mm 15mm 18mm 17mm,clip,width=\textwidth,height=\dimexpr\textheight-4pt\relax,keepaspectratio]{classification_typeI_embed.pdf}\hfil}\vfil}\par
\clearpage\thispagestyle{plain}
\pdfbookmark[2]{P-4n2 (118)}{table-typeI-sg118}
\noindent\vbox to\dimexpr\textheight-4pt\relax{\vfil\hbox to\textwidth{\hfil\includegraphics[page=45,trim=18mm 15mm 18mm 17mm,clip,width=\textwidth,height=\dimexpr\textheight-4pt\relax,keepaspectratio]{classification_typeI_embed.pdf}\hfil}\vfil}\par
\clearpage\thispagestyle{plain}
\pdfbookmark[2]{I-4m2 (119)}{table-typeI-sg119}
\pdfbookmark[2]{I-4c2 (120)}{table-typeI-sg120}
\noindent\vbox to\dimexpr\textheight-4pt\relax{\vfil\hbox to\textwidth{\hfil\includegraphics[page=46,trim=18mm 15mm 18mm 17mm,clip,width=\textwidth,height=\dimexpr\textheight-4pt\relax,keepaspectratio]{classification_typeI_embed.pdf}\hfil}\vfil}\par
\clearpage\thispagestyle{plain}
\pdfbookmark[2]{I-42m (121)}{table-typeI-sg121}
\pdfbookmark[2]{I-42d (122)}{table-typeI-sg122}
\noindent\vbox to\dimexpr\textheight-4pt\relax{\vfil\hbox to\textwidth{\hfil\includegraphics[page=47,trim=18mm 15mm 18mm 17mm,clip,width=\textwidth,height=\dimexpr\textheight-4pt\relax,keepaspectratio]{classification_typeI_embed.pdf}\hfil}\vfil}\par
\clearpage\thispagestyle{plain}
\pdfbookmark[2]{P4/mmm (123)}{table-typeI-sg123}
\noindent\vbox to\dimexpr\textheight-4pt\relax{\vfil\hbox to\textwidth{\hfil\includegraphics[page=48,trim=18mm 15mm 18mm 17mm,clip,width=\textwidth,height=\dimexpr\textheight-4pt\relax,keepaspectratio]{classification_typeI_embed.pdf}\hfil}\vfil}\par
\clearpage\thispagestyle{plain}
\pdfbookmark[2]{P4/mcc (124)}{table-typeI-sg124}
\noindent\vbox to\dimexpr\textheight-4pt\relax{\vfil\hbox to\textwidth{\hfil\includegraphics[page=49,trim=18mm 15mm 18mm 17mm,clip,width=\textwidth,height=\dimexpr\textheight-4pt\relax,keepaspectratio]{classification_typeI_embed.pdf}\hfil}\vfil}\par
\clearpage\thispagestyle{plain}
\pdfbookmark[2]{P4/nbm (125)}{table-typeI-sg125}
\noindent\vbox to\dimexpr\textheight-4pt\relax{\vfil\hbox to\textwidth{\hfil\includegraphics[page=50,trim=18mm 15mm 18mm 17mm,clip,width=\textwidth,height=\dimexpr\textheight-4pt\relax,keepaspectratio]{classification_typeI_embed.pdf}\hfil}\vfil}\par
\clearpage\thispagestyle{plain}
\pdfbookmark[2]{P4/nnc (126)}{table-typeI-sg126}
\noindent\vbox to\dimexpr\textheight-4pt\relax{\vfil\hbox to\textwidth{\hfil\includegraphics[page=51,trim=18mm 15mm 18mm 17mm,clip,width=\textwidth,height=\dimexpr\textheight-4pt\relax,keepaspectratio]{classification_typeI_embed.pdf}\hfil}\vfil}\par
\clearpage\thispagestyle{plain}
\pdfbookmark[2]{P4/mbm (127)}{table-typeI-sg127}
\noindent\vbox to\dimexpr\textheight-4pt\relax{\vfil\hbox to\textwidth{\hfil\includegraphics[page=52,trim=18mm 15mm 18mm 17mm,clip,width=\textwidth,height=\dimexpr\textheight-4pt\relax,keepaspectratio]{classification_typeI_embed.pdf}\hfil}\vfil}\par
\clearpage\thispagestyle{plain}
\noindent\vbox to\dimexpr\textheight-4pt\relax{\vfil\hbox to\textwidth{\hfil\includegraphics[page=53,trim=18mm 15mm 18mm 17mm,clip,width=\textwidth,height=\dimexpr\textheight-4pt\relax,keepaspectratio]{classification_typeI_embed.pdf}\hfil}\vfil}\par
\clearpage\thispagestyle{plain}
\pdfbookmark[2]{P4/mnc (128)}{table-typeI-sg128}
\pdfbookmark[2]{P4/nmm (129)}{table-typeI-sg129}
\noindent\vbox to\dimexpr\textheight-4pt\relax{\vfil\hbox to\textwidth{\hfil\includegraphics[page=54,trim=18mm 15mm 18mm 17mm,clip,width=\textwidth,height=\dimexpr\textheight-4pt\relax,keepaspectratio]{classification_typeI_embed.pdf}\hfil}\vfil}\par
\clearpage\thispagestyle{plain}
\noindent\vbox to\dimexpr\textheight-4pt\relax{\vfil\hbox to\textwidth{\hfil\includegraphics[page=55,trim=18mm 15mm 18mm 17mm,clip,width=\textwidth,height=\dimexpr\textheight-4pt\relax,keepaspectratio]{classification_typeI_embed.pdf}\hfil}\vfil}\par
\clearpage\thispagestyle{plain}
\pdfbookmark[2]{P4/ncc (130)}{table-typeI-sg130}
\noindent\vbox to\dimexpr\textheight-4pt\relax{\vfil\hbox to\textwidth{\hfil\includegraphics[page=56,trim=18mm 15mm 18mm 17mm,clip,width=\textwidth,height=\dimexpr\textheight-4pt\relax,keepaspectratio]{classification_typeI_embed.pdf}\hfil}\vfil}\par
\clearpage\thispagestyle{plain}
\pdfbookmark[2]{P42/mmc (131)}{table-typeI-sg131}
\noindent\vbox to\dimexpr\textheight-4pt\relax{\vfil\hbox to\textwidth{\hfil\includegraphics[page=57,trim=18mm 15mm 18mm 17mm,clip,width=\textwidth,height=\dimexpr\textheight-4pt\relax,keepaspectratio]{classification_typeI_embed.pdf}\hfil}\vfil}\par
\clearpage\thispagestyle{plain}
\pdfbookmark[2]{P42/mcm (132)}{table-typeI-sg132}
\noindent\vbox to\dimexpr\textheight-4pt\relax{\vfil\hbox to\textwidth{\hfil\includegraphics[page=58,trim=18mm 15mm 18mm 17mm,clip,width=\textwidth,height=\dimexpr\textheight-4pt\relax,keepaspectratio]{classification_typeI_embed.pdf}\hfil}\vfil}\par
\clearpage\thispagestyle{plain}
\pdfbookmark[2]{P42/nbc (133)}{table-typeI-sg133}
\noindent\vbox to\dimexpr\textheight-4pt\relax{\vfil\hbox to\textwidth{\hfil\includegraphics[page=59,trim=18mm 15mm 18mm 17mm,clip,width=\textwidth,height=\dimexpr\textheight-4pt\relax,keepaspectratio]{classification_typeI_embed.pdf}\hfil}\vfil}\par
\clearpage\thispagestyle{plain}
\pdfbookmark[2]{P42/nnm (134)}{table-typeI-sg134}
\noindent\vbox to\dimexpr\textheight-4pt\relax{\vfil\hbox to\textwidth{\hfil\includegraphics[page=60,trim=18mm 15mm 18mm 17mm,clip,width=\textwidth,height=\dimexpr\textheight-4pt\relax,keepaspectratio]{classification_typeI_embed.pdf}\hfil}\vfil}\par
\clearpage\thispagestyle{plain}
\noindent\vbox to\dimexpr\textheight-4pt\relax{\vfil\hbox to\textwidth{\hfil\includegraphics[page=61,trim=18mm 15mm 18mm 17mm,clip,width=\textwidth,height=\dimexpr\textheight-4pt\relax,keepaspectratio]{classification_typeI_embed.pdf}\hfil}\vfil}\par
\clearpage\thispagestyle{plain}
\pdfbookmark[2]{P42/mbc (135)}{table-typeI-sg135}
\noindent\vbox to\dimexpr\textheight-4pt\relax{\vfil\hbox to\textwidth{\hfil\includegraphics[page=62,trim=18mm 15mm 18mm 17mm,clip,width=\textwidth,height=\dimexpr\textheight-4pt\relax,keepaspectratio]{classification_typeI_embed.pdf}\hfil}\vfil}\par
\clearpage\thispagestyle{plain}
\pdfbookmark[2]{P42/mnm (136)}{table-typeI-sg136}
\noindent\vbox to\dimexpr\textheight-4pt\relax{\vfil\hbox to\textwidth{\hfil\includegraphics[page=63,trim=18mm 15mm 18mm 17mm,clip,width=\textwidth,height=\dimexpr\textheight-4pt\relax,keepaspectratio]{classification_typeI_embed.pdf}\hfil}\vfil}\par
\clearpage\thispagestyle{plain}
\pdfbookmark[2]{P42/nmc (137)}{table-typeI-sg137}
\noindent\vbox to\dimexpr\textheight-4pt\relax{\vfil\hbox to\textwidth{\hfil\includegraphics[page=64,trim=18mm 15mm 18mm 17mm,clip,width=\textwidth,height=\dimexpr\textheight-4pt\relax,keepaspectratio]{classification_typeI_embed.pdf}\hfil}\vfil}\par
\clearpage\thispagestyle{plain}
\pdfbookmark[2]{P42/ncm (138)}{table-typeI-sg138}
\noindent\vbox to\dimexpr\textheight-4pt\relax{\vfil\hbox to\textwidth{\hfil\includegraphics[page=65,trim=18mm 15mm 18mm 17mm,clip,width=\textwidth,height=\dimexpr\textheight-4pt\relax,keepaspectratio]{classification_typeI_embed.pdf}\hfil}\vfil}\par
\clearpage\thispagestyle{plain}
\noindent\vbox to\dimexpr\textheight-4pt\relax{\vfil\hbox to\textwidth{\hfil\includegraphics[page=66,trim=18mm 15mm 18mm 17mm,clip,width=\textwidth,height=\dimexpr\textheight-4pt\relax,keepaspectratio]{classification_typeI_embed.pdf}\hfil}\vfil}\par
\clearpage\thispagestyle{plain}
\pdfbookmark[2]{I4/mmm (139)}{table-typeI-sg139}
\noindent\vbox to\dimexpr\textheight-4pt\relax{\vfil\hbox to\textwidth{\hfil\includegraphics[page=67,trim=18mm 15mm 18mm 17mm,clip,width=\textwidth,height=\dimexpr\textheight-4pt\relax,keepaspectratio]{classification_typeI_embed.pdf}\hfil}\vfil}\par
\clearpage\thispagestyle{plain}
\pdfbookmark[2]{I4/mcm (140)}{table-typeI-sg140}
\noindent\vbox to\dimexpr\textheight-4pt\relax{\vfil\hbox to\textwidth{\hfil\includegraphics[page=68,trim=18mm 15mm 18mm 17mm,clip,width=\textwidth,height=\dimexpr\textheight-4pt\relax,keepaspectratio]{classification_typeI_embed.pdf}\hfil}\vfil}\par
\clearpage\thispagestyle{plain}
\noindent\vbox to\dimexpr\textheight-4pt\relax{\vfil\hbox to\textwidth{\hfil\includegraphics[page=69,trim=18mm 15mm 18mm 17mm,clip,width=\textwidth,height=\dimexpr\textheight-4pt\relax,keepaspectratio]{classification_typeI_embed.pdf}\hfil}\vfil}\par
\clearpage\thispagestyle{plain}
\pdfbookmark[2]{I41/amd (141)}{table-typeI-sg141}
\noindent\vbox to\dimexpr\textheight-4pt\relax{\vfil\hbox to\textwidth{\hfil\includegraphics[page=70,trim=18mm 15mm 18mm 17mm,clip,width=\textwidth,height=\dimexpr\textheight-4pt\relax,keepaspectratio]{classification_typeI_embed.pdf}\hfil}\vfil}\par
\clearpage\thispagestyle{plain}
\pdfbookmark[2]{I41/acd (142)}{table-typeI-sg142}
\pdfbookmark[2]{P3 (143)}{table-typeI-sg143}
\pdfbookmark[2]{P31 (144)}{table-typeI-sg144}
\noindent\vbox to\dimexpr\textheight-4pt\relax{\vfil\hbox to\textwidth{\hfil\includegraphics[page=71,trim=18mm 15mm 18mm 17mm,clip,width=\textwidth,height=\dimexpr\textheight-4pt\relax,keepaspectratio]{classification_typeI_embed.pdf}\hfil}\vfil}\par
\clearpage\thispagestyle{plain}
\pdfbookmark[2]{P32 (145)}{table-typeI-sg145}
\pdfbookmark[2]{R3 (146)}{table-typeI-sg146}
\pdfbookmark[2]{P-3 (147)}{table-typeI-sg147}
\pdfbookmark[2]{R-3 (148)}{table-typeI-sg148}
\pdfbookmark[2]{P312 (149)}{table-typeI-sg149}
\noindent\vbox to\dimexpr\textheight-4pt\relax{\vfil\hbox to\textwidth{\hfil\includegraphics[page=72,trim=18mm 15mm 18mm 17mm,clip,width=\textwidth,height=\dimexpr\textheight-4pt\relax,keepaspectratio]{classification_typeI_embed.pdf}\hfil}\vfil}\par
\clearpage\thispagestyle{plain}
\pdfbookmark[2]{P321 (150)}{table-typeI-sg150}
\pdfbookmark[2]{P3112 (151)}{table-typeI-sg151}
\pdfbookmark[2]{P3121 (152)}{table-typeI-sg152}
\noindent\vbox to\dimexpr\textheight-4pt\relax{\vfil\hbox to\textwidth{\hfil\includegraphics[page=73,trim=18mm 15mm 18mm 17mm,clip,width=\textwidth,height=\dimexpr\textheight-4pt\relax,keepaspectratio]{classification_typeI_embed.pdf}\hfil}\vfil}\par
\clearpage\thispagestyle{plain}
\pdfbookmark[2]{P3212 (153)}{table-typeI-sg153}
\pdfbookmark[2]{P3221 (154)}{table-typeI-sg154}
\pdfbookmark[2]{R32 (155)}{table-typeI-sg155}
\noindent\vbox to\dimexpr\textheight-4pt\relax{\vfil\hbox to\textwidth{\hfil\includegraphics[page=74,trim=18mm 15mm 18mm 17mm,clip,width=\textwidth,height=\dimexpr\textheight-4pt\relax,keepaspectratio]{classification_typeI_embed.pdf}\hfil}\vfil}\par
\clearpage\thispagestyle{plain}
\pdfbookmark[2]{P3m1 (156)}{table-typeI-sg156}
\pdfbookmark[2]{P31m (157)}{table-typeI-sg157}
\pdfbookmark[2]{P3c1 (158)}{table-typeI-sg158}
\pdfbookmark[2]{P31c (159)}{table-typeI-sg159}
\noindent\vbox to\dimexpr\textheight-4pt\relax{\vfil\hbox to\textwidth{\hfil\includegraphics[page=75,trim=18mm 15mm 18mm 17mm,clip,width=\textwidth,height=\dimexpr\textheight-4pt\relax,keepaspectratio]{classification_typeI_embed.pdf}\hfil}\vfil}\par
\clearpage\thispagestyle{plain}
\pdfbookmark[2]{R3m (160)}{table-typeI-sg160}
\pdfbookmark[2]{R3c (161)}{table-typeI-sg161}
\pdfbookmark[2]{P-31m (162)}{table-typeI-sg162}
\noindent\vbox to\dimexpr\textheight-4pt\relax{\vfil\hbox to\textwidth{\hfil\includegraphics[page=76,trim=18mm 15mm 18mm 17mm,clip,width=\textwidth,height=\dimexpr\textheight-4pt\relax,keepaspectratio]{classification_typeI_embed.pdf}\hfil}\vfil}\par
\clearpage\thispagestyle{plain}
\pdfbookmark[2]{P-31c (163)}{table-typeI-sg163}
\pdfbookmark[2]{P-3m1 (164)}{table-typeI-sg164}
\noindent\vbox to\dimexpr\textheight-4pt\relax{\vfil\hbox to\textwidth{\hfil\includegraphics[page=77,trim=18mm 15mm 18mm 17mm,clip,width=\textwidth,height=\dimexpr\textheight-4pt\relax,keepaspectratio]{classification_typeI_embed.pdf}\hfil}\vfil}\par
\clearpage\thispagestyle{plain}
\pdfbookmark[2]{P-3c1 (165)}{table-typeI-sg165}
\noindent\vbox to\dimexpr\textheight-4pt\relax{\vfil\hbox to\textwidth{\hfil\includegraphics[page=78,trim=18mm 15mm 18mm 17mm,clip,width=\textwidth,height=\dimexpr\textheight-4pt\relax,keepaspectratio]{classification_typeI_embed.pdf}\hfil}\vfil}\par
\clearpage\thispagestyle{plain}
\pdfbookmark[2]{R-3m (166)}{table-typeI-sg166}
\pdfbookmark[2]{R-3c (167)}{table-typeI-sg167}
\noindent\vbox to\dimexpr\textheight-4pt\relax{\vfil\hbox to\textwidth{\hfil\includegraphics[page=79,trim=18mm 15mm 18mm 17mm,clip,width=\textwidth,height=\dimexpr\textheight-4pt\relax,keepaspectratio]{classification_typeI_embed.pdf}\hfil}\vfil}\par
\clearpage\thispagestyle{plain}
\pdfbookmark[2]{P6 (168)}{table-typeI-sg168}
\pdfbookmark[2]{P61 (169)}{table-typeI-sg169}
\pdfbookmark[2]{P65 (170)}{table-typeI-sg170}
\noindent\vbox to\dimexpr\textheight-4pt\relax{\vfil\hbox to\textwidth{\hfil\includegraphics[page=80,trim=18mm 15mm 18mm 17mm,clip,width=\textwidth,height=\dimexpr\textheight-4pt\relax,keepaspectratio]{classification_typeI_embed.pdf}\hfil}\vfil}\par
\clearpage\thispagestyle{plain}
\pdfbookmark[2]{P62 (171)}{table-typeI-sg171}
\pdfbookmark[2]{P64 (172)}{table-typeI-sg172}
\pdfbookmark[2]{P63 (173)}{table-typeI-sg173}
\pdfbookmark[2]{P-6 (174)}{table-typeI-sg174}
\noindent\vbox to\dimexpr\textheight-4pt\relax{\vfil\hbox to\textwidth{\hfil\includegraphics[page=81,trim=18mm 15mm 18mm 17mm,clip,width=\textwidth,height=\dimexpr\textheight-4pt\relax,keepaspectratio]{classification_typeI_embed.pdf}\hfil}\vfil}\par
\clearpage\thispagestyle{plain}
\pdfbookmark[2]{P6/m (175)}{table-typeI-sg175}
\pdfbookmark[2]{P63/m (176)}{table-typeI-sg176}
\noindent\vbox to\dimexpr\textheight-4pt\relax{\vfil\hbox to\textwidth{\hfil\includegraphics[page=82,trim=18mm 15mm 18mm 17mm,clip,width=\textwidth,height=\dimexpr\textheight-4pt\relax,keepaspectratio]{classification_typeI_embed.pdf}\hfil}\vfil}\par
\clearpage\thispagestyle{plain}
\pdfbookmark[2]{P622 (177)}{table-typeI-sg177}
\pdfbookmark[2]{P6122 (178)}{table-typeI-sg178}
\noindent\vbox to\dimexpr\textheight-4pt\relax{\vfil\hbox to\textwidth{\hfil\includegraphics[page=83,trim=18mm 15mm 18mm 17mm,clip,width=\textwidth,height=\dimexpr\textheight-4pt\relax,keepaspectratio]{classification_typeI_embed.pdf}\hfil}\vfil}\par
\clearpage\thispagestyle{plain}
\pdfbookmark[2]{P6522 (179)}{table-typeI-sg179}
\pdfbookmark[2]{P6222 (180)}{table-typeI-sg180}
\noindent\vbox to\dimexpr\textheight-4pt\relax{\vfil\hbox to\textwidth{\hfil\includegraphics[page=84,trim=18mm 15mm 18mm 17mm,clip,width=\textwidth,height=\dimexpr\textheight-4pt\relax,keepaspectratio]{classification_typeI_embed.pdf}\hfil}\vfil}\par
\clearpage\thispagestyle{plain}
\pdfbookmark[2]{P6422 (181)}{table-typeI-sg181}
\noindent\vbox to\dimexpr\textheight-4pt\relax{\vfil\hbox to\textwidth{\hfil\includegraphics[page=85,trim=18mm 15mm 18mm 17mm,clip,width=\textwidth,height=\dimexpr\textheight-4pt\relax,keepaspectratio]{classification_typeI_embed.pdf}\hfil}\vfil}\par
\clearpage\thispagestyle{plain}
\pdfbookmark[2]{P6322 (182)}{table-typeI-sg182}
\pdfbookmark[2]{P6mm (183)}{table-typeI-sg183}
\noindent\vbox to\dimexpr\textheight-4pt\relax{\vfil\hbox to\textwidth{\hfil\includegraphics[page=86,trim=18mm 15mm 18mm 17mm,clip,width=\textwidth,height=\dimexpr\textheight-4pt\relax,keepaspectratio]{classification_typeI_embed.pdf}\hfil}\vfil}\par
\clearpage\thispagestyle{plain}
\pdfbookmark[2]{P6cc (184)}{table-typeI-sg184}
\pdfbookmark[2]{P63cm (185)}{table-typeI-sg185}
\noindent\vbox to\dimexpr\textheight-4pt\relax{\vfil\hbox to\textwidth{\hfil\includegraphics[page=87,trim=18mm 15mm 18mm 17mm,clip,width=\textwidth,height=\dimexpr\textheight-4pt\relax,keepaspectratio]{classification_typeI_embed.pdf}\hfil}\vfil}\par
\clearpage\thispagestyle{plain}
\pdfbookmark[2]{P63mc (186)}{table-typeI-sg186}
\noindent\vbox to\dimexpr\textheight-4pt\relax{\vfil\hbox to\textwidth{\hfil\includegraphics[page=88,trim=18mm 15mm 18mm 17mm,clip,width=\textwidth,height=\dimexpr\textheight-4pt\relax,keepaspectratio]{classification_typeI_embed.pdf}\hfil}\vfil}\par
\clearpage\thispagestyle{plain}
\pdfbookmark[2]{P-6m2 (187)}{table-typeI-sg187}
\pdfbookmark[2]{P-6c2 (188)}{table-typeI-sg188}
\noindent\vbox to\dimexpr\textheight-4pt\relax{\vfil\hbox to\textwidth{\hfil\includegraphics[page=89,trim=18mm 15mm 18mm 17mm,clip,width=\textwidth,height=\dimexpr\textheight-4pt\relax,keepaspectratio]{classification_typeI_embed.pdf}\hfil}\vfil}\par
\clearpage\thispagestyle{plain}
\pdfbookmark[2]{P-62m (189)}{table-typeI-sg189}
\noindent\vbox to\dimexpr\textheight-4pt\relax{\vfil\hbox to\textwidth{\hfil\includegraphics[page=90,trim=18mm 15mm 18mm 17mm,clip,width=\textwidth,height=\dimexpr\textheight-4pt\relax,keepaspectratio]{classification_typeI_embed.pdf}\hfil}\vfil}\par
\clearpage\thispagestyle{plain}
\pdfbookmark[2]{P-62c (190)}{table-typeI-sg190}
\pdfbookmark[2]{P6/mmm (191)}{table-typeI-sg191}
\noindent\vbox to\dimexpr\textheight-4pt\relax{\vfil\hbox to\textwidth{\hfil\includegraphics[page=91,trim=18mm 15mm 18mm 17mm,clip,width=\textwidth,height=\dimexpr\textheight-4pt\relax,keepaspectratio]{classification_typeI_embed.pdf}\hfil}\vfil}\par
\clearpage\thispagestyle{plain}
\noindent\vbox to\dimexpr\textheight-4pt\relax{\vfil\hbox to\textwidth{\hfil\includegraphics[page=92,trim=18mm 15mm 18mm 17mm,clip,width=\textwidth,height=\dimexpr\textheight-4pt\relax,keepaspectratio]{classification_typeI_embed.pdf}\hfil}\vfil}\par
\clearpage\thispagestyle{plain}
\pdfbookmark[2]{P6/mcc (192)}{table-typeI-sg192}
\noindent\vbox to\dimexpr\textheight-4pt\relax{\vfil\hbox to\textwidth{\hfil\includegraphics[page=93,trim=18mm 15mm 18mm 17mm,clip,width=\textwidth,height=\dimexpr\textheight-4pt\relax,keepaspectratio]{classification_typeI_embed.pdf}\hfil}\vfil}\par
\clearpage\thispagestyle{plain}
\pdfbookmark[2]{P63/mcm (193)}{table-typeI-sg193}
\noindent\vbox to\dimexpr\textheight-4pt\relax{\vfil\hbox to\textwidth{\hfil\includegraphics[page=94,trim=18mm 15mm 18mm 17mm,clip,width=\textwidth,height=\dimexpr\textheight-4pt\relax,keepaspectratio]{classification_typeI_embed.pdf}\hfil}\vfil}\par
\clearpage\thispagestyle{plain}
\noindent\vbox to\dimexpr\textheight-4pt\relax{\vfil\hbox to\textwidth{\hfil\includegraphics[page=95,trim=18mm 15mm 18mm 17mm,clip,width=\textwidth,height=\dimexpr\textheight-4pt\relax,keepaspectratio]{classification_typeI_embed.pdf}\hfil}\vfil}\par
\clearpage\thispagestyle{plain}
\pdfbookmark[2]{P63/mmc (194)}{table-typeI-sg194}
\noindent\vbox to\dimexpr\textheight-4pt\relax{\vfil\hbox to\textwidth{\hfil\includegraphics[page=96,trim=18mm 15mm 18mm 17mm,clip,width=\textwidth,height=\dimexpr\textheight-4pt\relax,keepaspectratio]{classification_typeI_embed.pdf}\hfil}\vfil}\par
\clearpage\thispagestyle{plain}
\pdfbookmark[2]{P23 (195)}{table-typeI-sg195}
\noindent\vbox to\dimexpr\textheight-4pt\relax{\vfil\hbox to\textwidth{\hfil\includegraphics[page=97,trim=18mm 15mm 18mm 17mm,clip,width=\textwidth,height=\dimexpr\textheight-4pt\relax,keepaspectratio]{classification_typeI_embed.pdf}\hfil}\vfil}\par
\clearpage\thispagestyle{plain}
\pdfbookmark[2]{F23 (196)}{table-typeI-sg196}
\pdfbookmark[2]{I23 (197)}{table-typeI-sg197}
\pdfbookmark[2]{P213 (198)}{table-typeI-sg198}
\noindent\vbox to\dimexpr\textheight-4pt\relax{\vfil\hbox to\textwidth{\hfil\includegraphics[page=98,trim=18mm 15mm 18mm 17mm,clip,width=\textwidth,height=\dimexpr\textheight-4pt\relax,keepaspectratio]{classification_typeI_embed.pdf}\hfil}\vfil}\par
\clearpage\thispagestyle{plain}
\pdfbookmark[2]{I213 (199)}{table-typeI-sg199}
\pdfbookmark[2]{Pm-3 (200)}{table-typeI-sg200}
\noindent\vbox to\dimexpr\textheight-4pt\relax{\vfil\hbox to\textwidth{\hfil\includegraphics[page=99,trim=18mm 15mm 18mm 17mm,clip,width=\textwidth,height=\dimexpr\textheight-4pt\relax,keepaspectratio]{classification_typeI_embed.pdf}\hfil}\vfil}\par
\clearpage\thispagestyle{plain}
\pdfbookmark[2]{Pn-3 (201)}{table-typeI-sg201}
\noindent\vbox to\dimexpr\textheight-4pt\relax{\vfil\hbox to\textwidth{\hfil\includegraphics[page=100,trim=18mm 15mm 18mm 17mm,clip,width=\textwidth,height=\dimexpr\textheight-4pt\relax,keepaspectratio]{classification_typeI_embed.pdf}\hfil}\vfil}\par
\clearpage\thispagestyle{plain}
\pdfbookmark[2]{Fm-3 (202)}{table-typeI-sg202}
\noindent\vbox to\dimexpr\textheight-4pt\relax{\vfil\hbox to\textwidth{\hfil\includegraphics[page=101,trim=18mm 15mm 18mm 17mm,clip,width=\textwidth,height=\dimexpr\textheight-4pt\relax,keepaspectratio]{classification_typeI_embed.pdf}\hfil}\vfil}\par
\clearpage\thispagestyle{plain}
\noindent\vbox to\dimexpr\textheight-4pt\relax{\vfil\hbox to\textwidth{\hfil\includegraphics[page=102,trim=18mm 15mm 18mm 17mm,clip,width=\textwidth,height=\dimexpr\textheight-4pt\relax,keepaspectratio]{classification_typeI_embed.pdf}\hfil}\vfil}\par
\clearpage\thispagestyle{plain}
\pdfbookmark[2]{Fd-3 (203)}{table-typeI-sg203}
\pdfbookmark[2]{Im-3 (204)}{table-typeI-sg204}
\noindent\vbox to\dimexpr\textheight-4pt\relax{\vfil\hbox to\textwidth{\hfil\includegraphics[page=103,trim=18mm 15mm 18mm 17mm,clip,width=\textwidth,height=\dimexpr\textheight-4pt\relax,keepaspectratio]{classification_typeI_embed.pdf}\hfil}\vfil}\par
\clearpage\thispagestyle{plain}
\pdfbookmark[2]{Pa-3 (205)}{table-typeI-sg205}
\pdfbookmark[2]{Ia-3 (206)}{table-typeI-sg206}
\noindent\vbox to\dimexpr\textheight-4pt\relax{\vfil\hbox to\textwidth{\hfil\includegraphics[page=104,trim=18mm 15mm 18mm 17mm,clip,width=\textwidth,height=\dimexpr\textheight-4pt\relax,keepaspectratio]{classification_typeI_embed.pdf}\hfil}\vfil}\par
\clearpage\thispagestyle{plain}
\pdfbookmark[2]{P432 (207)}{table-typeI-sg207}
\noindent\vbox to\dimexpr\textheight-4pt\relax{\vfil\hbox to\textwidth{\hfil\includegraphics[page=105,trim=18mm 15mm 18mm 17mm,clip,width=\textwidth,height=\dimexpr\textheight-4pt\relax,keepaspectratio]{classification_typeI_embed.pdf}\hfil}\vfil}\par
\clearpage\thispagestyle{plain}
\pdfbookmark[2]{P4232 (208)}{table-typeI-sg208}
\noindent\vbox to\dimexpr\textheight-4pt\relax{\vfil\hbox to\textwidth{\hfil\includegraphics[page=106,trim=18mm 15mm 18mm 17mm,clip,width=\textwidth,height=\dimexpr\textheight-4pt\relax,keepaspectratio]{classification_typeI_embed.pdf}\hfil}\vfil}\par
\clearpage\thispagestyle{plain}
\pdfbookmark[2]{F432 (209)}{table-typeI-sg209}
\pdfbookmark[2]{F4132 (210)}{table-typeI-sg210}
\noindent\vbox to\dimexpr\textheight-4pt\relax{\vfil\hbox to\textwidth{\hfil\includegraphics[page=107,trim=18mm 15mm 18mm 17mm,clip,width=\textwidth,height=\dimexpr\textheight-4pt\relax,keepaspectratio]{classification_typeI_embed.pdf}\hfil}\vfil}\par
\clearpage\thispagestyle{plain}
\pdfbookmark[2]{I432 (211)}{table-typeI-sg211}
\noindent\vbox to\dimexpr\textheight-4pt\relax{\vfil\hbox to\textwidth{\hfil\includegraphics[page=108,trim=18mm 15mm 18mm 17mm,clip,width=\textwidth,height=\dimexpr\textheight-4pt\relax,keepaspectratio]{classification_typeI_embed.pdf}\hfil}\vfil}\par
\clearpage\thispagestyle{plain}
\pdfbookmark[2]{P4332 (212)}{table-typeI-sg212}
\pdfbookmark[2]{P4132 (213)}{table-typeI-sg213}
\noindent\vbox to\dimexpr\textheight-4pt\relax{\vfil\hbox to\textwidth{\hfil\includegraphics[page=109,trim=18mm 15mm 18mm 17mm,clip,width=\textwidth,height=\dimexpr\textheight-4pt\relax,keepaspectratio]{classification_typeI_embed.pdf}\hfil}\vfil}\par
\clearpage\thispagestyle{plain}
\pdfbookmark[2]{I4132 (214)}{table-typeI-sg214}
\pdfbookmark[2]{P-43m (215)}{table-typeI-sg215}
\noindent\vbox to\dimexpr\textheight-4pt\relax{\vfil\hbox to\textwidth{\hfil\includegraphics[page=110,trim=18mm 15mm 18mm 17mm,clip,width=\textwidth,height=\dimexpr\textheight-4pt\relax,keepaspectratio]{classification_typeI_embed.pdf}\hfil}\vfil}\par
\clearpage\thispagestyle{plain}
\pdfbookmark[2]{F-43m (216)}{table-typeI-sg216}
\noindent\vbox to\dimexpr\textheight-4pt\relax{\vfil\hbox to\textwidth{\hfil\includegraphics[page=111,trim=18mm 15mm 18mm 17mm,clip,width=\textwidth,height=\dimexpr\textheight-4pt\relax,keepaspectratio]{classification_typeI_embed.pdf}\hfil}\vfil}\par
\clearpage\thispagestyle{plain}
\pdfbookmark[2]{I-43m (217)}{table-typeI-sg217}
\noindent\vbox to\dimexpr\textheight-4pt\relax{\vfil\hbox to\textwidth{\hfil\includegraphics[page=112,trim=18mm 15mm 18mm 17mm,clip,width=\textwidth,height=\dimexpr\textheight-4pt\relax,keepaspectratio]{classification_typeI_embed.pdf}\hfil}\vfil}\par
\clearpage\thispagestyle{plain}
\pdfbookmark[2]{P-43n (218)}{table-typeI-sg218}
\pdfbookmark[2]{F-43c (219)}{table-typeI-sg219}
\noindent\vbox to\dimexpr\textheight-4pt\relax{\vfil\hbox to\textwidth{\hfil\includegraphics[page=113,trim=18mm 15mm 18mm 17mm,clip,width=\textwidth,height=\dimexpr\textheight-4pt\relax,keepaspectratio]{classification_typeI_embed.pdf}\hfil}\vfil}\par
\clearpage\thispagestyle{plain}
\pdfbookmark[2]{I-43d (220)}{table-typeI-sg220}
\pdfbookmark[2]{Pm-3m (221)}{table-typeI-sg221}
\noindent\vbox to\dimexpr\textheight-4pt\relax{\vfil\hbox to\textwidth{\hfil\includegraphics[page=114,trim=18mm 15mm 18mm 17mm,clip,width=\textwidth,height=\dimexpr\textheight-4pt\relax,keepaspectratio]{classification_typeI_embed.pdf}\hfil}\vfil}\par
\clearpage\thispagestyle{plain}
\noindent\vbox to\dimexpr\textheight-4pt\relax{\vfil\hbox to\textwidth{\hfil\includegraphics[page=115,trim=18mm 15mm 18mm 17mm,clip,width=\textwidth,height=\dimexpr\textheight-4pt\relax,keepaspectratio]{classification_typeI_embed.pdf}\hfil}\vfil}\par
\clearpage\thispagestyle{plain}
\noindent\vbox to\dimexpr\textheight-4pt\relax{\vfil\hbox to\textwidth{\hfil\includegraphics[page=116,trim=18mm 15mm 18mm 17mm,clip,width=\textwidth,height=\dimexpr\textheight-4pt\relax,keepaspectratio]{classification_typeI_embed.pdf}\hfil}\vfil}\par
\clearpage\thispagestyle{plain}
\noindent\vbox to\dimexpr\textheight-4pt\relax{\vfil\hbox to\textwidth{\hfil\includegraphics[page=117,trim=18mm 15mm 18mm 17mm,clip,width=\textwidth,height=\dimexpr\textheight-4pt\relax,keepaspectratio]{classification_typeI_embed.pdf}\hfil}\vfil}\par
\clearpage\thispagestyle{plain}
\pdfbookmark[2]{Pn-3n (222)}{table-typeI-sg222}
\noindent\vbox to\dimexpr\textheight-4pt\relax{\vfil\hbox to\textwidth{\hfil\includegraphics[page=118,trim=18mm 15mm 18mm 17mm,clip,width=\textwidth,height=\dimexpr\textheight-4pt\relax,keepaspectratio]{classification_typeI_embed.pdf}\hfil}\vfil}\par
\clearpage\thispagestyle{plain}
\pdfbookmark[2]{Pm-3n (223)}{table-typeI-sg223}
\noindent\vbox to\dimexpr\textheight-4pt\relax{\vfil\hbox to\textwidth{\hfil\includegraphics[page=119,trim=18mm 15mm 18mm 17mm,clip,width=\textwidth,height=\dimexpr\textheight-4pt\relax,keepaspectratio]{classification_typeI_embed.pdf}\hfil}\vfil}\par
\clearpage\thispagestyle{plain}
\noindent\vbox to\dimexpr\textheight-4pt\relax{\vfil\hbox to\textwidth{\hfil\includegraphics[page=120,trim=18mm 15mm 18mm 17mm,clip,width=\textwidth,height=\dimexpr\textheight-4pt\relax,keepaspectratio]{classification_typeI_embed.pdf}\hfil}\vfil}\par
\clearpage\thispagestyle{plain}
\pdfbookmark[2]{Pn-3m (224)}{table-typeI-sg224}
\noindent\vbox to\dimexpr\textheight-4pt\relax{\vfil\hbox to\textwidth{\hfil\includegraphics[page=121,trim=18mm 15mm 18mm 17mm,clip,width=\textwidth,height=\dimexpr\textheight-4pt\relax,keepaspectratio]{classification_typeI_embed.pdf}\hfil}\vfil}\par
\clearpage\thispagestyle{plain}
\noindent\vbox to\dimexpr\textheight-4pt\relax{\vfil\hbox to\textwidth{\hfil\includegraphics[page=122,trim=18mm 15mm 18mm 17mm,clip,width=\textwidth,height=\dimexpr\textheight-4pt\relax,keepaspectratio]{classification_typeI_embed.pdf}\hfil}\vfil}\par
\clearpage\thispagestyle{plain}
\noindent\vbox to\dimexpr\textheight-4pt\relax{\vfil\hbox to\textwidth{\hfil\includegraphics[page=123,trim=18mm 15mm 18mm 17mm,clip,width=\textwidth,height=\dimexpr\textheight-4pt\relax,keepaspectratio]{classification_typeI_embed.pdf}\hfil}\vfil}\par
\clearpage\thispagestyle{plain}
\pdfbookmark[2]{Fm-3m (225)}{table-typeI-sg225}
\noindent\vbox to\dimexpr\textheight-4pt\relax{\vfil\hbox to\textwidth{\hfil\includegraphics[page=124,trim=18mm 15mm 18mm 17mm,clip,width=\textwidth,height=\dimexpr\textheight-4pt\relax,keepaspectratio]{classification_typeI_embed.pdf}\hfil}\vfil}\par
\clearpage\thispagestyle{plain}
\noindent\vbox to\dimexpr\textheight-4pt\relax{\vfil\hbox to\textwidth{\hfil\includegraphics[page=125,trim=18mm 15mm 18mm 17mm,clip,width=\textwidth,height=\dimexpr\textheight-4pt\relax,keepaspectratio]{classification_typeI_embed.pdf}\hfil}\vfil}\par
\clearpage\thispagestyle{plain}
\noindent\vbox to\dimexpr\textheight-4pt\relax{\vfil\hbox to\textwidth{\hfil\includegraphics[page=126,trim=18mm 15mm 18mm 17mm,clip,width=\textwidth,height=\dimexpr\textheight-4pt\relax,keepaspectratio]{classification_typeI_embed.pdf}\hfil}\vfil}\par
\clearpage\thispagestyle{plain}
\pdfbookmark[2]{Fm-3c (226)}{table-typeI-sg226}
\noindent\vbox to\dimexpr\textheight-4pt\relax{\vfil\hbox to\textwidth{\hfil\includegraphics[page=127,trim=18mm 15mm 18mm 17mm,clip,width=\textwidth,height=\dimexpr\textheight-4pt\relax,keepaspectratio]{classification_typeI_embed.pdf}\hfil}\vfil}\par
\clearpage\thispagestyle{plain}
\noindent\vbox to\dimexpr\textheight-4pt\relax{\vfil\hbox to\textwidth{\hfil\includegraphics[page=128,trim=18mm 15mm 18mm 17mm,clip,width=\textwidth,height=\dimexpr\textheight-4pt\relax,keepaspectratio]{classification_typeI_embed.pdf}\hfil}\vfil}\par
\clearpage\thispagestyle{plain}
\pdfbookmark[2]{Fd-3m (227)}{table-typeI-sg227}
\noindent\vbox to\dimexpr\textheight-4pt\relax{\vfil\hbox to\textwidth{\hfil\includegraphics[page=129,trim=18mm 15mm 18mm 17mm,clip,width=\textwidth,height=\dimexpr\textheight-4pt\relax,keepaspectratio]{classification_typeI_embed.pdf}\hfil}\vfil}\par
\clearpage\thispagestyle{plain}
\pdfbookmark[2]{Fd-3c (228)}{table-typeI-sg228}
\noindent\vbox to\dimexpr\textheight-4pt\relax{\vfil\hbox to\textwidth{\hfil\includegraphics[page=130,trim=18mm 15mm 18mm 17mm,clip,width=\textwidth,height=\dimexpr\textheight-4pt\relax,keepaspectratio]{classification_typeI_embed.pdf}\hfil}\vfil}\par
\clearpage\thispagestyle{plain}
\noindent\vbox to\dimexpr\textheight-4pt\relax{\vfil\hbox to\textwidth{\hfil\includegraphics[page=131,trim=18mm 15mm 18mm 17mm,clip,width=\textwidth,height=\dimexpr\textheight-4pt\relax,keepaspectratio]{classification_typeI_embed.pdf}\hfil}\vfil}\par
\clearpage\thispagestyle{plain}
\pdfbookmark[2]{Im-3m (229)}{table-typeI-sg229}
\noindent\vbox to\dimexpr\textheight-4pt\relax{\vfil\hbox to\textwidth{\hfil\includegraphics[page=132,trim=18mm 15mm 18mm 17mm,clip,width=\textwidth,height=\dimexpr\textheight-4pt\relax,keepaspectratio]{classification_typeI_embed.pdf}\hfil}\vfil}\par
\clearpage\thispagestyle{plain}
\noindent\vbox to\dimexpr\textheight-4pt\relax{\vfil\hbox to\textwidth{\hfil\includegraphics[page=133,trim=18mm 15mm 18mm 17mm,clip,width=\textwidth,height=\dimexpr\textheight-4pt\relax,keepaspectratio]{classification_typeI_embed.pdf}\hfil}\vfil}\par
\clearpage\thispagestyle{plain}
\pdfbookmark[2]{Ia-3d (230)}{table-typeI-sg230}
\noindent\vbox to\dimexpr\textheight-4pt\relax{\vfil\hbox to\textwidth{\hfil\includegraphics[page=134,trim=18mm 15mm 18mm 17mm,clip,width=\textwidth,height=\dimexpr\textheight-4pt\relax,keepaspectratio]{classification_typeI_embed.pdf}\hfil}\vfil}\par
\clearpage\thispagestyle{plain}
\noindent\vbox to\dimexpr\textheight-4pt\relax{\vfil\hbox to\textwidth{\hfil\includegraphics[page=135,trim=18mm 15mm 18mm 17mm,clip,width=\textwidth,height=\dimexpr\textheight-4pt\relax,keepaspectratio]{classification_typeI_embed.pdf}\hfil}\vfil}\par
\clearpage\thispagestyle{plain}
\refstepcounter{table}\label{tab:classification-typeII}
\addcontentsline{toc}{section}{Table S6. Type-II generator extensions}
\pdfbookmark[2]{P1 (1)}{table-typeII-sg1}
\noindent\vbox to\dimexpr\textheight-4pt\relax{\vfil\hbox to\textwidth{\hfil\includegraphics[page=1,trim=18mm 15mm 18mm 17mm,clip,width=\textwidth,height=\dimexpr\textheight-4pt\relax,keepaspectratio]{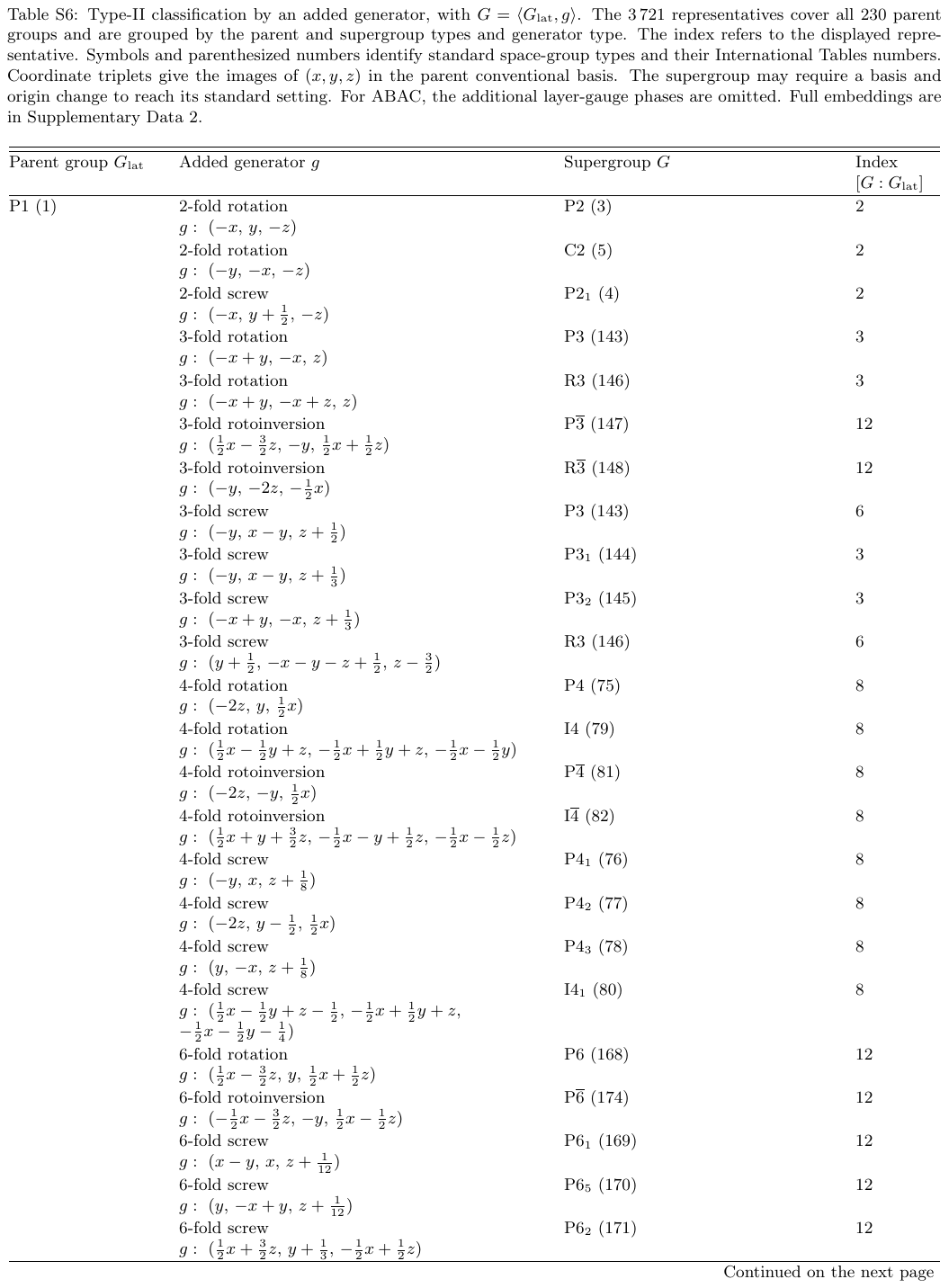}\hfil}\vfil}\par
\clearpage\thispagestyle{plain}
\pdfbookmark[2]{P-1 (2)}{table-typeII-sg2}
\noindent\vbox to\dimexpr\textheight-4pt\relax{\vfil\hbox to\textwidth{\hfil\includegraphics[page=2,trim=18mm 15mm 18mm 17mm,clip,width=\textwidth,height=\dimexpr\textheight-4pt\relax,keepaspectratio]{classification_typeII_embed.pdf}\hfil}\vfil}\par
\clearpage\thispagestyle{plain}
\pdfbookmark[2]{P2 (3)}{table-typeII-sg3}
\noindent\vbox to\dimexpr\textheight-4pt\relax{\vfil\hbox to\textwidth{\hfil\includegraphics[page=3,trim=18mm 15mm 18mm 17mm,clip,width=\textwidth,height=\dimexpr\textheight-4pt\relax,keepaspectratio]{classification_typeII_embed.pdf}\hfil}\vfil}\par
\clearpage\thispagestyle{plain}
\noindent\vbox to\dimexpr\textheight-4pt\relax{\vfil\hbox to\textwidth{\hfil\includegraphics[page=4,trim=18mm 15mm 18mm 17mm,clip,width=\textwidth,height=\dimexpr\textheight-4pt\relax,keepaspectratio]{classification_typeII_embed.pdf}\hfil}\vfil}\par
\clearpage\thispagestyle{plain}
\noindent\vbox to\dimexpr\textheight-4pt\relax{\vfil\hbox to\textwidth{\hfil\includegraphics[page=5,trim=18mm 15mm 18mm 17mm,clip,width=\textwidth,height=\dimexpr\textheight-4pt\relax,keepaspectratio]{classification_typeII_embed.pdf}\hfil}\vfil}\par
\clearpage\thispagestyle{plain}
\pdfbookmark[2]{P21 (4)}{table-typeII-sg4}
\noindent\vbox to\dimexpr\textheight-4pt\relax{\vfil\hbox to\textwidth{\hfil\includegraphics[page=6,trim=18mm 15mm 18mm 17mm,clip,width=\textwidth,height=\dimexpr\textheight-4pt\relax,keepaspectratio]{classification_typeII_embed.pdf}\hfil}\vfil}\par
\clearpage\thispagestyle{plain}
\noindent\vbox to\dimexpr\textheight-4pt\relax{\vfil\hbox to\textwidth{\hfil\includegraphics[page=7,trim=18mm 15mm 18mm 17mm,clip,width=\textwidth,height=\dimexpr\textheight-4pt\relax,keepaspectratio]{classification_typeII_embed.pdf}\hfil}\vfil}\par
\clearpage\thispagestyle{plain}
\noindent\vbox to\dimexpr\textheight-4pt\relax{\vfil\hbox to\textwidth{\hfil\includegraphics[page=8,trim=18mm 15mm 18mm 17mm,clip,width=\textwidth,height=\dimexpr\textheight-4pt\relax,keepaspectratio]{classification_typeII_embed.pdf}\hfil}\vfil}\par
\clearpage\thispagestyle{plain}
\noindent\vbox to\dimexpr\textheight-4pt\relax{\vfil\hbox to\textwidth{\hfil\includegraphics[page=9,trim=18mm 15mm 18mm 17mm,clip,width=\textwidth,height=\dimexpr\textheight-4pt\relax,keepaspectratio]{classification_typeII_embed.pdf}\hfil}\vfil}\par
\clearpage\thispagestyle{plain}
\noindent\vbox to\dimexpr\textheight-4pt\relax{\vfil\hbox to\textwidth{\hfil\includegraphics[page=10,trim=18mm 15mm 18mm 17mm,clip,width=\textwidth,height=\dimexpr\textheight-4pt\relax,keepaspectratio]{classification_typeII_embed.pdf}\hfil}\vfil}\par
\clearpage\thispagestyle{plain}
\pdfbookmark[2]{C2 (5)}{table-typeII-sg5}
\noindent\vbox to\dimexpr\textheight-4pt\relax{\vfil\hbox to\textwidth{\hfil\includegraphics[page=11,trim=18mm 15mm 18mm 17mm,clip,width=\textwidth,height=\dimexpr\textheight-4pt\relax,keepaspectratio]{classification_typeII_embed.pdf}\hfil}\vfil}\par
\clearpage\thispagestyle{plain}
\noindent\vbox to\dimexpr\textheight-4pt\relax{\vfil\hbox to\textwidth{\hfil\includegraphics[page=12,trim=18mm 15mm 18mm 17mm,clip,width=\textwidth,height=\dimexpr\textheight-4pt\relax,keepaspectratio]{classification_typeII_embed.pdf}\hfil}\vfil}\par
\clearpage\thispagestyle{plain}
\noindent\vbox to\dimexpr\textheight-4pt\relax{\vfil\hbox to\textwidth{\hfil\includegraphics[page=13,trim=18mm 15mm 18mm 17mm,clip,width=\textwidth,height=\dimexpr\textheight-4pt\relax,keepaspectratio]{classification_typeII_embed.pdf}\hfil}\vfil}\par
\clearpage\thispagestyle{plain}
\noindent\vbox to\dimexpr\textheight-4pt\relax{\vfil\hbox to\textwidth{\hfil\includegraphics[page=14,trim=18mm 15mm 18mm 17mm,clip,width=\textwidth,height=\dimexpr\textheight-4pt\relax,keepaspectratio]{classification_typeII_embed.pdf}\hfil}\vfil}\par
\clearpage\thispagestyle{plain}
\noindent\vbox to\dimexpr\textheight-4pt\relax{\vfil\hbox to\textwidth{\hfil\includegraphics[page=15,trim=18mm 15mm 18mm 17mm,clip,width=\textwidth,height=\dimexpr\textheight-4pt\relax,keepaspectratio]{classification_typeII_embed.pdf}\hfil}\vfil}\par
\clearpage\thispagestyle{plain}
\pdfbookmark[2]{Pm (6)}{table-typeII-sg6}
\noindent\vbox to\dimexpr\textheight-4pt\relax{\vfil\hbox to\textwidth{\hfil\includegraphics[page=16,trim=18mm 15mm 18mm 17mm,clip,width=\textwidth,height=\dimexpr\textheight-4pt\relax,keepaspectratio]{classification_typeII_embed.pdf}\hfil}\vfil}\par
\clearpage\thispagestyle{plain}
\noindent\vbox to\dimexpr\textheight-4pt\relax{\vfil\hbox to\textwidth{\hfil\includegraphics[page=17,trim=18mm 15mm 18mm 17mm,clip,width=\textwidth,height=\dimexpr\textheight-4pt\relax,keepaspectratio]{classification_typeII_embed.pdf}\hfil}\vfil}\par
\clearpage\thispagestyle{plain}
\pdfbookmark[2]{Pc (7)}{table-typeII-sg7}
\noindent\vbox to\dimexpr\textheight-4pt\relax{\vfil\hbox to\textwidth{\hfil\includegraphics[page=18,trim=18mm 15mm 18mm 17mm,clip,width=\textwidth,height=\dimexpr\textheight-4pt\relax,keepaspectratio]{classification_typeII_embed.pdf}\hfil}\vfil}\par
\clearpage\thispagestyle{plain}
\noindent\vbox to\dimexpr\textheight-4pt\relax{\vfil\hbox to\textwidth{\hfil\includegraphics[page=19,trim=18mm 15mm 18mm 17mm,clip,width=\textwidth,height=\dimexpr\textheight-4pt\relax,keepaspectratio]{classification_typeII_embed.pdf}\hfil}\vfil}\par
\clearpage\thispagestyle{plain}
\noindent\vbox to\dimexpr\textheight-4pt\relax{\vfil\hbox to\textwidth{\hfil\includegraphics[page=20,trim=18mm 15mm 18mm 17mm,clip,width=\textwidth,height=\dimexpr\textheight-4pt\relax,keepaspectratio]{classification_typeII_embed.pdf}\hfil}\vfil}\par
\clearpage\thispagestyle{plain}
\pdfbookmark[2]{Cm (8)}{table-typeII-sg8}
\noindent\vbox to\dimexpr\textheight-4pt\relax{\vfil\hbox to\textwidth{\hfil\includegraphics[page=21,trim=18mm 15mm 18mm 17mm,clip,width=\textwidth,height=\dimexpr\textheight-4pt\relax,keepaspectratio]{classification_typeII_embed.pdf}\hfil}\vfil}\par
\clearpage\thispagestyle{plain}
\noindent\vbox to\dimexpr\textheight-4pt\relax{\vfil\hbox to\textwidth{\hfil\includegraphics[page=22,trim=18mm 15mm 18mm 17mm,clip,width=\textwidth,height=\dimexpr\textheight-4pt\relax,keepaspectratio]{classification_typeII_embed.pdf}\hfil}\vfil}\par
\clearpage\thispagestyle{plain}
\noindent\vbox to\dimexpr\textheight-4pt\relax{\vfil\hbox to\textwidth{\hfil\includegraphics[page=23,trim=18mm 15mm 18mm 17mm,clip,width=\textwidth,height=\dimexpr\textheight-4pt\relax,keepaspectratio]{classification_typeII_embed.pdf}\hfil}\vfil}\par
\clearpage\thispagestyle{plain}
\pdfbookmark[2]{Cc (9)}{table-typeII-sg9}
\noindent\vbox to\dimexpr\textheight-4pt\relax{\vfil\hbox to\textwidth{\hfil\includegraphics[page=24,trim=18mm 15mm 18mm 17mm,clip,width=\textwidth,height=\dimexpr\textheight-4pt\relax,keepaspectratio]{classification_typeII_embed.pdf}\hfil}\vfil}\par
\clearpage\thispagestyle{plain}
\noindent\vbox to\dimexpr\textheight-4pt\relax{\vfil\hbox to\textwidth{\hfil\includegraphics[page=25,trim=18mm 15mm 18mm 17mm,clip,width=\textwidth,height=\dimexpr\textheight-4pt\relax,keepaspectratio]{classification_typeII_embed.pdf}\hfil}\vfil}\par
\clearpage\thispagestyle{plain}
\pdfbookmark[2]{P2/m (10)}{table-typeII-sg10}
\noindent\vbox to\dimexpr\textheight-4pt\relax{\vfil\hbox to\textwidth{\hfil\includegraphics[page=26,trim=18mm 15mm 18mm 17mm,clip,width=\textwidth,height=\dimexpr\textheight-4pt\relax,keepaspectratio]{classification_typeII_embed.pdf}\hfil}\vfil}\par
\clearpage\thispagestyle{plain}
\noindent\vbox to\dimexpr\textheight-4pt\relax{\vfil\hbox to\textwidth{\hfil\includegraphics[page=27,trim=18mm 15mm 18mm 17mm,clip,width=\textwidth,height=\dimexpr\textheight-4pt\relax,keepaspectratio]{classification_typeII_embed.pdf}\hfil}\vfil}\par
\clearpage\thispagestyle{plain}
\pdfbookmark[2]{P21/m (11)}{table-typeII-sg11}
\noindent\vbox to\dimexpr\textheight-4pt\relax{\vfil\hbox to\textwidth{\hfil\includegraphics[page=28,trim=18mm 15mm 18mm 17mm,clip,width=\textwidth,height=\dimexpr\textheight-4pt\relax,keepaspectratio]{classification_typeII_embed.pdf}\hfil}\vfil}\par
\clearpage\thispagestyle{plain}
\pdfbookmark[2]{C2/m (12)}{table-typeII-sg12}
\noindent\vbox to\dimexpr\textheight-4pt\relax{\vfil\hbox to\textwidth{\hfil\includegraphics[page=29,trim=18mm 15mm 18mm 17mm,clip,width=\textwidth,height=\dimexpr\textheight-4pt\relax,keepaspectratio]{classification_typeII_embed.pdf}\hfil}\vfil}\par
\clearpage\thispagestyle{plain}
\noindent\vbox to\dimexpr\textheight-4pt\relax{\vfil\hbox to\textwidth{\hfil\includegraphics[page=30,trim=18mm 15mm 18mm 17mm,clip,width=\textwidth,height=\dimexpr\textheight-4pt\relax,keepaspectratio]{classification_typeII_embed.pdf}\hfil}\vfil}\par
\clearpage\thispagestyle{plain}
\pdfbookmark[2]{P2/c (13)}{table-typeII-sg13}
\noindent\vbox to\dimexpr\textheight-4pt\relax{\vfil\hbox to\textwidth{\hfil\includegraphics[page=31,trim=18mm 15mm 18mm 17mm,clip,width=\textwidth,height=\dimexpr\textheight-4pt\relax,keepaspectratio]{classification_typeII_embed.pdf}\hfil}\vfil}\par
\clearpage\thispagestyle{plain}
\noindent\vbox to\dimexpr\textheight-4pt\relax{\vfil\hbox to\textwidth{\hfil\includegraphics[page=32,trim=18mm 15mm 18mm 17mm,clip,width=\textwidth,height=\dimexpr\textheight-4pt\relax,keepaspectratio]{classification_typeII_embed.pdf}\hfil}\vfil}\par
\clearpage\thispagestyle{plain}
\noindent\vbox to\dimexpr\textheight-4pt\relax{\vfil\hbox to\textwidth{\hfil\includegraphics[page=33,trim=18mm 15mm 18mm 17mm,clip,width=\textwidth,height=\dimexpr\textheight-4pt\relax,keepaspectratio]{classification_typeII_embed.pdf}\hfil}\vfil}\par
\clearpage\thispagestyle{plain}
\pdfbookmark[2]{P21/c (14)}{table-typeII-sg14}
\noindent\vbox to\dimexpr\textheight-4pt\relax{\vfil\hbox to\textwidth{\hfil\includegraphics[page=34,trim=18mm 15mm 18mm 17mm,clip,width=\textwidth,height=\dimexpr\textheight-4pt\relax,keepaspectratio]{classification_typeII_embed.pdf}\hfil}\vfil}\par
\clearpage\thispagestyle{plain}
\noindent\vbox to\dimexpr\textheight-4pt\relax{\vfil\hbox to\textwidth{\hfil\includegraphics[page=35,trim=18mm 15mm 18mm 17mm,clip,width=\textwidth,height=\dimexpr\textheight-4pt\relax,keepaspectratio]{classification_typeII_embed.pdf}\hfil}\vfil}\par
\clearpage\thispagestyle{plain}
\noindent\vbox to\dimexpr\textheight-4pt\relax{\vfil\hbox to\textwidth{\hfil\includegraphics[page=36,trim=18mm 15mm 18mm 17mm,clip,width=\textwidth,height=\dimexpr\textheight-4pt\relax,keepaspectratio]{classification_typeII_embed.pdf}\hfil}\vfil}\par
\clearpage\thispagestyle{plain}
\pdfbookmark[2]{C2/c (15)}{table-typeII-sg15}
\noindent\vbox to\dimexpr\textheight-4pt\relax{\vfil\hbox to\textwidth{\hfil\includegraphics[page=37,trim=18mm 15mm 18mm 17mm,clip,width=\textwidth,height=\dimexpr\textheight-4pt\relax,keepaspectratio]{classification_typeII_embed.pdf}\hfil}\vfil}\par
\clearpage\thispagestyle{plain}
\noindent\vbox to\dimexpr\textheight-4pt\relax{\vfil\hbox to\textwidth{\hfil\includegraphics[page=38,trim=18mm 15mm 18mm 17mm,clip,width=\textwidth,height=\dimexpr\textheight-4pt\relax,keepaspectratio]{classification_typeII_embed.pdf}\hfil}\vfil}\par
\clearpage\thispagestyle{plain}
\noindent\vbox to\dimexpr\textheight-4pt\relax{\vfil\hbox to\textwidth{\hfil\includegraphics[page=39,trim=18mm 15mm 18mm 17mm,clip,width=\textwidth,height=\dimexpr\textheight-4pt\relax,keepaspectratio]{classification_typeII_embed.pdf}\hfil}\vfil}\par
\clearpage\thispagestyle{plain}
\pdfbookmark[2]{P222 (16)}{table-typeII-sg16}
\noindent\vbox to\dimexpr\textheight-4pt\relax{\vfil\hbox to\textwidth{\hfil\includegraphics[page=40,trim=18mm 15mm 18mm 17mm,clip,width=\textwidth,height=\dimexpr\textheight-4pt\relax,keepaspectratio]{classification_typeII_embed.pdf}\hfil}\vfil}\par
\clearpage\thispagestyle{plain}
\pdfbookmark[2]{P2221 (17)}{table-typeII-sg17}
\noindent\vbox to\dimexpr\textheight-4pt\relax{\vfil\hbox to\textwidth{\hfil\includegraphics[page=41,trim=18mm 15mm 18mm 17mm,clip,width=\textwidth,height=\dimexpr\textheight-4pt\relax,keepaspectratio]{classification_typeII_embed.pdf}\hfil}\vfil}\par
\clearpage\thispagestyle{plain}
\noindent\vbox to\dimexpr\textheight-4pt\relax{\vfil\hbox to\textwidth{\hfil\includegraphics[page=42,trim=18mm 15mm 18mm 17mm,clip,width=\textwidth,height=\dimexpr\textheight-4pt\relax,keepaspectratio]{classification_typeII_embed.pdf}\hfil}\vfil}\par
\clearpage\thispagestyle{plain}
\pdfbookmark[2]{P21212 (18)}{table-typeII-sg18}
\noindent\vbox to\dimexpr\textheight-4pt\relax{\vfil\hbox to\textwidth{\hfil\includegraphics[page=43,trim=18mm 15mm 18mm 17mm,clip,width=\textwidth,height=\dimexpr\textheight-4pt\relax,keepaspectratio]{classification_typeII_embed.pdf}\hfil}\vfil}\par
\clearpage\thispagestyle{plain}
\noindent\vbox to\dimexpr\textheight-4pt\relax{\vfil\hbox to\textwidth{\hfil\includegraphics[page=44,trim=18mm 15mm 18mm 17mm,clip,width=\textwidth,height=\dimexpr\textheight-4pt\relax,keepaspectratio]{classification_typeII_embed.pdf}\hfil}\vfil}\par
\clearpage\thispagestyle{plain}
\pdfbookmark[2]{P212121 (19)}{table-typeII-sg19}
\noindent\vbox to\dimexpr\textheight-4pt\relax{\vfil\hbox to\textwidth{\hfil\includegraphics[page=45,trim=18mm 15mm 18mm 17mm,clip,width=\textwidth,height=\dimexpr\textheight-4pt\relax,keepaspectratio]{classification_typeII_embed.pdf}\hfil}\vfil}\par
\clearpage\thispagestyle{plain}
\noindent\vbox to\dimexpr\textheight-4pt\relax{\vfil\hbox to\textwidth{\hfil\includegraphics[page=46,trim=18mm 15mm 18mm 17mm,clip,width=\textwidth,height=\dimexpr\textheight-4pt\relax,keepaspectratio]{classification_typeII_embed.pdf}\hfil}\vfil}\par
\clearpage\thispagestyle{plain}
\pdfbookmark[2]{C2221 (20)}{table-typeII-sg20}
\noindent\vbox to\dimexpr\textheight-4pt\relax{\vfil\hbox to\textwidth{\hfil\includegraphics[page=47,trim=18mm 15mm 18mm 17mm,clip,width=\textwidth,height=\dimexpr\textheight-4pt\relax,keepaspectratio]{classification_typeII_embed.pdf}\hfil}\vfil}\par
\clearpage\thispagestyle{plain}
\pdfbookmark[2]{C222 (21)}{table-typeII-sg21}
\noindent\vbox to\dimexpr\textheight-4pt\relax{\vfil\hbox to\textwidth{\hfil\includegraphics[page=48,trim=18mm 15mm 18mm 17mm,clip,width=\textwidth,height=\dimexpr\textheight-4pt\relax,keepaspectratio]{classification_typeII_embed.pdf}\hfil}\vfil}\par
\clearpage\thispagestyle{plain}
\noindent\vbox to\dimexpr\textheight-4pt\relax{\vfil\hbox to\textwidth{\hfil\includegraphics[page=49,trim=18mm 15mm 18mm 17mm,clip,width=\textwidth,height=\dimexpr\textheight-4pt\relax,keepaspectratio]{classification_typeII_embed.pdf}\hfil}\vfil}\par
\clearpage\thispagestyle{plain}
\noindent\vbox to\dimexpr\textheight-4pt\relax{\vfil\hbox to\textwidth{\hfil\includegraphics[page=50,trim=18mm 15mm 18mm 17mm,clip,width=\textwidth,height=\dimexpr\textheight-4pt\relax,keepaspectratio]{classification_typeII_embed.pdf}\hfil}\vfil}\par
\clearpage\thispagestyle{plain}
\pdfbookmark[2]{F222 (22)}{table-typeII-sg22}
\pdfbookmark[2]{I222 (23)}{table-typeII-sg23}
\noindent\vbox to\dimexpr\textheight-4pt\relax{\vfil\hbox to\textwidth{\hfil\includegraphics[page=51,trim=18mm 15mm 18mm 17mm,clip,width=\textwidth,height=\dimexpr\textheight-4pt\relax,keepaspectratio]{classification_typeII_embed.pdf}\hfil}\vfil}\par
\clearpage\thispagestyle{plain}
\pdfbookmark[2]{I212121 (24)}{table-typeII-sg24}
\noindent\vbox to\dimexpr\textheight-4pt\relax{\vfil\hbox to\textwidth{\hfil\includegraphics[page=52,trim=18mm 15mm 18mm 17mm,clip,width=\textwidth,height=\dimexpr\textheight-4pt\relax,keepaspectratio]{classification_typeII_embed.pdf}\hfil}\vfil}\par
\clearpage\thispagestyle{plain}
\pdfbookmark[2]{Pmm2 (25)}{table-typeII-sg25}
\noindent\vbox to\dimexpr\textheight-4pt\relax{\vfil\hbox to\textwidth{\hfil\includegraphics[page=53,trim=18mm 15mm 18mm 17mm,clip,width=\textwidth,height=\dimexpr\textheight-4pt\relax,keepaspectratio]{classification_typeII_embed.pdf}\hfil}\vfil}\par
\clearpage\thispagestyle{plain}
\noindent\vbox to\dimexpr\textheight-4pt\relax{\vfil\hbox to\textwidth{\hfil\includegraphics[page=54,trim=18mm 15mm 18mm 17mm,clip,width=\textwidth,height=\dimexpr\textheight-4pt\relax,keepaspectratio]{classification_typeII_embed.pdf}\hfil}\vfil}\par
\clearpage\thispagestyle{plain}
\pdfbookmark[2]{Pmc21 (26)}{table-typeII-sg26}
\noindent\vbox to\dimexpr\textheight-4pt\relax{\vfil\hbox to\textwidth{\hfil\includegraphics[page=55,trim=18mm 15mm 18mm 17mm,clip,width=\textwidth,height=\dimexpr\textheight-4pt\relax,keepaspectratio]{classification_typeII_embed.pdf}\hfil}\vfil}\par
\clearpage\thispagestyle{plain}
\pdfbookmark[2]{Pcc2 (27)}{table-typeII-sg27}
\noindent\vbox to\dimexpr\textheight-4pt\relax{\vfil\hbox to\textwidth{\hfil\includegraphics[page=56,trim=18mm 15mm 18mm 17mm,clip,width=\textwidth,height=\dimexpr\textheight-4pt\relax,keepaspectratio]{classification_typeII_embed.pdf}\hfil}\vfil}\par
\clearpage\thispagestyle{plain}
\pdfbookmark[2]{Pma2 (28)}{table-typeII-sg28}
\noindent\vbox to\dimexpr\textheight-4pt\relax{\vfil\hbox to\textwidth{\hfil\includegraphics[page=57,trim=18mm 15mm 18mm 17mm,clip,width=\textwidth,height=\dimexpr\textheight-4pt\relax,keepaspectratio]{classification_typeII_embed.pdf}\hfil}\vfil}\par
\clearpage\thispagestyle{plain}
\noindent\vbox to\dimexpr\textheight-4pt\relax{\vfil\hbox to\textwidth{\hfil\includegraphics[page=58,trim=18mm 15mm 18mm 17mm,clip,width=\textwidth,height=\dimexpr\textheight-4pt\relax,keepaspectratio]{classification_typeII_embed.pdf}\hfil}\vfil}\par
\clearpage\thispagestyle{plain}
\pdfbookmark[2]{Pca21 (29)}{table-typeII-sg29}
\noindent\vbox to\dimexpr\textheight-4pt\relax{\vfil\hbox to\textwidth{\hfil\includegraphics[page=59,trim=18mm 15mm 18mm 17mm,clip,width=\textwidth,height=\dimexpr\textheight-4pt\relax,keepaspectratio]{classification_typeII_embed.pdf}\hfil}\vfil}\par
\clearpage\thispagestyle{plain}
\noindent\vbox to\dimexpr\textheight-4pt\relax{\vfil\hbox to\textwidth{\hfil\includegraphics[page=60,trim=18mm 15mm 18mm 17mm,clip,width=\textwidth,height=\dimexpr\textheight-4pt\relax,keepaspectratio]{classification_typeII_embed.pdf}\hfil}\vfil}\par
\clearpage\thispagestyle{plain}
\pdfbookmark[2]{Pnc2 (30)}{table-typeII-sg30}
\noindent\vbox to\dimexpr\textheight-4pt\relax{\vfil\hbox to\textwidth{\hfil\includegraphics[page=61,trim=18mm 15mm 18mm 17mm,clip,width=\textwidth,height=\dimexpr\textheight-4pt\relax,keepaspectratio]{classification_typeII_embed.pdf}\hfil}\vfil}\par
\clearpage\thispagestyle{plain}
\pdfbookmark[2]{Pmn21 (31)}{table-typeII-sg31}
\noindent\vbox to\dimexpr\textheight-4pt\relax{\vfil\hbox to\textwidth{\hfil\includegraphics[page=62,trim=18mm 15mm 18mm 17mm,clip,width=\textwidth,height=\dimexpr\textheight-4pt\relax,keepaspectratio]{classification_typeII_embed.pdf}\hfil}\vfil}\par
\clearpage\thispagestyle{plain}
\noindent\vbox to\dimexpr\textheight-4pt\relax{\vfil\hbox to\textwidth{\hfil\includegraphics[page=63,trim=18mm 15mm 18mm 17mm,clip,width=\textwidth,height=\dimexpr\textheight-4pt\relax,keepaspectratio]{classification_typeII_embed.pdf}\hfil}\vfil}\par
\clearpage\thispagestyle{plain}
\pdfbookmark[2]{Pba2 (32)}{table-typeII-sg32}
\noindent\vbox to\dimexpr\textheight-4pt\relax{\vfil\hbox to\textwidth{\hfil\includegraphics[page=64,trim=18mm 15mm 18mm 17mm,clip,width=\textwidth,height=\dimexpr\textheight-4pt\relax,keepaspectratio]{classification_typeII_embed.pdf}\hfil}\vfil}\par
\clearpage\thispagestyle{plain}
\pdfbookmark[2]{Pna21 (33)}{table-typeII-sg33}
\noindent\vbox to\dimexpr\textheight-4pt\relax{\vfil\hbox to\textwidth{\hfil\includegraphics[page=65,trim=18mm 15mm 18mm 17mm,clip,width=\textwidth,height=\dimexpr\textheight-4pt\relax,keepaspectratio]{classification_typeII_embed.pdf}\hfil}\vfil}\par
\clearpage\thispagestyle{plain}
\noindent\vbox to\dimexpr\textheight-4pt\relax{\vfil\hbox to\textwidth{\hfil\includegraphics[page=66,trim=18mm 15mm 18mm 17mm,clip,width=\textwidth,height=\dimexpr\textheight-4pt\relax,keepaspectratio]{classification_typeII_embed.pdf}\hfil}\vfil}\par
\clearpage\thispagestyle{plain}
\pdfbookmark[2]{Pnn2 (34)}{table-typeII-sg34}
\noindent\vbox to\dimexpr\textheight-4pt\relax{\vfil\hbox to\textwidth{\hfil\includegraphics[page=67,trim=18mm 15mm 18mm 17mm,clip,width=\textwidth,height=\dimexpr\textheight-4pt\relax,keepaspectratio]{classification_typeII_embed.pdf}\hfil}\vfil}\par
\clearpage\thispagestyle{plain}
\pdfbookmark[2]{Cmm2 (35)}{table-typeII-sg35}
\noindent\vbox to\dimexpr\textheight-4pt\relax{\vfil\hbox to\textwidth{\hfil\includegraphics[page=68,trim=18mm 15mm 18mm 17mm,clip,width=\textwidth,height=\dimexpr\textheight-4pt\relax,keepaspectratio]{classification_typeII_embed.pdf}\hfil}\vfil}\par
\clearpage\thispagestyle{plain}
\pdfbookmark[2]{Cmc21 (36)}{table-typeII-sg36}
\noindent\vbox to\dimexpr\textheight-4pt\relax{\vfil\hbox to\textwidth{\hfil\includegraphics[page=69,trim=18mm 15mm 18mm 17mm,clip,width=\textwidth,height=\dimexpr\textheight-4pt\relax,keepaspectratio]{classification_typeII_embed.pdf}\hfil}\vfil}\par
\clearpage\thispagestyle{plain}
\pdfbookmark[2]{Ccc2 (37)}{table-typeII-sg37}
\noindent\vbox to\dimexpr\textheight-4pt\relax{\vfil\hbox to\textwidth{\hfil\includegraphics[page=70,trim=18mm 15mm 18mm 17mm,clip,width=\textwidth,height=\dimexpr\textheight-4pt\relax,keepaspectratio]{classification_typeII_embed.pdf}\hfil}\vfil}\par
\clearpage\thispagestyle{plain}
\pdfbookmark[2]{Amm2 (38)}{table-typeII-sg38}
\noindent\vbox to\dimexpr\textheight-4pt\relax{\vfil\hbox to\textwidth{\hfil\includegraphics[page=71,trim=18mm 15mm 18mm 17mm,clip,width=\textwidth,height=\dimexpr\textheight-4pt\relax,keepaspectratio]{classification_typeII_embed.pdf}\hfil}\vfil}\par
\clearpage\thispagestyle{plain}
\noindent\vbox to\dimexpr\textheight-4pt\relax{\vfil\hbox to\textwidth{\hfil\includegraphics[page=72,trim=18mm 15mm 18mm 17mm,clip,width=\textwidth,height=\dimexpr\textheight-4pt\relax,keepaspectratio]{classification_typeII_embed.pdf}\hfil}\vfil}\par
\clearpage\thispagestyle{plain}
\pdfbookmark[2]{Aem2 (39)}{table-typeII-sg39}
\noindent\vbox to\dimexpr\textheight-4pt\relax{\vfil\hbox to\textwidth{\hfil\includegraphics[page=73,trim=18mm 15mm 18mm 17mm,clip,width=\textwidth,height=\dimexpr\textheight-4pt\relax,keepaspectratio]{classification_typeII_embed.pdf}\hfil}\vfil}\par
\clearpage\thispagestyle{plain}
\pdfbookmark[2]{Ama2 (40)}{table-typeII-sg40}
\noindent\vbox to\dimexpr\textheight-4pt\relax{\vfil\hbox to\textwidth{\hfil\includegraphics[page=74,trim=18mm 15mm 18mm 17mm,clip,width=\textwidth,height=\dimexpr\textheight-4pt\relax,keepaspectratio]{classification_typeII_embed.pdf}\hfil}\vfil}\par
\clearpage\thispagestyle{plain}
\pdfbookmark[2]{Aea2 (41)}{table-typeII-sg41}
\pdfbookmark[2]{Fmm2 (42)}{table-typeII-sg42}
\noindent\vbox to\dimexpr\textheight-4pt\relax{\vfil\hbox to\textwidth{\hfil\includegraphics[page=75,trim=18mm 15mm 18mm 17mm,clip,width=\textwidth,height=\dimexpr\textheight-4pt\relax,keepaspectratio]{classification_typeII_embed.pdf}\hfil}\vfil}\par
\clearpage\thispagestyle{plain}
\pdfbookmark[2]{Fdd2 (43)}{table-typeII-sg43}
\pdfbookmark[2]{Imm2 (44)}{table-typeII-sg44}
\noindent\vbox to\dimexpr\textheight-4pt\relax{\vfil\hbox to\textwidth{\hfil\includegraphics[page=76,trim=18mm 15mm 18mm 17mm,clip,width=\textwidth,height=\dimexpr\textheight-4pt\relax,keepaspectratio]{classification_typeII_embed.pdf}\hfil}\vfil}\par
\clearpage\thispagestyle{plain}
\pdfbookmark[2]{Iba2 (45)}{table-typeII-sg45}
\noindent\vbox to\dimexpr\textheight-4pt\relax{\vfil\hbox to\textwidth{\hfil\includegraphics[page=77,trim=18mm 15mm 18mm 17mm,clip,width=\textwidth,height=\dimexpr\textheight-4pt\relax,keepaspectratio]{classification_typeII_embed.pdf}\hfil}\vfil}\par
\clearpage\thispagestyle{plain}
\pdfbookmark[2]{Ima2 (46)}{table-typeII-sg46}
\noindent\vbox to\dimexpr\textheight-4pt\relax{\vfil\hbox to\textwidth{\hfil\includegraphics[page=78,trim=18mm 15mm 18mm 17mm,clip,width=\textwidth,height=\dimexpr\textheight-4pt\relax,keepaspectratio]{classification_typeII_embed.pdf}\hfil}\vfil}\par
\clearpage\thispagestyle{plain}
\pdfbookmark[2]{Pmmm (47)}{table-typeII-sg47}
\noindent\vbox to\dimexpr\textheight-4pt\relax{\vfil\hbox to\textwidth{\hfil\includegraphics[page=79,trim=18mm 15mm 18mm 17mm,clip,width=\textwidth,height=\dimexpr\textheight-4pt\relax,keepaspectratio]{classification_typeII_embed.pdf}\hfil}\vfil}\par
\clearpage\thispagestyle{plain}
\pdfbookmark[2]{Pnnn (48)}{table-typeII-sg48}
\pdfbookmark[2]{Pccm (49)}{table-typeII-sg49}
\noindent\vbox to\dimexpr\textheight-4pt\relax{\vfil\hbox to\textwidth{\hfil\includegraphics[page=80,trim=18mm 15mm 18mm 17mm,clip,width=\textwidth,height=\dimexpr\textheight-4pt\relax,keepaspectratio]{classification_typeII_embed.pdf}\hfil}\vfil}\par
\clearpage\thispagestyle{plain}
\pdfbookmark[2]{Pban (50)}{table-typeII-sg50}
\noindent\vbox to\dimexpr\textheight-4pt\relax{\vfil\hbox to\textwidth{\hfil\includegraphics[page=81,trim=18mm 15mm 18mm 17mm,clip,width=\textwidth,height=\dimexpr\textheight-4pt\relax,keepaspectratio]{classification_typeII_embed.pdf}\hfil}\vfil}\par
\clearpage\thispagestyle{plain}
\pdfbookmark[2]{Pmma (51)}{table-typeII-sg51}
\noindent\vbox to\dimexpr\textheight-4pt\relax{\vfil\hbox to\textwidth{\hfil\includegraphics[page=82,trim=18mm 15mm 18mm 17mm,clip,width=\textwidth,height=\dimexpr\textheight-4pt\relax,keepaspectratio]{classification_typeII_embed.pdf}\hfil}\vfil}\par
\clearpage\thispagestyle{plain}
\pdfbookmark[2]{Pnna (52)}{table-typeII-sg52}
\noindent\vbox to\dimexpr\textheight-4pt\relax{\vfil\hbox to\textwidth{\hfil\includegraphics[page=83,trim=18mm 15mm 18mm 17mm,clip,width=\textwidth,height=\dimexpr\textheight-4pt\relax,keepaspectratio]{classification_typeII_embed.pdf}\hfil}\vfil}\par
\clearpage\thispagestyle{plain}
\pdfbookmark[2]{Pmna (53)}{table-typeII-sg53}
\pdfbookmark[2]{Pcca (54)}{table-typeII-sg54}
\noindent\vbox to\dimexpr\textheight-4pt\relax{\vfil\hbox to\textwidth{\hfil\includegraphics[page=84,trim=18mm 15mm 18mm 17mm,clip,width=\textwidth,height=\dimexpr\textheight-4pt\relax,keepaspectratio]{classification_typeII_embed.pdf}\hfil}\vfil}\par
\clearpage\thispagestyle{plain}
\pdfbookmark[2]{Pbam (55)}{table-typeII-sg55}
\noindent\vbox to\dimexpr\textheight-4pt\relax{\vfil\hbox to\textwidth{\hfil\includegraphics[page=85,trim=18mm 15mm 18mm 17mm,clip,width=\textwidth,height=\dimexpr\textheight-4pt\relax,keepaspectratio]{classification_typeII_embed.pdf}\hfil}\vfil}\par
\clearpage\thispagestyle{plain}
\pdfbookmark[2]{Pccn (56)}{table-typeII-sg56}
\pdfbookmark[2]{Pbcm (57)}{table-typeII-sg57}
\noindent\vbox to\dimexpr\textheight-4pt\relax{\vfil\hbox to\textwidth{\hfil\includegraphics[page=86,trim=18mm 15mm 18mm 17mm,clip,width=\textwidth,height=\dimexpr\textheight-4pt\relax,keepaspectratio]{classification_typeII_embed.pdf}\hfil}\vfil}\par
\clearpage\thispagestyle{plain}
\pdfbookmark[2]{Pnnm (58)}{table-typeII-sg58}
\pdfbookmark[2]{Pmmn (59)}{table-typeII-sg59}
\noindent\vbox to\dimexpr\textheight-4pt\relax{\vfil\hbox to\textwidth{\hfil\includegraphics[page=87,trim=18mm 15mm 18mm 17mm,clip,width=\textwidth,height=\dimexpr\textheight-4pt\relax,keepaspectratio]{classification_typeII_embed.pdf}\hfil}\vfil}\par
\clearpage\thispagestyle{plain}
\pdfbookmark[2]{Pbcn (60)}{table-typeII-sg60}
\noindent\vbox to\dimexpr\textheight-4pt\relax{\vfil\hbox to\textwidth{\hfil\includegraphics[page=88,trim=18mm 15mm 18mm 17mm,clip,width=\textwidth,height=\dimexpr\textheight-4pt\relax,keepaspectratio]{classification_typeII_embed.pdf}\hfil}\vfil}\par
\clearpage\thispagestyle{plain}
\pdfbookmark[2]{Pbca (61)}{table-typeII-sg61}
\pdfbookmark[2]{Pnma (62)}{table-typeII-sg62}
\pdfbookmark[2]{Cmcm (63)}{table-typeII-sg63}
\noindent\vbox to\dimexpr\textheight-4pt\relax{\vfil\hbox to\textwidth{\hfil\includegraphics[page=89,trim=18mm 15mm 18mm 17mm,clip,width=\textwidth,height=\dimexpr\textheight-4pt\relax,keepaspectratio]{classification_typeII_embed.pdf}\hfil}\vfil}\par
\clearpage\thispagestyle{plain}
\pdfbookmark[2]{Cmce (64)}{table-typeII-sg64}
\pdfbookmark[2]{Cmmm (65)}{table-typeII-sg65}
\noindent\vbox to\dimexpr\textheight-4pt\relax{\vfil\hbox to\textwidth{\hfil\includegraphics[page=90,trim=18mm 15mm 18mm 17mm,clip,width=\textwidth,height=\dimexpr\textheight-4pt\relax,keepaspectratio]{classification_typeII_embed.pdf}\hfil}\vfil}\par
\clearpage\thispagestyle{plain}
\pdfbookmark[2]{Cccm (66)}{table-typeII-sg66}
\noindent\vbox to\dimexpr\textheight-4pt\relax{\vfil\hbox to\textwidth{\hfil\includegraphics[page=91,trim=18mm 15mm 18mm 17mm,clip,width=\textwidth,height=\dimexpr\textheight-4pt\relax,keepaspectratio]{classification_typeII_embed.pdf}\hfil}\vfil}\par
\clearpage\thispagestyle{plain}
\pdfbookmark[2]{Cmme (67)}{table-typeII-sg67}
\pdfbookmark[2]{Ccce (68)}{table-typeII-sg68}
\noindent\vbox to\dimexpr\textheight-4pt\relax{\vfil\hbox to\textwidth{\hfil\includegraphics[page=92,trim=18mm 15mm 18mm 17mm,clip,width=\textwidth,height=\dimexpr\textheight-4pt\relax,keepaspectratio]{classification_typeII_embed.pdf}\hfil}\vfil}\par
\clearpage\thispagestyle{plain}
\pdfbookmark[2]{Fmmm (69)}{table-typeII-sg69}
\pdfbookmark[2]{Fddd (70)}{table-typeII-sg70}
\noindent\vbox to\dimexpr\textheight-4pt\relax{\vfil\hbox to\textwidth{\hfil\includegraphics[page=93,trim=18mm 15mm 18mm 17mm,clip,width=\textwidth,height=\dimexpr\textheight-4pt\relax,keepaspectratio]{classification_typeII_embed.pdf}\hfil}\vfil}\par
\clearpage\thispagestyle{plain}
\pdfbookmark[2]{Immm (71)}{table-typeII-sg71}
\pdfbookmark[2]{Ibam (72)}{table-typeII-sg72}
\noindent\vbox to\dimexpr\textheight-4pt\relax{\vfil\hbox to\textwidth{\hfil\includegraphics[page=94,trim=18mm 15mm 18mm 17mm,clip,width=\textwidth,height=\dimexpr\textheight-4pt\relax,keepaspectratio]{classification_typeII_embed.pdf}\hfil}\vfil}\par
\clearpage\thispagestyle{plain}
\pdfbookmark[2]{Ibca (73)}{table-typeII-sg73}
\pdfbookmark[2]{Imma (74)}{table-typeII-sg74}
\pdfbookmark[2]{P4 (75)}{table-typeII-sg75}
\noindent\vbox to\dimexpr\textheight-4pt\relax{\vfil\hbox to\textwidth{\hfil\includegraphics[page=95,trim=18mm 15mm 18mm 17mm,clip,width=\textwidth,height=\dimexpr\textheight-4pt\relax,keepaspectratio]{classification_typeII_embed.pdf}\hfil}\vfil}\par
\clearpage\thispagestyle{plain}
\pdfbookmark[2]{P41 (76)}{table-typeII-sg76}
\noindent\vbox to\dimexpr\textheight-4pt\relax{\vfil\hbox to\textwidth{\hfil\includegraphics[page=96,trim=18mm 15mm 18mm 17mm,clip,width=\textwidth,height=\dimexpr\textheight-4pt\relax,keepaspectratio]{classification_typeII_embed.pdf}\hfil}\vfil}\par
\clearpage\thispagestyle{plain}
\pdfbookmark[2]{P42 (77)}{table-typeII-sg77}
\noindent\vbox to\dimexpr\textheight-4pt\relax{\vfil\hbox to\textwidth{\hfil\includegraphics[page=97,trim=18mm 15mm 18mm 17mm,clip,width=\textwidth,height=\dimexpr\textheight-4pt\relax,keepaspectratio]{classification_typeII_embed.pdf}\hfil}\vfil}\par
\clearpage\thispagestyle{plain}
\pdfbookmark[2]{P43 (78)}{table-typeII-sg78}
\pdfbookmark[2]{I4 (79)}{table-typeII-sg79}
\noindent\vbox to\dimexpr\textheight-4pt\relax{\vfil\hbox to\textwidth{\hfil\includegraphics[page=98,trim=18mm 15mm 18mm 17mm,clip,width=\textwidth,height=\dimexpr\textheight-4pt\relax,keepaspectratio]{classification_typeII_embed.pdf}\hfil}\vfil}\par
\clearpage\thispagestyle{plain}
\pdfbookmark[2]{I41 (80)}{table-typeII-sg80}
\pdfbookmark[2]{P-4 (81)}{table-typeII-sg81}
\noindent\vbox to\dimexpr\textheight-4pt\relax{\vfil\hbox to\textwidth{\hfil\includegraphics[page=99,trim=18mm 15mm 18mm 17mm,clip,width=\textwidth,height=\dimexpr\textheight-4pt\relax,keepaspectratio]{classification_typeII_embed.pdf}\hfil}\vfil}\par
\clearpage\thispagestyle{plain}
\noindent\vbox to\dimexpr\textheight-4pt\relax{\vfil\hbox to\textwidth{\hfil\includegraphics[page=100,trim=18mm 15mm 18mm 17mm,clip,width=\textwidth,height=\dimexpr\textheight-4pt\relax,keepaspectratio]{classification_typeII_embed.pdf}\hfil}\vfil}\par
\clearpage\thispagestyle{plain}
\pdfbookmark[2]{I-4 (82)}{table-typeII-sg82}
\noindent\vbox to\dimexpr\textheight-4pt\relax{\vfil\hbox to\textwidth{\hfil\includegraphics[page=101,trim=18mm 15mm 18mm 17mm,clip,width=\textwidth,height=\dimexpr\textheight-4pt\relax,keepaspectratio]{classification_typeII_embed.pdf}\hfil}\vfil}\par
\clearpage\thispagestyle{plain}
\pdfbookmark[2]{P4/m (83)}{table-typeII-sg83}
\pdfbookmark[2]{P42/m (84)}{table-typeII-sg84}
\pdfbookmark[2]{P4/n (85)}{table-typeII-sg85}
\noindent\vbox to\dimexpr\textheight-4pt\relax{\vfil\hbox to\textwidth{\hfil\includegraphics[page=102,trim=18mm 15mm 18mm 17mm,clip,width=\textwidth,height=\dimexpr\textheight-4pt\relax,keepaspectratio]{classification_typeII_embed.pdf}\hfil}\vfil}\par
\clearpage\thispagestyle{plain}
\pdfbookmark[2]{P42/n (86)}{table-typeII-sg86}
\noindent\vbox to\dimexpr\textheight-4pt\relax{\vfil\hbox to\textwidth{\hfil\includegraphics[page=103,trim=18mm 15mm 18mm 17mm,clip,width=\textwidth,height=\dimexpr\textheight-4pt\relax,keepaspectratio]{classification_typeII_embed.pdf}\hfil}\vfil}\par
\clearpage\thispagestyle{plain}
\pdfbookmark[2]{I4/m (87)}{table-typeII-sg87}
\pdfbookmark[2]{I41/a (88)}{table-typeII-sg88}
\pdfbookmark[2]{P422 (89)}{table-typeII-sg89}
\noindent\vbox to\dimexpr\textheight-4pt\relax{\vfil\hbox to\textwidth{\hfil\includegraphics[page=104,trim=18mm 15mm 18mm 17mm,clip,width=\textwidth,height=\dimexpr\textheight-4pt\relax,keepaspectratio]{classification_typeII_embed.pdf}\hfil}\vfil}\par
\clearpage\thispagestyle{plain}
\pdfbookmark[2]{P4212 (90)}{table-typeII-sg90}
\noindent\vbox to\dimexpr\textheight-4pt\relax{\vfil\hbox to\textwidth{\hfil\includegraphics[page=105,trim=18mm 15mm 18mm 17mm,clip,width=\textwidth,height=\dimexpr\textheight-4pt\relax,keepaspectratio]{classification_typeII_embed.pdf}\hfil}\vfil}\par
\clearpage\thispagestyle{plain}
\pdfbookmark[2]{P4122 (91)}{table-typeII-sg91}
\pdfbookmark[2]{P41212 (92)}{table-typeII-sg92}
\noindent\vbox to\dimexpr\textheight-4pt\relax{\vfil\hbox to\textwidth{\hfil\includegraphics[page=106,trim=18mm 15mm 18mm 17mm,clip,width=\textwidth,height=\dimexpr\textheight-4pt\relax,keepaspectratio]{classification_typeII_embed.pdf}\hfil}\vfil}\par
\clearpage\thispagestyle{plain}
\pdfbookmark[2]{P4222 (93)}{table-typeII-sg93}
\noindent\vbox to\dimexpr\textheight-4pt\relax{\vfil\hbox to\textwidth{\hfil\includegraphics[page=107,trim=18mm 15mm 18mm 17mm,clip,width=\textwidth,height=\dimexpr\textheight-4pt\relax,keepaspectratio]{classification_typeII_embed.pdf}\hfil}\vfil}\par
\clearpage\thispagestyle{plain}
\pdfbookmark[2]{P42212 (94)}{table-typeII-sg94}
\pdfbookmark[2]{P4322 (95)}{table-typeII-sg95}
\noindent\vbox to\dimexpr\textheight-4pt\relax{\vfil\hbox to\textwidth{\hfil\includegraphics[page=108,trim=18mm 15mm 18mm 17mm,clip,width=\textwidth,height=\dimexpr\textheight-4pt\relax,keepaspectratio]{classification_typeII_embed.pdf}\hfil}\vfil}\par
\clearpage\thispagestyle{plain}
\pdfbookmark[2]{P43212 (96)}{table-typeII-sg96}
\noindent\vbox to\dimexpr\textheight-4pt\relax{\vfil\hbox to\textwidth{\hfil\includegraphics[page=109,trim=18mm 15mm 18mm 17mm,clip,width=\textwidth,height=\dimexpr\textheight-4pt\relax,keepaspectratio]{classification_typeII_embed.pdf}\hfil}\vfil}\par
\clearpage\thispagestyle{plain}
\pdfbookmark[2]{I422 (97)}{table-typeII-sg97}
\pdfbookmark[2]{I4122 (98)}{table-typeII-sg98}
\noindent\vbox to\dimexpr\textheight-4pt\relax{\vfil\hbox to\textwidth{\hfil\includegraphics[page=110,trim=18mm 15mm 18mm 17mm,clip,width=\textwidth,height=\dimexpr\textheight-4pt\relax,keepaspectratio]{classification_typeII_embed.pdf}\hfil}\vfil}\par
\clearpage\thispagestyle{plain}
\pdfbookmark[2]{P4mm (99)}{table-typeII-sg99}
\pdfbookmark[2]{P4bm (100)}{table-typeII-sg100}
\pdfbookmark[2]{P42cm (101)}{table-typeII-sg101}
\pdfbookmark[2]{P42nm (102)}{table-typeII-sg102}
\noindent\vbox to\dimexpr\textheight-4pt\relax{\vfil\hbox to\textwidth{\hfil\includegraphics[page=111,trim=18mm 15mm 18mm 17mm,clip,width=\textwidth,height=\dimexpr\textheight-4pt\relax,keepaspectratio]{classification_typeII_embed.pdf}\hfil}\vfil}\par
\clearpage\thispagestyle{plain}
\pdfbookmark[2]{P4cc (103)}{table-typeII-sg103}
\pdfbookmark[2]{P4nc (104)}{table-typeII-sg104}
\pdfbookmark[2]{P42mc (105)}{table-typeII-sg105}
\noindent\vbox to\dimexpr\textheight-4pt\relax{\vfil\hbox to\textwidth{\hfil\includegraphics[page=112,trim=18mm 15mm 18mm 17mm,clip,width=\textwidth,height=\dimexpr\textheight-4pt\relax,keepaspectratio]{classification_typeII_embed.pdf}\hfil}\vfil}\par
\clearpage\thispagestyle{plain}
\pdfbookmark[2]{P42bc (106)}{table-typeII-sg106}
\pdfbookmark[2]{I4mm (107)}{table-typeII-sg107}
\pdfbookmark[2]{I4cm (108)}{table-typeII-sg108}
\noindent\vbox to\dimexpr\textheight-4pt\relax{\vfil\hbox to\textwidth{\hfil\includegraphics[page=113,trim=18mm 15mm 18mm 17mm,clip,width=\textwidth,height=\dimexpr\textheight-4pt\relax,keepaspectratio]{classification_typeII_embed.pdf}\hfil}\vfil}\par
\clearpage\thispagestyle{plain}
\pdfbookmark[2]{I41md (109)}{table-typeII-sg109}
\pdfbookmark[2]{I41cd (110)}{table-typeII-sg110}
\pdfbookmark[2]{P-42m (111)}{table-typeII-sg111}
\noindent\vbox to\dimexpr\textheight-4pt\relax{\vfil\hbox to\textwidth{\hfil\includegraphics[page=114,trim=18mm 15mm 18mm 17mm,clip,width=\textwidth,height=\dimexpr\textheight-4pt\relax,keepaspectratio]{classification_typeII_embed.pdf}\hfil}\vfil}\par
\clearpage\thispagestyle{plain}
\pdfbookmark[2]{P-42c (112)}{table-typeII-sg112}
\pdfbookmark[2]{P-421m (113)}{table-typeII-sg113}
\noindent\vbox to\dimexpr\textheight-4pt\relax{\vfil\hbox to\textwidth{\hfil\includegraphics[page=115,trim=18mm 15mm 18mm 17mm,clip,width=\textwidth,height=\dimexpr\textheight-4pt\relax,keepaspectratio]{classification_typeII_embed.pdf}\hfil}\vfil}\par
\clearpage\thispagestyle{plain}
\pdfbookmark[2]{P-421c (114)}{table-typeII-sg114}
\noindent\vbox to\dimexpr\textheight-4pt\relax{\vfil\hbox to\textwidth{\hfil\includegraphics[page=116,trim=18mm 15mm 18mm 17mm,clip,width=\textwidth,height=\dimexpr\textheight-4pt\relax,keepaspectratio]{classification_typeII_embed.pdf}\hfil}\vfil}\par
\clearpage\thispagestyle{plain}
\pdfbookmark[2]{P-4m2 (115)}{table-typeII-sg115}
\pdfbookmark[2]{P-4c2 (116)}{table-typeII-sg116}
\noindent\vbox to\dimexpr\textheight-4pt\relax{\vfil\hbox to\textwidth{\hfil\includegraphics[page=117,trim=18mm 15mm 18mm 17mm,clip,width=\textwidth,height=\dimexpr\textheight-4pt\relax,keepaspectratio]{classification_typeII_embed.pdf}\hfil}\vfil}\par
\clearpage\thispagestyle{plain}
\pdfbookmark[2]{P-4b2 (117)}{table-typeII-sg117}
\noindent\vbox to\dimexpr\textheight-4pt\relax{\vfil\hbox to\textwidth{\hfil\includegraphics[page=118,trim=18mm 15mm 18mm 17mm,clip,width=\textwidth,height=\dimexpr\textheight-4pt\relax,keepaspectratio]{classification_typeII_embed.pdf}\hfil}\vfil}\par
\clearpage\thispagestyle{plain}
\pdfbookmark[2]{P-4n2 (118)}{table-typeII-sg118}
\noindent\vbox to\dimexpr\textheight-4pt\relax{\vfil\hbox to\textwidth{\hfil\includegraphics[page=119,trim=18mm 15mm 18mm 17mm,clip,width=\textwidth,height=\dimexpr\textheight-4pt\relax,keepaspectratio]{classification_typeII_embed.pdf}\hfil}\vfil}\par
\clearpage\thispagestyle{plain}
\pdfbookmark[2]{I-4m2 (119)}{table-typeII-sg119}
\pdfbookmark[2]{I-4c2 (120)}{table-typeII-sg120}
\noindent\vbox to\dimexpr\textheight-4pt\relax{\vfil\hbox to\textwidth{\hfil\includegraphics[page=120,trim=18mm 15mm 18mm 17mm,clip,width=\textwidth,height=\dimexpr\textheight-4pt\relax,keepaspectratio]{classification_typeII_embed.pdf}\hfil}\vfil}\par
\clearpage\thispagestyle{plain}
\pdfbookmark[2]{I-42m (121)}{table-typeII-sg121}
\pdfbookmark[2]{I-42d (122)}{table-typeII-sg122}
\pdfbookmark[2]{P4/mmm (123)}{table-typeII-sg123}
\noindent\vbox to\dimexpr\textheight-4pt\relax{\vfil\hbox to\textwidth{\hfil\includegraphics[page=121,trim=18mm 15mm 18mm 17mm,clip,width=\textwidth,height=\dimexpr\textheight-4pt\relax,keepaspectratio]{classification_typeII_embed.pdf}\hfil}\vfil}\par
\clearpage\thispagestyle{plain}
\pdfbookmark[2]{P4/mcc (124)}{table-typeII-sg124}
\pdfbookmark[2]{P4/nbm (125)}{table-typeII-sg125}
\pdfbookmark[2]{P4/nnc (126)}{table-typeII-sg126}
\pdfbookmark[2]{P4/mbm (127)}{table-typeII-sg127}
\pdfbookmark[2]{P4/mnc (128)}{table-typeII-sg128}
\noindent\vbox to\dimexpr\textheight-4pt\relax{\vfil\hbox to\textwidth{\hfil\includegraphics[page=122,trim=18mm 15mm 18mm 17mm,clip,width=\textwidth,height=\dimexpr\textheight-4pt\relax,keepaspectratio]{classification_typeII_embed.pdf}\hfil}\vfil}\par
\clearpage\thispagestyle{plain}
\pdfbookmark[2]{P4/nmm (129)}{table-typeII-sg129}
\pdfbookmark[2]{P4/ncc (130)}{table-typeII-sg130}
\pdfbookmark[2]{P42/mmc (131)}{table-typeII-sg131}
\pdfbookmark[2]{P42/mcm (132)}{table-typeII-sg132}
\noindent\vbox to\dimexpr\textheight-4pt\relax{\vfil\hbox to\textwidth{\hfil\includegraphics[page=123,trim=18mm 15mm 18mm 17mm,clip,width=\textwidth,height=\dimexpr\textheight-4pt\relax,keepaspectratio]{classification_typeII_embed.pdf}\hfil}\vfil}\par
\clearpage\thispagestyle{plain}
\pdfbookmark[2]{P42/nbc (133)}{table-typeII-sg133}
\pdfbookmark[2]{P42/nnm (134)}{table-typeII-sg134}
\pdfbookmark[2]{P42/mbc (135)}{table-typeII-sg135}
\pdfbookmark[2]{P42/mnm (136)}{table-typeII-sg136}
\noindent\vbox to\dimexpr\textheight-4pt\relax{\vfil\hbox to\textwidth{\hfil\includegraphics[page=124,trim=18mm 15mm 18mm 17mm,clip,width=\textwidth,height=\dimexpr\textheight-4pt\relax,keepaspectratio]{classification_typeII_embed.pdf}\hfil}\vfil}\par
\clearpage\thispagestyle{plain}
\pdfbookmark[2]{P42/nmc (137)}{table-typeII-sg137}
\pdfbookmark[2]{P42/ncm (138)}{table-typeII-sg138}
\pdfbookmark[2]{I4/mmm (139)}{table-typeII-sg139}
\pdfbookmark[2]{I4/mcm (140)}{table-typeII-sg140}
\noindent\vbox to\dimexpr\textheight-4pt\relax{\vfil\hbox to\textwidth{\hfil\includegraphics[page=125,trim=18mm 15mm 18mm 17mm,clip,width=\textwidth,height=\dimexpr\textheight-4pt\relax,keepaspectratio]{classification_typeII_embed.pdf}\hfil}\vfil}\par
\clearpage\thispagestyle{plain}
\pdfbookmark[2]{I41/amd (141)}{table-typeII-sg141}
\pdfbookmark[2]{I41/acd (142)}{table-typeII-sg142}
\pdfbookmark[2]{P3 (143)}{table-typeII-sg143}
\pdfbookmark[2]{P31 (144)}{table-typeII-sg144}
\noindent\vbox to\dimexpr\textheight-4pt\relax{\vfil\hbox to\textwidth{\hfil\includegraphics[page=126,trim=18mm 15mm 18mm 17mm,clip,width=\textwidth,height=\dimexpr\textheight-4pt\relax,keepaspectratio]{classification_typeII_embed.pdf}\hfil}\vfil}\par
\clearpage\thispagestyle{plain}
\pdfbookmark[2]{P32 (145)}{table-typeII-sg145}
\pdfbookmark[2]{R3 (146)}{table-typeII-sg146}
\noindent\vbox to\dimexpr\textheight-4pt\relax{\vfil\hbox to\textwidth{\hfil\includegraphics[page=127,trim=18mm 15mm 18mm 17mm,clip,width=\textwidth,height=\dimexpr\textheight-4pt\relax,keepaspectratio]{classification_typeII_embed.pdf}\hfil}\vfil}\par
\clearpage\thispagestyle{plain}
\pdfbookmark[2]{P-3 (147)}{table-typeII-sg147}
\pdfbookmark[2]{R-3 (148)}{table-typeII-sg148}
\noindent\vbox to\dimexpr\textheight-4pt\relax{\vfil\hbox to\textwidth{\hfil\includegraphics[page=128,trim=18mm 15mm 18mm 17mm,clip,width=\textwidth,height=\dimexpr\textheight-4pt\relax,keepaspectratio]{classification_typeII_embed.pdf}\hfil}\vfil}\par
\clearpage\thispagestyle{plain}
\pdfbookmark[2]{P312 (149)}{table-typeII-sg149}
\noindent\vbox to\dimexpr\textheight-4pt\relax{\vfil\hbox to\textwidth{\hfil\includegraphics[page=129,trim=18mm 15mm 18mm 17mm,clip,width=\textwidth,height=\dimexpr\textheight-4pt\relax,keepaspectratio]{classification_typeII_embed.pdf}\hfil}\vfil}\par
\clearpage\thispagestyle{plain}
\pdfbookmark[2]{P321 (150)}{table-typeII-sg150}
\pdfbookmark[2]{P3112 (151)}{table-typeII-sg151}
\pdfbookmark[2]{P3121 (152)}{table-typeII-sg152}
\noindent\vbox to\dimexpr\textheight-4pt\relax{\vfil\hbox to\textwidth{\hfil\includegraphics[page=130,trim=18mm 15mm 18mm 17mm,clip,width=\textwidth,height=\dimexpr\textheight-4pt\relax,keepaspectratio]{classification_typeII_embed.pdf}\hfil}\vfil}\par
\clearpage\thispagestyle{plain}
\pdfbookmark[2]{P3212 (153)}{table-typeII-sg153}
\pdfbookmark[2]{P3221 (154)}{table-typeII-sg154}
\pdfbookmark[2]{R32 (155)}{table-typeII-sg155}
\noindent\vbox to\dimexpr\textheight-4pt\relax{\vfil\hbox to\textwidth{\hfil\includegraphics[page=131,trim=18mm 15mm 18mm 17mm,clip,width=\textwidth,height=\dimexpr\textheight-4pt\relax,keepaspectratio]{classification_typeII_embed.pdf}\hfil}\vfil}\par
\clearpage\thispagestyle{plain}
\pdfbookmark[2]{P3m1 (156)}{table-typeII-sg156}
\noindent\vbox to\dimexpr\textheight-4pt\relax{\vfil\hbox to\textwidth{\hfil\includegraphics[page=132,trim=18mm 15mm 18mm 17mm,clip,width=\textwidth,height=\dimexpr\textheight-4pt\relax,keepaspectratio]{classification_typeII_embed.pdf}\hfil}\vfil}\par
\clearpage\thispagestyle{plain}
\pdfbookmark[2]{P31m (157)}{table-typeII-sg157}
\pdfbookmark[2]{P3c1 (158)}{table-typeII-sg158}
\pdfbookmark[2]{P31c (159)}{table-typeII-sg159}
\pdfbookmark[2]{R3m (160)}{table-typeII-sg160}
\noindent\vbox to\dimexpr\textheight-4pt\relax{\vfil\hbox to\textwidth{\hfil\includegraphics[page=133,trim=18mm 15mm 18mm 17mm,clip,width=\textwidth,height=\dimexpr\textheight-4pt\relax,keepaspectratio]{classification_typeII_embed.pdf}\hfil}\vfil}\par
\clearpage\thispagestyle{plain}
\pdfbookmark[2]{R3c (161)}{table-typeII-sg161}
\pdfbookmark[2]{P-31m (162)}{table-typeII-sg162}
\pdfbookmark[2]{P-31c (163)}{table-typeII-sg163}
\noindent\vbox to\dimexpr\textheight-4pt\relax{\vfil\hbox to\textwidth{\hfil\includegraphics[page=134,trim=18mm 15mm 18mm 17mm,clip,width=\textwidth,height=\dimexpr\textheight-4pt\relax,keepaspectratio]{classification_typeII_embed.pdf}\hfil}\vfil}\par
\clearpage\thispagestyle{plain}
\pdfbookmark[2]{P-3m1 (164)}{table-typeII-sg164}
\pdfbookmark[2]{P-3c1 (165)}{table-typeII-sg165}
\pdfbookmark[2]{R-3m (166)}{table-typeII-sg166}
\noindent\vbox to\dimexpr\textheight-4pt\relax{\vfil\hbox to\textwidth{\hfil\includegraphics[page=135,trim=18mm 15mm 18mm 17mm,clip,width=\textwidth,height=\dimexpr\textheight-4pt\relax,keepaspectratio]{classification_typeII_embed.pdf}\hfil}\vfil}\par
\clearpage\thispagestyle{plain}
\pdfbookmark[2]{R-3c (167)}{table-typeII-sg167}
\pdfbookmark[2]{P6 (168)}{table-typeII-sg168}
\pdfbookmark[2]{P61 (169)}{table-typeII-sg169}
\noindent\vbox to\dimexpr\textheight-4pt\relax{\vfil\hbox to\textwidth{\hfil\includegraphics[page=136,trim=18mm 15mm 18mm 17mm,clip,width=\textwidth,height=\dimexpr\textheight-4pt\relax,keepaspectratio]{classification_typeII_embed.pdf}\hfil}\vfil}\par
\clearpage\thispagestyle{plain}
\pdfbookmark[2]{P65 (170)}{table-typeII-sg170}
\pdfbookmark[2]{P62 (171)}{table-typeII-sg171}
\pdfbookmark[2]{P64 (172)}{table-typeII-sg172}
\pdfbookmark[2]{P63 (173)}{table-typeII-sg173}
\pdfbookmark[2]{P-6 (174)}{table-typeII-sg174}
\noindent\vbox to\dimexpr\textheight-4pt\relax{\vfil\hbox to\textwidth{\hfil\includegraphics[page=137,trim=18mm 15mm 18mm 17mm,clip,width=\textwidth,height=\dimexpr\textheight-4pt\relax,keepaspectratio]{classification_typeII_embed.pdf}\hfil}\vfil}\par
\clearpage\thispagestyle{plain}
\pdfbookmark[2]{P6/m (175)}{table-typeII-sg175}
\pdfbookmark[2]{P63/m (176)}{table-typeII-sg176}
\pdfbookmark[2]{P622 (177)}{table-typeII-sg177}
\pdfbookmark[2]{P6122 (178)}{table-typeII-sg178}
\pdfbookmark[2]{P6522 (179)}{table-typeII-sg179}
\pdfbookmark[2]{P6222 (180)}{table-typeII-sg180}
\pdfbookmark[2]{P6422 (181)}{table-typeII-sg181}
\noindent\vbox to\dimexpr\textheight-4pt\relax{\vfil\hbox to\textwidth{\hfil\includegraphics[page=138,trim=18mm 15mm 18mm 17mm,clip,width=\textwidth,height=\dimexpr\textheight-4pt\relax,keepaspectratio]{classification_typeII_embed.pdf}\hfil}\vfil}\par
\clearpage\thispagestyle{plain}
\pdfbookmark[2]{P6322 (182)}{table-typeII-sg182}
\pdfbookmark[2]{P6mm (183)}{table-typeII-sg183}
\pdfbookmark[2]{P6cc (184)}{table-typeII-sg184}
\pdfbookmark[2]{P63cm (185)}{table-typeII-sg185}
\pdfbookmark[2]{P63mc (186)}{table-typeII-sg186}
\pdfbookmark[2]{P-6m2 (187)}{table-typeII-sg187}
\pdfbookmark[2]{P-6c2 (188)}{table-typeII-sg188}
\noindent\vbox to\dimexpr\textheight-4pt\relax{\vfil\hbox to\textwidth{\hfil\includegraphics[page=139,trim=18mm 15mm 18mm 17mm,clip,width=\textwidth,height=\dimexpr\textheight-4pt\relax,keepaspectratio]{classification_typeII_embed.pdf}\hfil}\vfil}\par
\clearpage\thispagestyle{plain}
\pdfbookmark[2]{P-62m (189)}{table-typeII-sg189}
\pdfbookmark[2]{P-62c (190)}{table-typeII-sg190}
\pdfbookmark[2]{P6/mmm (191)}{table-typeII-sg191}
\pdfbookmark[2]{P6/mcc (192)}{table-typeII-sg192}
\pdfbookmark[2]{P63/mcm (193)}{table-typeII-sg193}
\pdfbookmark[2]{P63/mmc (194)}{table-typeII-sg194}
\pdfbookmark[2]{P23 (195)}{table-typeII-sg195}
\noindent\vbox to\dimexpr\textheight-4pt\relax{\vfil\hbox to\textwidth{\hfil\includegraphics[page=140,trim=18mm 15mm 18mm 17mm,clip,width=\textwidth,height=\dimexpr\textheight-4pt\relax,keepaspectratio]{classification_typeII_embed.pdf}\hfil}\vfil}\par
\clearpage\thispagestyle{plain}
\pdfbookmark[2]{F23 (196)}{table-typeII-sg196}
\pdfbookmark[2]{I23 (197)}{table-typeII-sg197}
\pdfbookmark[2]{P213 (198)}{table-typeII-sg198}
\pdfbookmark[2]{I213 (199)}{table-typeII-sg199}
\pdfbookmark[2]{Pm-3 (200)}{table-typeII-sg200}
\noindent\vbox to\dimexpr\textheight-4pt\relax{\vfil\hbox to\textwidth{\hfil\includegraphics[page=141,trim=18mm 15mm 18mm 17mm,clip,width=\textwidth,height=\dimexpr\textheight-4pt\relax,keepaspectratio]{classification_typeII_embed.pdf}\hfil}\vfil}\par
\clearpage\thispagestyle{plain}
\pdfbookmark[2]{Pn-3 (201)}{table-typeII-sg201}
\pdfbookmark[2]{Fm-3 (202)}{table-typeII-sg202}
\pdfbookmark[2]{Fd-3 (203)}{table-typeII-sg203}
\pdfbookmark[2]{Im-3 (204)}{table-typeII-sg204}
\pdfbookmark[2]{Pa-3 (205)}{table-typeII-sg205}
\pdfbookmark[2]{Ia-3 (206)}{table-typeII-sg206}
\pdfbookmark[2]{P432 (207)}{table-typeII-sg207}
\pdfbookmark[2]{P4232 (208)}{table-typeII-sg208}
\noindent\vbox to\dimexpr\textheight-4pt\relax{\vfil\hbox to\textwidth{\hfil\includegraphics[page=142,trim=18mm 15mm 18mm 17mm,clip,width=\textwidth,height=\dimexpr\textheight-4pt\relax,keepaspectratio]{classification_typeII_embed.pdf}\hfil}\vfil}\par
\clearpage\thispagestyle{plain}
\pdfbookmark[2]{F432 (209)}{table-typeII-sg209}
\pdfbookmark[2]{F4132 (210)}{table-typeII-sg210}
\pdfbookmark[2]{I432 (211)}{table-typeII-sg211}
\pdfbookmark[2]{P4332 (212)}{table-typeII-sg212}
\pdfbookmark[2]{P4132 (213)}{table-typeII-sg213}
\pdfbookmark[2]{I4132 (214)}{table-typeII-sg214}
\pdfbookmark[2]{P-43m (215)}{table-typeII-sg215}
\pdfbookmark[2]{F-43m (216)}{table-typeII-sg216}
\pdfbookmark[2]{I-43m (217)}{table-typeII-sg217}
\noindent\vbox to\dimexpr\textheight-4pt\relax{\vfil\hbox to\textwidth{\hfil\includegraphics[page=143,trim=18mm 15mm 18mm 17mm,clip,width=\textwidth,height=\dimexpr\textheight-4pt\relax,keepaspectratio]{classification_typeII_embed.pdf}\hfil}\vfil}\par
\clearpage\thispagestyle{plain}
\pdfbookmark[2]{P-43n (218)}{table-typeII-sg218}
\pdfbookmark[2]{F-43c (219)}{table-typeII-sg219}
\pdfbookmark[2]{I-43d (220)}{table-typeII-sg220}
\pdfbookmark[2]{Pm-3m (221)}{table-typeII-sg221}
\pdfbookmark[2]{Pn-3n (222)}{table-typeII-sg222}
\pdfbookmark[2]{Pm-3n (223)}{table-typeII-sg223}
\pdfbookmark[2]{Pn-3m (224)}{table-typeII-sg224}
\pdfbookmark[2]{Fm-3m (225)}{table-typeII-sg225}
\pdfbookmark[2]{Fm-3c (226)}{table-typeII-sg226}
\pdfbookmark[2]{Fd-3m (227)}{table-typeII-sg227}
\pdfbookmark[2]{Fd-3c (228)}{table-typeII-sg228}
\pdfbookmark[2]{Im-3m (229)}{table-typeII-sg229}
\pdfbookmark[2]{Ia-3d (230)}{table-typeII-sg230}
\noindent\vbox to\dimexpr\textheight-4pt\relax{\vfil\hbox to\textwidth{\hfil\includegraphics[page=144,trim=18mm 15mm 18mm 17mm,clip,width=\textwidth,height=\dimexpr\textheight-4pt\relax,keepaspectratio]{classification_typeII_embed.pdf}\hfil}\vfil}\par